% ...
\documentclass[aps, prc, tightenlines,superscriptaddress, letterpaper, amsmath, amssymb, preprintnumbers, floatfix, longbibliography, nofootinbib]{revtex4-2}
\usepackage[T1]{fontenc}
\usepackage{bm}
\usepackage{xspace}
\usepackage[dvipsnames]{xcolor}
\usepackage{graphicx}
\usepackage{tabularx}
\usepackage{multirow}
\usepackage{enumitem}
\usepackage{qcircuit}
\usepackage{braket}
\usepackage{dsfont}
\usepackage{placeins}
\usepackage[normalem]{ulem}
\usepackage{soul}
\usepackage{diagbox}
\usepackage{physics}
\usepackage{amsmath, nccmath}
\usepackage{lipsum}
\usepackage{cancel}
\usepackage{bm}
\usepackage{listings}

\lstdefinelanguage{Python}{
  keywords={for, typeof, new, true, false, catch, function, return, null, catch, switch, var, if, in, while, do, else, case, break},
  ndkeywords={class, export, boolean, throw, implements, import, this},
  sensitive=false,
  comment=[l]{//},
  morecomment=[s]{/*}{*/},
  morestring=[b]',
  morestring=[b]"
}

\usepackage[hypertexnames=false]{hyperref}
\hypersetup{
    colorlinks=true,       
    linkcolor=blue,          
    citecolor=blue,        
    filecolor=blue,      
    urlcolor=blue           
}

\newcolumntype{Y}{>{\centering\arraybackslash}X}

\usepackage{mathtools}
\usepackage{orcidlink}

\definecolor{lavenderindigo}{rgb}{0.58, 0.34, 0.92}

\usepackage{stackengine}
\stackMath
\newcommand\tenq[2][1]{%
 \def\useanchorwidth{T}%
  \ifnum#1>1%
    \stackunder[0pt]{\tenq[\numexpr#1-1\relax]{#2}}{\scriptscriptstyle\sim}%
  \else%
    \stackunder[1pt]{#2}{\scriptscriptstyle\sim}%
  \fi%
}

\newcommand\scalemath[2]{\scalebox{#1}{\mbox{\ensuremath{\displaystyle #2}}}}

\begin{document}

\begin{figure}
\vskip -1.cm
\leftline{\includegraphics[width=0.15\textwidth]{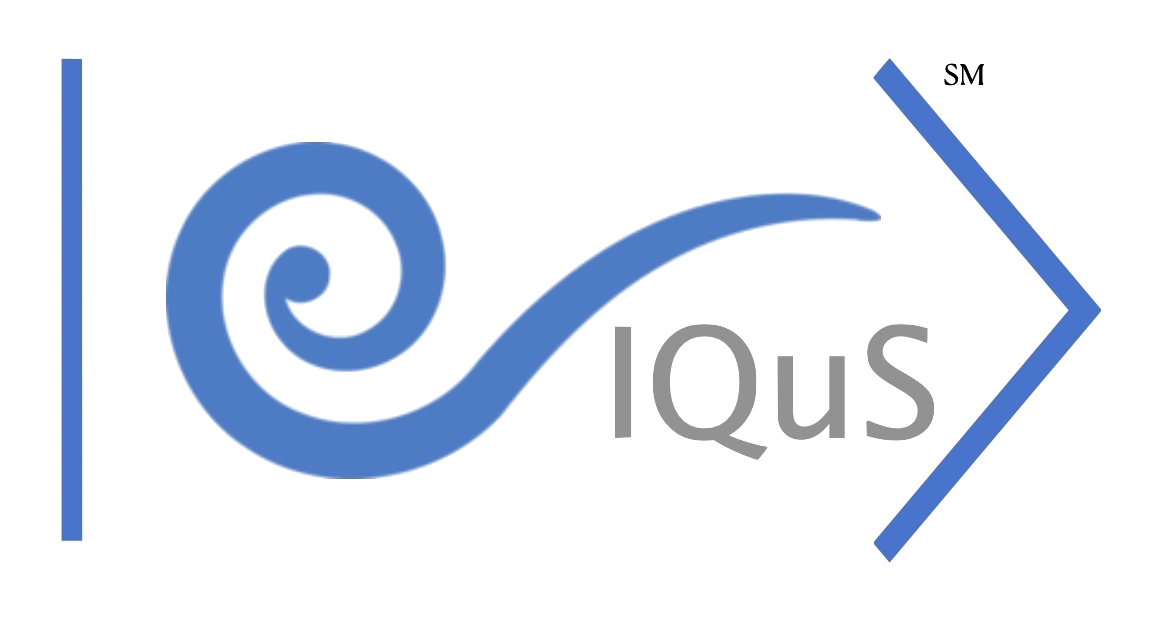}}
\vskip -0.5cm
\end{figure}

\title{Quantum Simulations of Hadron Dynamics in the Schwinger Model using 112 Qubits}

\author{Roland C.~Farrell\,\orcidlink{0000-0001-7189-0424
}}
\email{rolanf2@uw.edu}
\affiliation{InQubator for Quantum Simulation (IQuS), Department of Physics, University of Washington, Seattle, WA 98195, USA.}
\author{Marc Illa\,\orcidlink{0000-0003-3570-2849}}
\email{marcilla@uw.edu}
\affiliation{InQubator for Quantum Simulation (IQuS), Department of Physics, University of Washington, Seattle, WA 98195, USA.}
\author{Anthony N. Ciavarella\,\orcidlink{0000-0003-3918-4110}}
\email{ANCiavarella@lbl.gov}
\affiliation{InQubator for Quantum Simulation (IQuS), Department of Physics, University of Washington, Seattle, WA 98195, USA.}
\affiliation{Physics Division, Lawrence Berkeley National Laboratory, Berkeley, California 94720, USA}
\author{Martin J.~Savage\,\orcidlink{0000-0001-6502-7106}}
\email{mjs5@uw.edu}
\thanks{On leave from the Institute for Nuclear Theory.}
\affiliation{InQubator for Quantum Simulation (IQuS), Department of Physics, University of Washington, Seattle, WA 98195, USA.}

\preprint{IQuS@UW-21-069, NT@UW-24-1}
\date{\today}

\begin{abstract}
\noindent
Hadron wavepackets are prepared and time evolved in the Schwinger model using 112 qubits of IBM's 133-qubit Heron quantum computer {\tt ibm\_torino}.
The initialization of the hadron wavepacket is performed in two steps.
First, the vacuum is prepared across the whole lattice using the recently developed SC-ADAPT-VQE algorithm and workflow.
SC-ADAPT-VQE is then extended to the preparation of localized states, and used to establish a hadron wavepacket on top of the vacuum.
This is done by adaptively constructing low-depth circuits that maximize the overlap with an adiabatically prepared hadron wavepacket.
Due to the localized nature of the wavepacket, these circuits can be determined on a sequence of small lattices using {\it classical computers}, and then robustly scaled to prepare wavepackets on large lattices for simulations using {\it quantum computers}.
Time evolution is implemented with a second-order Trotterization.
To reduce both the required qubit connectivity and circuit depth, an approximate quasi-local interaction is introduced.
This approximation is made possible by the emergence of confinement at long distances, and converges exponentially with increasing distance of the interactions.
Using multiple error-mitigation strategies, up to 14 Trotter steps of time evolution are performed, employing 13,858 two-qubit gates (with a CNOT depth of 370).
The propagation of hadrons is clearly identified, with results that compare favorably with Matrix Product State simulations.
Prospects for a near-term quantum advantage in simulations of hadron scattering are discussed.
\end{abstract}

\maketitle
\newpage{}
\tableofcontents
\newpage{}

\section{Introduction}
\label{sec:intro}
\noindent
The highest-energy collisions of particles, such as those that take place in colliders and cosmic-ray events, reveal and provide insights into the underlying laws of nature.
They tighten constraints on the content, symmetries and parameters of the Standard Model (SM)~\cite{Glashow:1961tr,Higgs:1964pj,Weinberg:1967tq,Salam:1968rm,Politzer:1973fx,Gross:1973id}, and provide opportunities to discover what may lie beyond.
In searching for new physics and emergent phenomena in exotic states of matter, contributions from known physics must be reliably predicted with a complete quantification of  uncertainties.
The associated complexities, particularly from the strong interactions described by quantum chromodynamics (QCD), provide challenges for phenomenological modeling and classical simulation. 
Many forefront research questions in nuclear and particle physics require simulations of systems of fundamental particles that lie far beyond the capabilities of classical computing.

In principle, the collisions of fundamental and composite particles (hadrons)
could be simulated, from the initial state through to the final state(s), with sufficiently capable quantum computers (for recent reviews, see e.g., Refs.~\cite{Banuls:2019bmf,Guan:2020bdl,Klco:2021lap,Delgado:2022tpc,Bauer:2022hpo,Bauer:2023qgm,Beck:2023xhh,DiMeglio:2023nsa}).
Well before that point, new insights and improvements in predictions for such processes may come from NISQ-era devices~\cite{Preskill:2018jim}.
Progress is beginning to be made toward these objectives, with current focus on advancing low-dimensional models of QCD and the electroweak sector.
The Schwinger model~\cite{Schwinger:1962tp} has emerged as one of the early workhorses for this effort.
It is $U(1)$ electromagnetism in one spatial dimension, 1+1D, and is a simplified model of QCD as it exhibits confinement, has a chiral condensate, and a spectrum of hadrons that bind to form nuclei. 
Soon after early proposals for quantum simulations of lattice gauge theories appeared~\cite{IgnacioCirac:2010us,Bermudez:2010da,Boada:2010sh,Zohar:2012ay,Tagliacozzo:2012vg,Banerjee:2012pg,Zohar:2012ts,Tagliacozzo:2012df,Zohar:2013zla,Marcos:2013aya,Hauke:2013jga},
the first digital quantum simulation of the Schwinger model using 
four-qubits of a trapped-ion quantum computer~\cite{Martinez:2016yna,Muschik:2016tws} was performed.
Since then, there has been considerable progress in simulating $U(1)$, $SU(2)$ and $SU(3)$ lattice gauge theories, predominantly on small lattices in 1+1D and 2+1D using quantum devices~\cite{Martinez:2016yna,Klco:2018kyo,Kokail:2018eiw,Lu:2018pjk,Klco:2019evd,Surace:2019dtp,Mil:2019pbt,Yang:2020yer,Bauer:2021gup,Zhou:2021kdl,Nguyen:2021hyk,Ciavarella:2021nmj,Gong:2021bcp,ARahman:2021ktn,Mazzola:2021hma,deJong:2021wsd,Riechert:2021ink,Atas:2021ext,Ciavarella:2021lel,Mildenberger:2022jqr,Ciavarella:2022zhe,ARahman:2022tkr,Asaduzzaman:2022bpi,Mueller:2022xbg,Farrell:2022wyt,Atas:2022dqm,Farrell:2022vyh,Charles:2023zbl,Pomarico:2023png, PhysRevResearch.5.023010,zhang2023observation, Ciavarella:2023mfc,Borzenkova:2023xaf,Schuster:2023klj,Angelides:2023noe}, classical simulations~\cite{Yang:2016hjn,Muschik:2016tws,Kasper:2016mzj,Notarnicola:2019wzb,Davoudi:2019bhy,Avkhadiev:2019niu,Verdel:2019chj,Luo:2019vmi,Chakraborty:2020uhf,Tran:2020azk,Ferguson:2020qyf,Surace:2020ycc,Karpov:2020pqe,Davoudi:2021ney,Yamamoto:2021vxp,Honda:2021aum,Bennewitz:2021jqi,Shen:2021zrg,Andrade:2021pil,Honda:2021ovk,Jensen:2022hyu,Vovrosh:2022bpj,Halimeh:2022pkw,Xie:2022jgj,Davoudi:2022uzo,Avkhadiev:2022ttx,Florio:2023dke,Nagano:2023uaq,Ikeda:2023zil,Nagano:2023kge,Popov:2023xft,Lee:2023urk,Oshima:2023rmj,Meth:2023wzd,Vary:2023ihk,Sakamoto:2023cxs,Chai:2023qpq} and tensor-networks~\cite{Byrnes:2002nv,Banuls:2013jaa,Rico:2013qya,Buyens:2013yza,Kuhn:2014rha,Banuls:2015sta,Pichler:2015yqa,Kuhn:2015zqa,Buyens:2015tea,Banuls:2016lkq,Buyens:2016hhu,Zapp:2017fcr,Ercolessi:2017jbi,Sala:2018dui,Damme:2019rts,Funcke:2019zna,Magnifico:2019kyj,Butt:2019uul,Milsted:2020jmf,Rigobello:2021fxw,Okuda:2022hsq,Honda:2022edn,Desaules:2023ghs,Angelides:2023bme,Belyansky:2023rgh,Rigobello:2023ype,Barata:2023jgd,Hayata:2023pkw,Su:2024uuc}, with exploratory efforts in 3+1D~\cite{Ciavarella:2023mfc}.
Included in this progress are the first efforts toward fragmentation and hadronization~\cite{Bauer:2021gup,Florio:2023dke}, as well as wavepacket initialization and evolution~\cite{Kuhn:2015zqa,Pichler:2015yqa,Damme:2019rts,Avkhadiev:2019niu,Surace:2020ycc,Karpov:2020pqe,Milsted:2020jmf,Vovrosh:2022bpj,Asaduzzaman:2022bpi,Avkhadiev:2022ttx,Farrell:2022vyh,Atas:2022dqm,Vary:2023ihk,Chai:2023qpq,Belyansky:2023rgh}.
The parallel development of quantum computers, software and algorithms has led to the first ``utility-scale'' quantum simulations being performed late last year~\cite{Yu:2022ivm,Kim:2023bwr,Shtanko:2023tjn,Farrell:2023fgd,Baumer:2023vrf,Chen:2023tfg,Liao:2023eug,Bluvstein:2023zmt,Chowdhury:2023vbk}.
Progress continues to be driven, to a significant degree, by technology company road maps for developing quantum computing architectures that are accessible to researchers (see, for example, Refs.~\cite{ibmweb,quantinuumweb,ionqweb,queraweb}).

In this work, the real-time dynamics of composite particles, ``hadrons'', in the lattice Schwinger model are simulated using IBM's superconducting-qubit quantum computers. 
This work serves as a proof-of-concept, and builds toward future simulations that will probe highly-inelastic scattering of hadrons and out-of-equilibrium behavior of strongly interacting matter.
Our quantum simulations proceed with the following steps:\footnote{As this work was being completed, similar developments in the Thirring model were reported in Ref.~\cite{Chai:2023qpq}.}
\begin{enumerate}
    \item Prepare the interacting ground state (vacuum);
    \item Establish a localized hadron wavepacket on this vacuum;
    \item Evolve the system forward in time, allowing the hadrons to propagate;
    \item Measure observables in the final state that detect hadron propagation.
\end{enumerate}
Crucial to the success of our quantum simulations is the development of comprehensive suites of scalable techniques that minimize circuit depth and two-qubit entangling gate counts. 
The methods presented here are informed by the symmetries and phenomenological features of the Schwinger model.
They are physics-aware techniques with potential applicability to a broad class of lattice theories.

\begin{figure}[!tb]
    \centering
    \includegraphics[width=0.95\columnwidth]{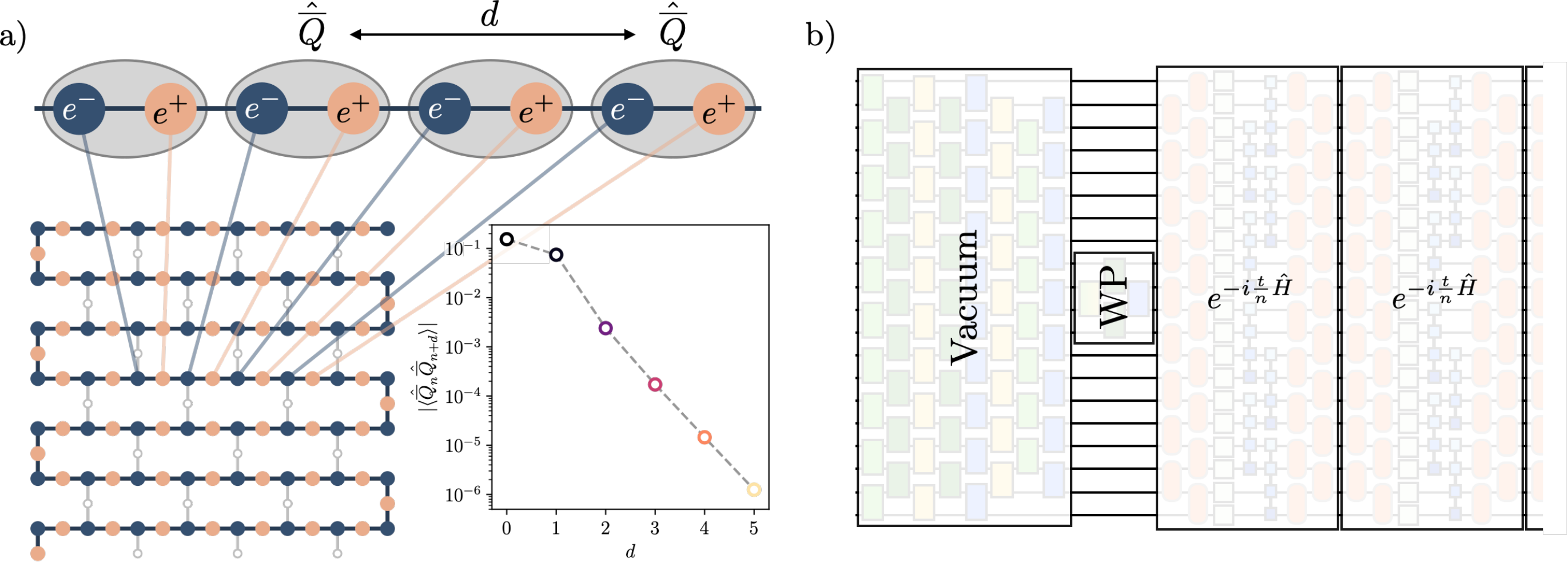}
    \caption{a) Mapping the $L=56$ lattice onto the qubits of IBM's quantum computer {\tt ibm\_torino} (bottom left). 
    The dynamical re-arrangement of charges in the vacuum screens the interactions between electric charges in the Schwinger model, giving rise to an exponential decay of correlations between spatial-site charges, $\langle \hat{\overline{Q}}_n \hat{\overline{Q}}_{n+d}\rangle$ (top and bottom right). 
    b) The charge screening informs an efficient construction of the quantum circuits used to simulate hadron dynamics.
    SC-ADAPT-VQE is used to prepare the vacuum and wavepacket, which are time-evolved using Trotterized circuits implementing $e^{-i t \hat{H} }$ 
    with a truncated electric interaction.
    }
    \label{fig:SimTricks}
\end{figure}

A significant challenge to performing quantum simulations of the Schwinger model is that, in axial gauge ($A_x=0$)~\cite{Farrell:2022wyt}, the electric interaction between fermions is all-to-all.\footnote{Working in Weyl gauge ($A_t=0$) eliminates the need for all-to-all connectivity, but requires additional qubits to encode the gauge field on the links of the lattice.}
This leads to an $\mathcal{O}(L^2)$ scaling in the number of quantum gates required for time evolution, where $L$ is the lattice volume.
It also requires quantum computers to have all-to-all connectivity between qubits for efficient simulation, a native feature in current
trapped-ion devices, but which has a large overhead on superconducting devices.
Fortunately, electric charges are screened in the Schwinger model, causing correlations between distant fermions to decay exponentially with separation; see Fig.~\ref{fig:SimTricks}a).
In Sec.~\ref{sec:HamTruncs}, this screening is used to truncate interactions between fermions beyond a distance, $\overline{\lambda}$, set by the correlation length and the desired level of precision of the simulation.
This improves the scaling of the number of gates required for time evolution to $\mathcal{O}(\overline{\lambda} L)$, with $\mathcal{O}(\overline{\lambda})$-nearest neighbor qubit connectivity.

The construction of low-depth quantum circuits for state preparation is another challenge addressed in this work.
In Ref.~\cite{Farrell:2023fgd}, building upon ADAPT-VQE~\cite{Grimsley_2019}, we introduced the SC-ADAPT-VQE algorithm, and applied it to the preparation of the Schwinger model vacuum on 100 qubits of {\tt ibm\_cusco}.
SC-ADAPT-VQE uses symmetries and hierarchies in length scales to determine low-depth quantum circuits for state preparation.
Using a hybrid workflow, quantum circuits are determined and optimized on a series of small and modest-sized systems using {\it classical computers}, and then systematically scaled to large systems to be executed on a {\it quantum computer}.
In Sec.~\ref{sec:SCADAPT}, SC-ADAPT-VQE is extended to the preparation of localized states, and used to establish a hadron wavepacket on top of the interacting vacuum; see Fig.~\ref{fig:SimTricks}b).
The wavepacket preparation circuits are optimized on a series of a small lattices by maximizing the overlap with an adiabatically prepared wavepacket.
The locality of the target state ensures that these circuits can be systematically extrapolated to prepare hadron wavepackets on large lattices.

Quantum 
circuits for state preparation and time evolution are developed in Sec.~\ref{sec:ClSim}.
The circuit design minimizes the two-qubit gate count for implementation on devices with nearest-neighbor connectivity, such as those available from IBM.
A building block for these circuits is a new gate decomposition for $R_{ZZ}$ rotations acting between all pairs of a set of qubits.
This nearest-neighbor decomposition uses the same number of two-qubit gates as decompositions for devices with all-to-all connectivity, at the cost of an increased circuit depth.
Results from classical simulations performed on small lattices are presented in Sec.~\ref{sec:Systematics}.
These simulations quantify the systematic errors originating from the approximations introduced in previous sections: preparation of the hadron wavepacket with SC-ADAPT-VQE, use of a truncated Hamiltonian for time evolution, and 
Trotterization of the time evolution operator.

In Sec.~\ref{sec:Qsim}, the techniques and ideas described in the previous paragraphs
are applied to quantum simulations of hadron dynamics on $L=56$ (112 qubit) lattices using IBM's quantum computer {\tt ibm\_torino}.
The initial state is prepared using SC-ADAPT-VQE, and time evolution is implemented with up to 14 Trotter steps, requiring 13,858 CNOTs (CNOT depth 370).
After applying a suite of error mitigation techniques, measurements of the local chiral condensate show clear signatures of hadron propagation.
The results obtained from {\tt ibm\_torino} are compared to classical simulations using the {\tt cuQuantum} Matrix Product State (MPS) simulator.
In these latter calculations, the bond dimension in the tensor network simulations grows with the simulation time, requiring increased classical computing overhead. 
Appendix~\ref{app:MPSSim} provides details about
the convergence of the MPS simulations, 
and App.~\ref{app:qSimDetails} provides details of
our error mitigation strategy, for our simulations using 112 qubits of IBM's quantum computers.
This work points to quantum simulations of more complex processes, such as inelastic collisions, fragmentation and hadronization, as being strong candidates for a near-term quantum advantage.

\section{Systematic Truncation of the Electric Interactions}
\label{sec:HamTruncs}
\noindent
The Schwinger model is quantum electrodynamics in 1+1D, the theory of electrons and positrons interacting via photon exchange.
In 1+1D, the photon is not a dynamical degree of freedom, as it is completely constrained by Gauss's law. 
As a result, the photon can be removed as an independent field, leaving a system of fermions interacting through a linear Coulomb potential.
In axial gauge with open boundary conditions (OBCs), zero background electric field, and using the Jordan-Wigner (JW) mapping, the Schwinger model Hamiltonian on a lattice with $L$ spatial sites ($2L$ staggered sites) is given by~\cite{Kogut:1974ag,Banks:1975gq}
\begin{align}
\hat H & \ =\  \hat H_m + \hat H_{kin} + \hat H_{el} \ = \ \frac{m}{ 2}\sum_{j=0}^{2L-1}\ \left[ (-1)^j \hat Z_j + \hat{I} \right] \ + \ \frac{1}{2}\sum_{j=0}^{2L-2}\ \left( \hat \sigma^+_j \hat\sigma^-_{j+1} + {\rm h.c.} \right) \ + \ \frac{g^2}{ 2}\sum_{j=0}^{2L-2}\bigg (\sum_{k\leq j} \hat Q_k \bigg )^2 
\ ,
\nonumber \\ 
\hat Q_k & \ = \ -\frac{1}{2}\left[ \hat Z_k + (-1)^k\hat{I} \right] \ .
\label{eq:Hgf}
\end{align}
The (bare) mass and coupling are $m$ and $g$, respectively, and the staggered lattice spacing has been set to one.
Due to the non-perturbative mechanism of confinement, all low-energy states (the vacuum and hadrons) have charge zero. 
The parameters $m=0.5,g=0.3$, which give rise to a mass of $m_{\text{hadron}}\approx1.1$ for the lowest-lying (vector) hadron, will be used throughout this work.
The conserved quantities and symmetries of this system are total charge, $\hat{Q} = \sum_{k} \hat{Q}_k$, time reversal and, due to the CPT theorem, the composition of charge conjugation and parity (CP).\footnote{The CP symmetry is realized in the $Q=0$ sector as the composition of a spin-flip and a reflection through the mid-point of the lattice.}

Due to the removal of the gauge degrees of freedom, the electric interactions are pair-wise between all of the fermions.
This is problematic for implementing time evolution $e^{-i t \hat{H}}$ on a quantum computer as it implies an $\mathcal{O}(L^2)$ scaling in the number of gates.
In addition, this interaction requires connectivity between every pair of qubits for efficient implementation.
Fortunately, charges are screened in  confining theories like the Schwinger model, and correlation functions decay exponentially between charges separated by more than approximately a correlation length, $\xi$.
The correlation length is a scale that emerges from the solution of the theory, and is naturally related to the hadron mass, $\xi \sim 1/m_{\text{hadron}}$.
This motivates the construction of an effective Hamiltonian where interactions between distant charges are removed.
Such an effective interaction is systematically improvable with exponentially suppressed errors, and only requires $\mathcal{O}(\xi L)$ gates acting between qubits with maximum separation $\sim\xi$.

To form the effective interactions, it is beneficial to first specialize to the $Q=0$ sector with zero background electric field.
There are many equivalent ways to express the interaction due to the freedom of integrating Gauss's law from the left or right side of the lattice 
when constraining the electric field.
However, the desire to preserve CP symmetry in the truncated theory motivates starting from a manifestly CP-symmetric interaction,
\begin{align}
\hat{H}_{el}^{(Q=0)} \ &= \ \frac{g^2}{2} \left \{\sum_{j=0}^{L-2}\left ( \sum_{k=0}^j \hat{Q}_k\right )^2\ + \ \sum_{j=L+1}^{2L-1}\left ( \sum_{k=j}^{2L-1} \hat{Q}_k\right )^2 \ + \ \frac{1}{2}\left [\left (\sum_{j=0}^{L-1} \hat{Q}_j \right )^2 + \left (\sum_{j=L}^{2L-1} \hat{Q}_j \right )^2 \right ] \right \}
\ .
\label{eq:HelCP}
\end{align}
This has decoupled the interactions between charges on different halves of the lattice.
The most straightforward way to form the effective interactions would be to remove $\hat{Q}_j \hat{Q}_{j+d}$ terms with $d\gtrsim \xi$.
However, this is ineffective because it is only the {\it connected} correlations that decay exponentially; on a staggered lattice, $\langle \hat{Q}_j \rangle \neq 0$ and $\langle \hat{Q}_j \hat{Q}_{j+d}\rangle = \langle \hat{Q}_j \rangle \langle \hat{Q}_{j+d} \rangle + {\cal O}(e^{-d/\xi})$.

\begin{figure}[!tb]
    \centering
    \includegraphics[width=\textwidth]{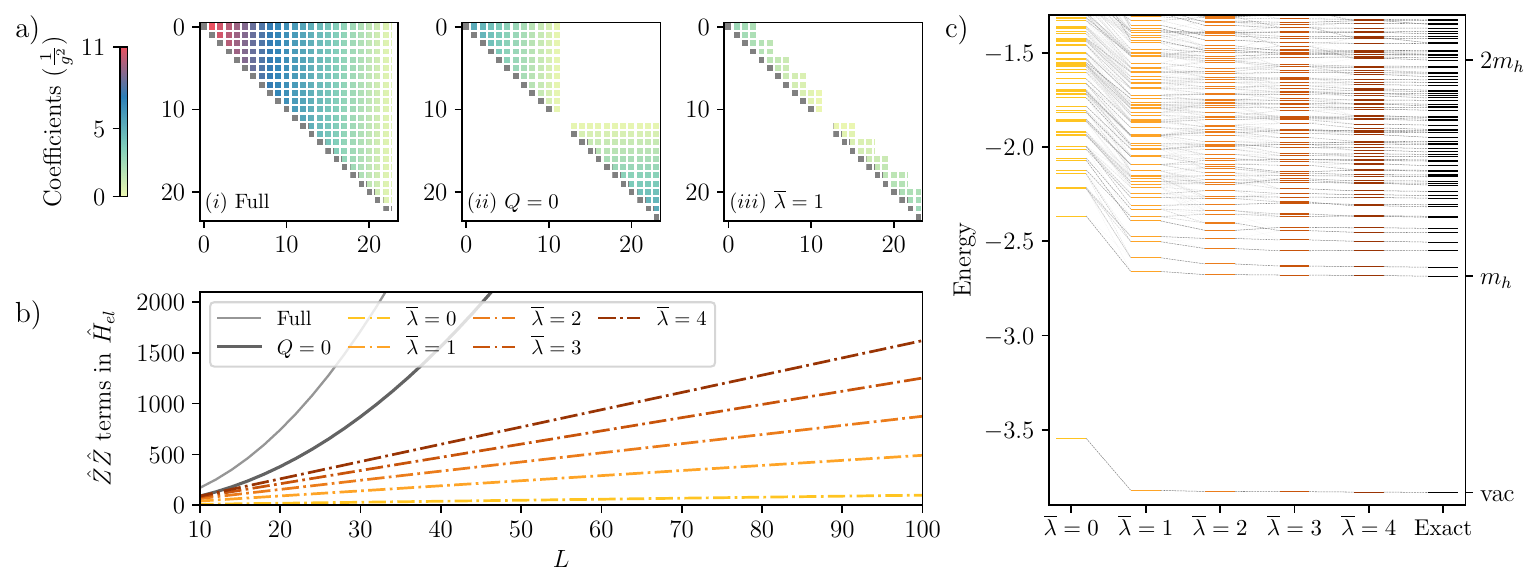}
    \caption{a) The qubit coupling matrix for select electric Hamiltonians with $L=12$:  $(i)$ shows the coupling matrix without truncation [Eq.~\eqref{eq:Hgf}],  
    $(ii)$ shows the impact of restricting to the $Q=0$ sector [Eq.~\eqref{eq:HelCP}], and 
    $(iii)$ corresponds to additionally truncating the interaction between charges separated by more than $\overline{\lambda}=1$ spatial sites [Eq.~\eqref{eq:spatTrun1}]. 
    b) The number of $\hat{Z}\hat{Z}$ terms in  different versions of the electric Hamiltonian as a function of $L$, showing the quadratic $L^2$ and linear $\overline{\lambda}L$ growth. 
    c) The effects of truncating the electric interaction on the low-lying CP-even and $Q=0$ spectrum as a function of $\overline{\lambda}$ for $L=12$. The transparency of the lines connecting energy levels is proportional to the overlap of their corresponding eigenstates.}
    \label{fig:ZZcomparison_2}
\end{figure}

In order to remove the effects of disconnected correlations, consider charges and dipole moments defined on {\it spatial} sites,
\begin{equation}
\hat{\overline{Q}}_n \ = \ \hat{Q}_{2n} + \hat{Q}_{2n+1} \ \ , \ \ \hat{\delta}_n \ = \ \hat{Q}_{2n} - \hat{Q}_{2n+1} \ .
\end{equation}
Unlike charges on staggered sites, the expectation value of a charge on a spatial site is zero, up to exponentially suppressed boundary effects, see App.~B of Ref.~\cite{Farrell:2023fgd}.
Of relevance to constructing the effective Hamiltonian is that correlations between spatial charges, and between spatial charges and dipole moments, decay exponentially,
\begin{equation}
\langle \hat{\overline{Q}}_n \hat{\overline{Q}}_{n+d}\rangle \sim e^{- d/\bar{\xi}} \ \ , \ \ \langle \hat{\overline{Q}}_n \hat{\delta}_{n+d}\rangle \sim e^{-d/\bar{\xi}} \ ,
\end{equation}
for $d\gtrsim \bar{\xi}$,\footnote{Dipole-dipole interactions between spatial sites vanish since the Coulomb potential is linear in one dimension.} where $\bar{\xi} = \xi/2$ is the correlation length in units of spatial sites.
Rewriting $\hat{H}_{el}^{(Q=0)}$ in terms of spatial charges and dipole moments, and truncating interactions beyond $\overline{\lambda}$ spatial sites, it is found that
\begin{align}
\hat{H}_{el}^{(Q=0)} & (\bar{\lambda}) \ = \   \ \frac{g^2}{2}\left\{ \sum_{n=0}^{\frac{L}{2}-1} \left[ \left( L - \frac{5}{4} - 2n \right) \hat{\overline{Q}}^2_n + \frac{1}{2} \hat{\overline{Q}}_n \hat{\delta}_n + \frac{1}{4} \hat{\delta}^2_n + \left( \frac{3}{4} + 2n \right) \hat{\overline{Q}}^2_{\frac{L}{2}+n} - \frac{1}{2} \hat{\overline{Q}}_{\frac{L}{2}+n} \hat{\delta}_{\frac{L}{2}+n} + \frac{1}{4} \hat{\delta}^2_{\frac{L}{2}+n} \right] \right.  \nonumber \\
&+ \left. 2\sum_{n=0}^{\frac{L}{2}-2} \ \ \sum_{m=n+1}^{\min(\frac{L}{2}-1,n+\bar{\lambda})} \left[ \left( L - 1 - 2m \right) \hat{\overline{Q}}_n\hat{\overline{Q}}_m + \frac{1}{2} \hat{\overline{Q}}_{n} \hat{\delta}_{m} + \left( 1 + 2n \right) \hat{\overline{Q}}_{\frac{L}{2}+n}\hat{\overline{Q}}_{\frac{L}{2}+m} - \frac{1}{2} \hat{\overline{Q}}_{\frac{L}{2}+m} \hat{\delta}_{\frac{L}{2}+n} \right] \right\} \ .
\label{eq:spatTrunce}
\end{align}
This expression holds for even $L$, and the analogous expression for odd $L$ can be found in App.~\ref{app:truncHam}.
For $m=0.5,g=0.3$, $\overline{\xi} \sim 0.5$, and $\overline{\lambda}=1$ will be used for demonstration purposes in the remainder of this work. 
Expressed in terms of spin operators, the $\overline{\lambda}=1$ interaction is
\begin{align}
\hat{H}_{el}^{(Q=0)}(1) 
\ = \ &\frac{g^2}{2}\Bigg\{ \sum_{n=0}^{\frac{L}{2}-1} \left[ \left( \frac{L}{2} - \frac{3}{4} - n \right)\hat{Z}_{2n}\hat{Z}_{2n+1}+\left (n+\frac{1}{4} \right )\hat{Z}_{L+2n}\hat{Z}_{L+2n+1}\right ] 
\nonumber \\
&+ \ \frac{1}{2}
\sum_{n=1}^{\frac{L}{2}-2} \left (2 \hat{Z}_{2n} + \hat{Z}_{2n+1}-\hat{Z}_{L+2n}-2\hat{Z}_{L+2n+1} \right )
\nonumber \\
&+ \ \frac{1}{2}\left (2\hat{Z}_{0}+\hat{Z}_{1}+\hat{Z}_{L-2}-\hat{Z}_{L+1}-\hat{Z}_{2L-2}-2\hat{Z}_{2L-1} \right )  \nonumber \\
&+ \sum_{n=0}^{\frac{L}{2}-2} \bigg[\left (\frac{L}{2}-\frac{5}{4}-n \right )(\hat{Z}_{2n}+\hat{Z}_{2n+1})\hat{Z}_{2n+2} + \left( \frac{L}{2} - \frac{7}{4} - n \right)(\hat{Z}_{2n}+\hat{Z}_{2n+1})\hat{Z}_{2n+3}
\nonumber \\
&+ \  \left (n+\frac{1}{4} \right )(\hat{Z}_{L+2n+2}+\hat{Z}_{L+2n+3})\hat{Z}_{L+2n} +\left (n+\frac{3}{4} \right )(\hat{Z}_{L+2n+2}+\hat{Z}_{L+2n+3})\hat{Z}_{L+2n+1} \bigg] \Bigg\} 
\ .
\label{eq:spatTrun1}
\end{align}
Factors of the identity have been dropped as they do not impact time evolution, and this expression only holds for even $L\geq 4$. 
The effects of these truncations on qubit connectivity, number of two-qubit $\hat{Z}\hat{Z}$ terms, and the low-lying spectrum are illustrated in Fig.~\ref{fig:ZZcomparison_2}.
The number of two-qubit operations required for time evolution now scales linearly with volume ${\cal O}(\overline{\lambda} L)$, and there are only operations between qubits separated by at most $(2\overline{\lambda}+1)$ staggered sites.
This interaction will be used to time evolve a wavepacket of single hadrons, and it is important that the impact of these truncations is small on the low-lying hadron states. 
This is illustrated in panel c) of Fig.~\ref{fig:ZZcomparison_2}, where the low-lying spectrum is shown to rapidly converge with increasing $\overline{\lambda}$.
There is some transient behavior presumably due to tunneling beyond the truncation range.
It is important to stress that the exponentially-converging truncations that are made possible by confinement are not obvious at the level of the spin Hamiltonian in Eq.~\eqref{eq:Hgf} due, in part, to $\hat{\overline{Q}}_n \hat{\overline{Q}}_m$ having conspiring single $\hat{Z}$ and double $\hat{Z}\hat{Z}$ terms.
%

\section{SC-ADAPT-VQE for State Preparation}
\label{sec:SCADAPT}
\noindent
In previous work~\cite{Farrell:2023fgd}, we introduced the Scalable Circuits-ADAPT-VQE (SC-ADAPT-VQE) algorithm and workflow, and used it to prepare the vacuum of the Schwinger model on 100 qubits of IBM's quantum computers. 
Here, SC-ADAPT-VQE will be detailed in general, and subsequent sections will apply it to prepare both the vacuum and a hadron wavepacket.
The goal of SC-ADAPT-VQE is to determine low-depth circuits for preparing a target wavefunction that are systematically scalable to any lattice size. 
This scalability enables a  hybrid workflow where circuits determined using {\it classical computers} are scaled and executed on a {\it quantum computer}. 
This eliminates the difficult task of optimizing parameterized quantum circuits on a quantum computer that has both statistical noise from a finite number of shots and device errors~\cite{Wang:2020yjh,Scriva:2023sgz,Cerezo:2023nqf}.

The initial steps of SC-ADAPT-VQE parallel those of ADAPT-VQE~\cite{Grimsley_2019}, and can be summarized as follows:
\begin{itemize}
    \item[1.] Define a pool of operators $\{ \hat{O} \}$ that respect the symmetries of the prepared state. 
    Scalability and phenomenological considerations are used to inform which operators are included in the pool. 
    \item[2.] Initialize a state $\lvert \psi_{{\rm ansatz}}\rangle$ with the quantum numbers of the target state $\lvert \psi_{{\rm target}}\rangle$. 
    \item[3.] Determine a quantity that measures the quality of the ansatz state. 
    For demonstration, consider the infidelity between the ansatz and target states, ${\cal I} = 1 - \vert \langle \psi_{{\rm target}} \vert \psi_{{\rm ansatz}} \rangle \vert ^2$. 
    \item[4.] For each operator in the pool $\hat{O}_i$ determine the gradient of the infidelity between the target and evolved ansatz states, $\frac{\partial}{\partial\theta_i}\left. {\cal I} \right|_{\theta_i=0} = \frac{\partial}{\partial\theta_i} \left.\left( 1 - |\langle\psi_{\rm target} | e^{i \theta_i \hat{O}_i } |\psi_{\rm ansatz}\rangle|^2 \right)\right|_{\theta_i=0}$. 
    This is one way of ranking the relative impact of $\hat{O}_i$ on the infidelity.
    \item[5.] Identify the operator $\hat{O}_n$ with the largest magnitude gradient.
    Update the ansatz with the parameterized evolution of the operator $\lvert \psi_{{\rm ansatz}} \rangle \to e^{i \theta_n \hat{O}_n}\lvert \psi_{{\rm ansatz}} \rangle$.
    \item[6.] Optimize the variational parameters to minimize the infidelity.
    The previously optimized values for $\theta_{1,...,n-1}$ and $\theta_n=0$, are used as initial conditions.
    \item[7.] Return to step 4 until the desired tolerance is achieved.
\end{itemize}
ADAPT-VQE returns an ordered sequence of unitary operators $\{ \hat{U}_i \} = \{ {\exp}(i \theta_i \hat{O}_i) \} $ that prepares the target state up to a desired tolerance.
For use on a quantum computer, the sequence of unitaries can be converted to a sequence of gates through, for example, Trotterization.
If this introduces Trotter errors, the unitaries in steps 4 and 5 should be replaced by their Trotterized versions, ${\exp}(i \theta_i \hat{O}_i) \rightarrow \prod\limits_j \hat{U}_j^{(i)}$.
In SC-ADAPT-VQE, the previous steps are supplemented with the following:
\begin{itemize}
    \item[8.] Repeat ADAPT-VQE for a series of lattice volumes $\{L_1, L_2, \ldots, L_N\}$ using a classical computer (or a small partition of a quantum computer).
    \item[9.] Extrapolate the sequence of unitary operators $\{\{\hat{U}_i\}_{L_1}, \{\hat{U}_i\}_{L_2}, \ldots, \{\hat{U}_i\}_{L_N}\}$ to the desired $L$.
    This sequence is expected to converge for states with localized correlations.
    $L$ can be arbitrarily large and beyond what is accessible
     using a classical computer.
\end{itemize}
The sequence of extrapolated unitaries $\{\hat{U}_i\}_L$ can then be used to prepare the target state on a quantum computer.
This provides an explicit implementation of systematically-localizable~\cite{Klco:2019yrb} and fixed-point~\cite{Klco:2020aud} quantum operators and circuits.

\subsection{Vacuum Preparation}
\noindent
This section will review the use of SC-ADAPT-VQE to prepare the vacuum presented in Ref.~\cite{Farrell:2023fgd}.
The operator pool is constrained by the symmetries and conserved charges of the Schwinger model vacuum; total charge, CP and time reversal.
In addition, there is an approximate translational symmetry in the volume if $L \gg \xi$, that is broken by the boundaries. 
This motivates organizing the pool into volume operators, $\hat{O}^V$, that are translationally invariant, and surface operators, $\hat{O}^S$, whose support is restricted to the boundary.
For the range of $m$ and $g$ explored in Ref.~\cite{Farrell:2023fgd}, including $m=0.5,g=0.3$, an effective pool of operators is
\begin{align}
\label{eq:poolComm}
\{ \hat{O} \}_{\text{vac}} &= \{ \hat O_{mh}^{V}(d) \ , \ \hat O_{mh}^{S}(0,d) \ , \ \hat O_{mh}^{S}(1,d)
 \}
\ ,\\
 \hat O_{mh}^{V}(d) & \equiv i\left [\hat{\Theta}_m^V, \hat{\Theta}_{h}^V(d)\right ]
 = 
\frac{1}{2}\sum_{n=0}^{2L-1-d}(-1)^n\left (
\hat X_n\hat Z^{d-1}\hat Y_{n+d} - 
\hat Y_n\hat Z^{d-1}\hat X_{n+d} 
\right )
\ ,\nonumber\\
\hat O_{mh}^{S}(0,d) & \equiv i\left [\hat{\Theta}_{m}^S(0), \hat{\Theta}_{h}^V(d) \right ]
 = 
\frac{1}{4}\left (\hat X_0\hat Z^{d-1}\hat Y_{d} - \hat Y_0\hat Z^{d-1}\hat X_{d} 
- \hat Y_{2L-1-d}\hat Z^{d-1}\hat X_{2L-1} + \hat X_{2L-1-d}\hat Z^{d-1}\hat Y_{2L-1}\right ) 
\ ,\nonumber \\
 \hat O_{mh}^{S}(1,d) & \equiv i\left [\hat{\Theta}_{m}^S(1), \hat{\Theta}_{h}^S(d) \right ]
 = 
\frac{1}{4}\left (\hat Y_1\hat Z^{d-1}\hat X_{d+1} - \hat X_1\hat Z^{d-1}\hat Y_{d+1} 
+ \hat Y_{2L-2-d}\hat Z^{d-1}\hat X_{2L-2} - \hat X_{2L-2-d}\hat Z^{d-1}\hat Y_{2L-2} \right )
\ . \nonumber
\end{align}
Time reversal invariance implies that the wavefunction is real, and constrains the pool operators to be  imaginary and anti-symmetric, e.g., $i$ times the commutators of orthogonal  operators $\hat{\Theta}$.
Here, $\hat{\Theta}_m^V$ ($\hat{\Theta}_m^S$) 
is a  volume (surface) 
mass term 
and $\hat{\Theta}_{h}^V(d)$ ($\hat{\Theta}_{h}^S(d)$) is a generalized volume (surface) hopping term that spans an odd-number of fermion sites, $d$.
Only $d$ odd is kept as $d$ even breaks CP.
Unlabeled $\hat Z$s act on the qubits between left-most and right-most Pauli operators.\footnote{For $\hat{O}_{mh}^V(d)$ and $\hat{O}_{mh}^S(0,d)$, the range of $d$ is $d\in \{1,3,\ldots 2L-3\}$, and for $\hat{O}_{mh}^S(1,d)$ it is $d\in \{1,3,\ldots 2L-5\}$.}

The individual terms in each operator do not all commute, and they are converted to gates through a first-order Trotterization, ${\exp}(i \theta_i \hat{O}_i) \rightarrow \prod\limits_j \hat{U}_j^{(i)}$, introducing (higher-order) systematic deviations from the target unitary operator.
The initial state for SC-ADAPT-VQE is chosen to be the strong-coupling vacuum $|\Omega_0\rangle = \lvert\uparrow\downarrow\uparrow\downarrow \ldots\uparrow\downarrow\rangle$ where every fermion site is unoccupied. 
To determine the quality of the ansatz state in step-3 of SC-ADAPT-VQE, the expectation value of the Hamiltonian $E = \langle \psi_{{\text{ansatz}}} \vert \hat{H} \vert \psi_{{\text{ansatz}}} \rangle$ is determined, with the gradient in step-4 being computed via $\frac{\partial}{\partial\theta_i}\left. E \right|_{\theta_i=0} = i \langle \psi_{{\text{ansatz}}} \vert [\hat{H}, \hat{O}_i ] \vert \psi_{{\text{ansatz}}} \rangle$.
The convergence of this algorithm and workflow was studied in detail in Ref.~\cite{Farrell:2023fgd} as a function of the number of SC-ADAPT-VQE steps. 
For $m=0.5,g=0.3$, it was found that  two steps of SC-ADAPT-VQE, was sufficient to achieve percent-level precision in relevant observables. 
Both 2-step (7.8 CNOTs/qubit) and 3-step (21 CNOTs/qubit) preparations have been performed on up to 100 qubits of IBM's quantum computers.

\subsection{Hadron Wavepacket Preparation}
\label{sec:SCADAPTWP}
\begin{figure}[!tb]
    \centering
    \includegraphics[width=\textwidth]{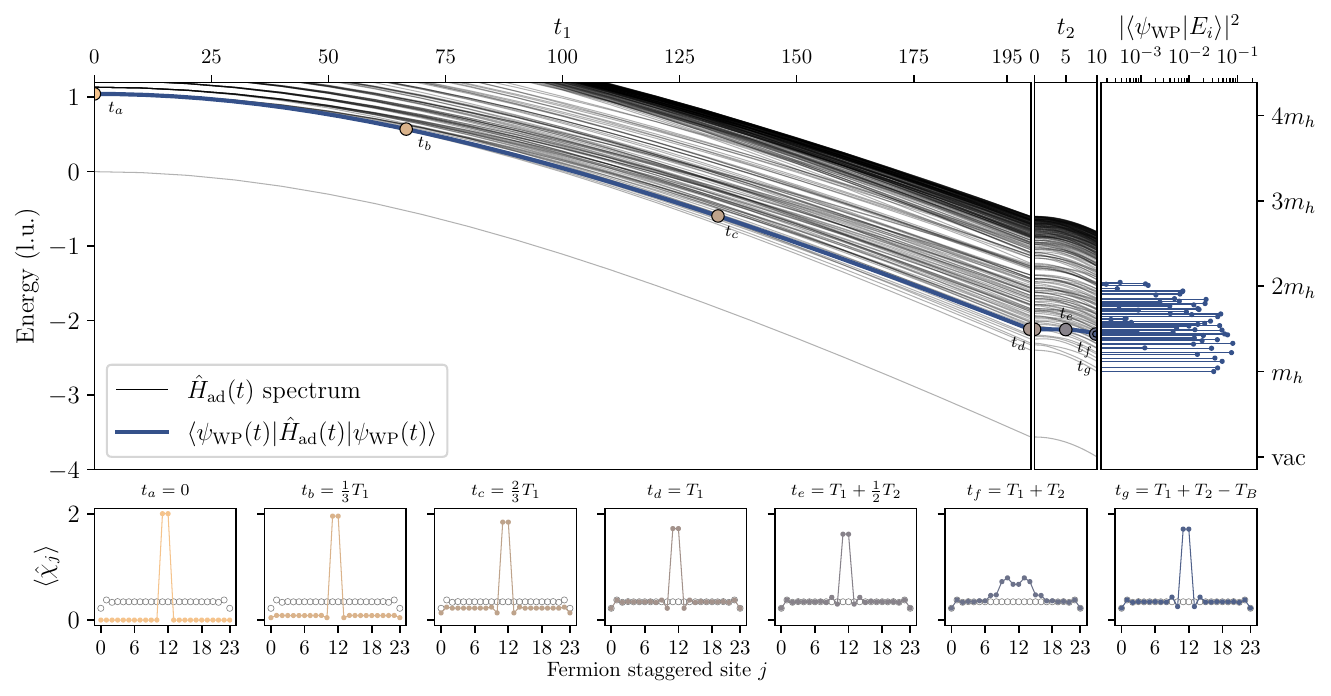}
    \caption{Adiabatic state preparation for $L=12$. Upper panels: the lowest 200 eigenenergies of $\hat{H}_{\text{ad}}(t)$ as a function of adiabatic turn-on time, the energy of the state $|\psi_{\rm WP}(t)\rangle$, and the final overlap between $|\psi_{\rm WP}\rangle$ and the eigenstates of $\hat{H}$, $|E_i\rangle$. Lower panels: evolution of the chiral condensate $\langle\hat{\chi}_j\rangle$, defined in Eq.~\eqref{eq:localCC}, of $|\psi_{\rm WP}(t)\rangle$ for a selection of times, 
    $t_a-t_g$, 
    with the empty markers showing the $\langle \hat{\chi}_j \rangle$ of the vacuum $|\psi_{\rm vac}\rangle$ of $\hat{H}$.}
    \label{fig:adiabatic_stateprep}
\end{figure}

\noindent
SC-ADAPT-VQE can be used to prepare a state that has large overlap with an adiabatically prepared hadron wavepacket.
An alternative method for preparing wavepackets is discussed in App.~\ref{app:chempot}.
In a lattice theory of interacting scalar fields, a complete procedure for preparing single particle wavepackets has been proposed by Jordan, Lee and Preskill~\cite{Jordan_2018,DBLP:journals/qic/JordanLP14}.\footnote{Other proposals for creating initial states and wavepackets can be found in Refs.~\cite{Kuhn:2015zqa,Pichler:2015yqa,Damme:2019rts,Avkhadiev:2019niu,Surace:2020ycc,Karpov:2020pqe,Milsted:2020jmf,Vovrosh:2022bpj,Asaduzzaman:2022bpi,Avkhadiev:2022ttx,Farrell:2022vyh,Atas:2022dqm,Vary:2023ihk,Chai:2023qpq,Roy:2023uil}, including recent work on creating hadronic sources in the bosonized form of the Schwinger model using circuit-QED~\cite{Belyansky:2023rgh}.}
In their method, wavepackets are first prepared in free scalar field theory, and then the $\lambda \phi^4$ interaction is adiabatically ``turned on''.
This method runs into difficulty in the Schwinger model because the single particle states (hadrons) of the interacting theory are non-perturbatively different from the single particle states of the non-interacting theory (electrons).
To overcome this, consider starting in the interacting theory with $m=0.5$ and $g=0.3$, and adiabatically turning on the kinetic term.
The initial Hamiltonian is diagonal in the computational $z$-basis, and the ground state is the same as the infinite coupling (anti-ferromagnetic) vacuum $\vert \Omega_0 \rangle$.
The infinite-coupling vacuum provides a suitable starting configuration upon which to build the wavepacket as it correctly encodes the long-distance correlations that characterize this confining theory.\footnote{The strong-coupling limit has been extensively studied, particularly in the context of lattice QCD. See, for example, Ref.~\cite{Fromm:2009xw} and references therein.}
On this vacuum, a hadron can be excited by creating an $e^-e^+$ pair on adjacent staggered sites.
By preparing a superposition of such hadrons at different locations, an arbitrary wavepacket can be prepared. 
Here, the focus will be on preparing a localized hadron wavepacket that is centered in the middle of the lattice to preserve CP and minimize boundary effects.
A suitable initial state is, 
\begin{equation}
|\psi_{{\rm WP}}\rangle_{\text{init}} = \hat{X}_{L-1}\hat{X}_L\vert \Omega_0\rangle \ .
\end{equation}
To transition to a hadron wavepacket in the full theory, this state is taken through two steps of adiabatic evolution with a time-dependent Hamiltonian (illustrated in Fig.~\ref{fig:adiabatic_stateprep}),
\begin{equation}
    \hat{H}_{\rm ad}(t) = 
    \begin{cases}
    \hat{H}_m + \hat{H}_{el} + \frac{t}{T_1} \left [ \hat{H}_{kin} \ - \ \frac{1}{2} (\sigma^+_{L-2}\sigma^-_{L-1} + \sigma^+_{L}\sigma^-_{L+1} + {\rm h.c.} ) \right ]  \ &0<t\leq T_1  \ , \\[5pt]
    \hat{H}_m + \hat{H}_{el} + \hat{H}_{kin} \ - \ \left (1-\frac{t-T_1}{T_2} \right ) \frac{1}{2} (\sigma^+_{L-2}\sigma^-_{L-1} + \sigma^+_{L}\sigma^-_{L+1} + {\rm h.c.} )   \ &T_1<t\leq T_1+T_2  \ . 
    \end{cases}
\end{equation}
For $t\in (0,T_1 ]$, the kinetic term is adiabatically turned on everywhere except for the links connecting the initial wavepacket to the rest of the lattice.
This mitigates spatial spreading of the initial wavepacket (see times $t_{a,b,c,d}$ in Fig.~\ref{fig:adiabatic_stateprep}).
Next, for $t\in (T_1,T_2 ]$, the remaining two links are adiabatically turned on. 
These remaining links are spatially localized (act over a pair of staggered sites), and therefore primarily couple to high-momentum (energy) states.
This implies that the energy gap relevant for the adiabatic evolution is large, and the second evolution can be performed much faster than the first evolution.
There is a small amount of wavepacket spreading (times $t_{e,f}$), which is undone by evolving backwards in time for a duration $T_B = T_2/2$ with the full Hamiltonian $\hat{H}$ from Eq.~\eqref{eq:Hgf} (time $t_g$).
Explicitly, the hadron wavepacket is given by,
\begin{equation}
\vert \psi_{{\rm WP}} \rangle \ = \  
 e^{i T_B \hat{H}} \, {\cal T} e^{-i \int_{0}^{T_1 + T_2}dt \hat{H}_{\rm ad}(t)} \vert \psi_{{\rm WP}}\rangle_{\text{init}} \ ,
\label{eq:adiabaticWP}
\end{equation}
where ${\cal T}$ denotes time-ordering.
For practical implementation, the evolution of the time-dependent Hamiltonian can be accomplished with Trotterization,
\begin{equation}
{\cal T} e^{-i \int_{0}^{T_1 + T_2}dt \hat{H}_{\rm ad}(t)} \ \approx \ {\cal T} e^{-i \sum_{n=0}^{N_T-1} \delta t \hat{H}_{\rm ad}\left [(n+0.5) \delta t\right ]} \ \approx \ {\cal T} 
 \prod_{n=0}^{N_T-1}e^{-i \left (\delta t \right ) \hat{H}_{\rm ad}\left [(n+0.5) \delta t\right ]} \ ,
\end{equation}
where $N_T$ is the number of Trotter steps and $\delta t = \frac{T_1+T_2}{N_T}$ is the step size.
For the simulation parameters chosen in this work, we find that $T_1 = 200$, $T_2=10$ and $\delta t=0.2$ are sufficient for adiabatic evolution.
The final state is localized (within a few sites), and primarily consists of single-hadron states (see overlaps in Fig.~\ref{fig:adiabatic_stateprep}).

In principle, this adiabatic procedure could be used to prepare a hadronic wavepacket on a quantum computer. 
In practice, the required circuits are too deep to run on current devices.
To address this, SC-ADAPT-VQE is used to find low-depth circuits that prepare an approximation to the adiabatically determined wavepacket.
These low-depth circuits act on the vacuum, whose preparation was outlined in the previous section.
Scalability of the state preparation circuits is expected because the constructed wavepacket is localized away from the boundaries, and is built on top of a vacuum state that has converged exponentially in $L$ to its infinite-volume form~\cite{Farrell:2023fgd}.
As both the initial state (vacuum) and target state (single hadron) are CP even and charge zero, the operators in the pool must conserve charge and CP.
An operator pool that is found to produce a wavefunction that converges exponentially fast in circuit depth is,
\begin{align}
    \{\hat{O} \}_{\text{WP}} \ &= \ 
    \{ \hat{O}_{mh}(n,d), \, \hat{O}_{h}(n,d),\, \hat{O}_{m}(n) \} \ ,\nonumber \\
    \hat{O}_{mh}(n,d) &=\frac{1}{2} \left[ 
    \hat{X}_{L-n} \hat{Z}^{d-1}  \hat{Y}_{L-n+d} 
    - \hat{Y}_{L-n}  \hat{Z}^{d-1} \hat{X}_{L-n+d}  
    \ + \ (-1)^{d+1}\left (1-\delta_{L-n,\gamma} \right  )\left (\hat{X}_{\gamma} \hat{Z}^{d-1}  \hat{Y}_{\gamma+d} - \hat{Y}_{\gamma}  \hat{Z}^{d-1} \hat{X}_{\gamma+d} \right ) \right] 
    \ ,\nonumber \\[4pt]
    \hat{O}_{h}(n,d) &=\frac{1}{2} \left[ 
    \hat{X}_{L-n} \hat{Z}^{d-1}  \hat{X}_{L-n+d} 
    + \hat{Y}_{L-n}  \hat{Z}^{d-1} \hat{Y}_{L-n+d}  
    \ + \ (-1)^{d+1}\left (1-\delta_{L-n,\gamma} \right  )
    \left (\hat{X}_{\gamma} \hat{Z}^{d-1}  \hat{X}_{\gamma+d} + \hat{Y}_{\gamma}  \hat{Z}^{d-1} \hat{Y}_{\gamma+d} \right ) \right] \ ,\nonumber \\[4pt]
    \hat{O}_{m}(n) & = \hat{Z}_{L-n}  \ - \ \hat{Z}_{L-1+n} \ ,
    \label{eq:PacketPool}
\end{align}
where $\gamma=L-1+n-d$, $n\in \{1,\ldots,L \}$, and the $\left (1-\delta_{L-n,\gamma} \right  )$ coefficients prevent double counting operators that are already CP-symmetric.
The pool operators are inspired by the Hamiltonian, with $\hat{O}_m(n)$ being a mass-like operator, $\hat{O}_h(n,d)$ a generalized hopping operator spanning $d$ staggered sites, and $\hat{O}_{mh}(n,d)$ being proportional to their commutator.
Note that unlike the operator pool used to prepare the vacuum, $\{ \hat{O} \}_{\text{WP}}$ is not constrained by time reversal or translational symmetry, and the individual terms in each operator commute.
Thus there are no Trotter errors when the corresponding unitaries are converted to circuits.

\begin{figure}[!tb]
    \centering
    \includegraphics[width=0.55\textwidth]{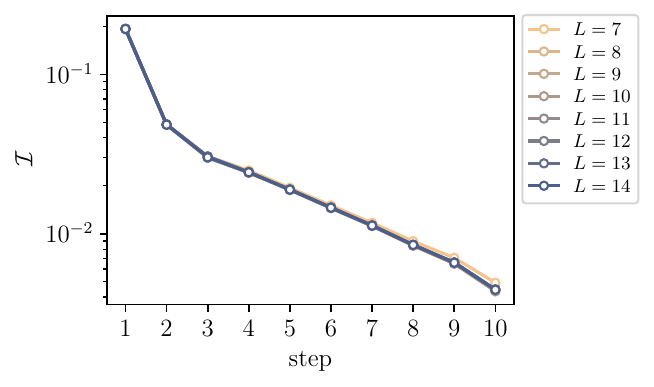}
    \caption{Infidelity of the wavepacket, defined in Eq.~\eqref{eq:inf_wp_scadapt}, prepared with multiple steps of SC-ADAPT-VQE for a range of $L$.}
    \label{fig:WP_adapt_inf}
\end{figure}

The initial state for SC-ADAPT-VQE is chosen to be $\lvert \psi_{\text{ansatz}} \rangle = \lvert \psi_{\text{vac}} \rangle$, as this correctly reproduces the vacuum outside of the support of the hadron wavepacket.
In this section, all calculations are performed with exact diagonalization, and the initial state is the exact vacuum. 
In Secs.~\ref{sec:ClSim} and~\ref{sec:Qsim}, the initial state will be the SC-ADAPT-VQE prepared vacuum. 
Using the exact vacuum instead of the SC-ADAPT-VQE vacuum prevents operators from being chosen that improve the vacuum but do not build out the local profile of the wavepacket.
The quality of the prepared state is determined by the infidelity of the ansatz state with the adiabatically prepared state from Eq.~\eqref{eq:adiabaticWP},
\begin{equation}
{\cal I} \ = \ 1 \ - \ \vert \langle \psi_{\text{WP}} \vert \psi_{\text{ansatz}}\rangle\vert^2 \ .
\label{eq:inf_wp_scadapt}
\end{equation}
Results obtained from performing the steps in SC-ADAPT-VQE (outlined in the introduction of Sec.~\ref{sec:SCADAPT}) for $L=7-14$ are shown in Fig.~\ref{fig:WP_adapt_inf} and Table~\ref{tab:AnglesWP10}.\footnote{The vacuum maximizes the infidelity (has ${\cal I}=1$) with the adiabatically determined state as there is no overlap between the vacuum and the single-hadron states that make up the wavepacket.
This presents a problem in step 4 of SC-ADAPT-VQE since $\frac{\partial}{\partial \theta_i} {\cal I}$ is zero for all operators in the pool.
To overcome this, for the first iteration of SC-ADAPT-VQE, the parameterized evolution of the ansatz with each operator is determined separately.
The operator that minimizes the infidelity is chosen for the first operator in the SC-ADAPT-VQE ansatz.}
Up to the tolerance of the optimizer, the variational parameters have converged in $L$, and therefore the $L=14$ parameters and operator ordering can be used 
to prepare a hadron wavepacket for any $L>14$.
Initially, short-range operators localized around the center of the wavepacket are 
selected by SC-ADAPT-VQE.\footnote{It is interesting to note the similarities between this wavepacket construction, and the construction of hadronic sources and sinks in Euclidean-space lattice QCD calculation.
Here, the initial interpolating operator for the hadronic wavepacket is being ``dressed'' by an increasing number of operators with exponentially improving precision. 
In Euclidean-space lattice QCD, a matrix of correlation functions between a set of sources and sinks is diagonalized to provide a set of correlators with extended plateaus toward shorter times, corresponding to the lowest-lying levels in the spectrum that have overlap with the operator set.
This ``variational method'', e.g., Refs.~\cite{Michael:1982gb,Luscher:1990ck,blossier2008efficient}, provides upper bounds to the energies of the states in the spectrum.
The sources and sinks for hadrons are operators constructed in terms of quark and gluon fields, and correlation functions are formed by contracting field operators of the sinks with those of the sources (or with themselves when both quark and anti-quark operators are present).
This becomes computationally challenging with increasingly complex operator structures, as required, for instance, to study nuclei, see for example Refs.~\cite{Beane:2009gs,Beane:2009py,Aoki:2009yy,Beane:2010em,NPLQCD:2012mex,Yamazaki:2012hi,Yamazaki:2015vjn,Yamazaki:2015asa,Drischler:2019xuo,Davoudi:2020ngi,Amarasinghe:2021lqa}.}
This is as expected for a wavepacket composed of single hadron states with short correlation lengths,  that is approximately a delta function in position space.\footnote{The variational parameters change sign between even- and odd-values of $L$ if $d$ is odd (even) in $\hat{O}_{mh}$ ($\hat{O}_h$). 
Also, note that $\hat{O}_m$ is not chosen until after step 10 in the SC-ADAPT-VQE ansatz.}
The convergence of the infidelity is found to be exponential in the 
step of the algorithm (circuit depth), and independent of $L$.
This is in agreement with previous discussions on localized states being built on top of an exponentially converged vacuum.
Note that the convergence in $L$ is smoother for the SC-ADAPT-VQE wavepacket than for the vacuum as the boundary effects are much smaller (see Fig.~5 in Ref.~\cite{Farrell:2023fgd}).
Two steps of SC-ADAPT-VQE reaches an infidelity of 0.05, and will be used in the remainder of the work to prepare the wavepacket.
\begin{table}[!tb]
\renewcommand{\arraystretch}{1.4}
\begin{tabularx}{\textwidth}{|c || Y | Y | Y | Y | Y | Y | Y | Y | Y | Y |}
 \hline
 \diagbox[height=23pt]{$L$}{$\theta_i$} & $\hat O_{mh}(1,1)$ & $\hat O_{mh}(2,2)$ & $\hat O_{mh}(3,2)$ & $\hat O_{mh}(3,1)$ & $\hat O_{mh}(5,4)$ &  $\hat O_{h}(2,2)$ & $\hat O_{mh}(4,4)$ & $\hat O_{mh}(4,5)$ & $\hat O_{h}(4,4)$ & $\hat O_{mh}(2,3)$ \\
 \hline\hline
 7 & 1.6370 & -0.3154 & -0.0978 & 0.0590 & -0.0513 & -0.0494 & -0.0518 & -0.0389 & 0.0359 & 0.0528\\
 \hline
 8 & -1.6371 & -0.3157 & -0.0976 & -0.0615 & -0.0499 & 0.0493 & -0.0515 & 0.0391 & -0.0360 & -0.0529\\
 \hline
 9 & 1.6370 & -0.3155 & -0.0980 & 0.0609 & -0.0509 & -0.0493 & -0.0515 & -0.0390 & 0.0361 & 0.0527\\
 \hline
 10 & -1.6370 & -0.3154 & -0.0984 & -0.0598 & -0.0501 & 0.0493 & -0.0515 & 0.0389 & -0.0360 & -0.0527\\
 \hline
 11 & 1.6370 & -0.3155 & -0.0984 & 0.0598 & -0.0507 & -0.0492 & -0.0515 & -0.0390 & 0.0360 & 0.0527\\
 \hline
 12 & -1.6371 & -0.3156 & -0.0975 & -0.0616 & -0.0505 & 0.0493 & -0.0516 & 0.0391 & -0.0361 & -0.0528\\
 \hline
 13 & 1.6371 & -0.3157 & -0.0973 &  0.0617 & -0.0506 & -0.0494 & -0.0516 & -0.0391 &  0.0359 &  0.0529\\
 \hline
 14 & -1.6370 & -0.3155 & -0.0981 & -0.0602 & -0.0506 & 0.0493 & -0.0515 &  0.0390 & -0.0359 & -0.0527\\
 \hline
\end{tabularx}
\caption{The operator ordering and variational parameters that prepare the 10 step SC-ADAPT-VQE hadron wavepacket. Results are shown for $L=7-14$, and were obtained from a classical simulation using exact exponentiation.}
 \label{tab:AnglesWP10}
\end{table}
%

\section{Quantum Circuits}
\label{sec:ClSim}
\noindent
In this section, the quantum circuits that prepare hadron wavepackets and implement time evolution are developed.
These circuits are constructed to minimize CNOT count and circuit depth in order to reduce the effects of device errors. 
In addition, with the goal of running on {\tt IBM}'s quantum computers, the circuits are optimized for nearest-neighbor connectivity.
These circuits are verified using the {\tt qiskit} classical simulator, and the systematic errors arising from the approximations used in this work are quantified.

\subsection{Quantum Circuits for Vacuum and Hadron Wavepacket Preparation}
\begin{figure}[!b]
    \centering
    \includegraphics[width=0.45\textwidth]{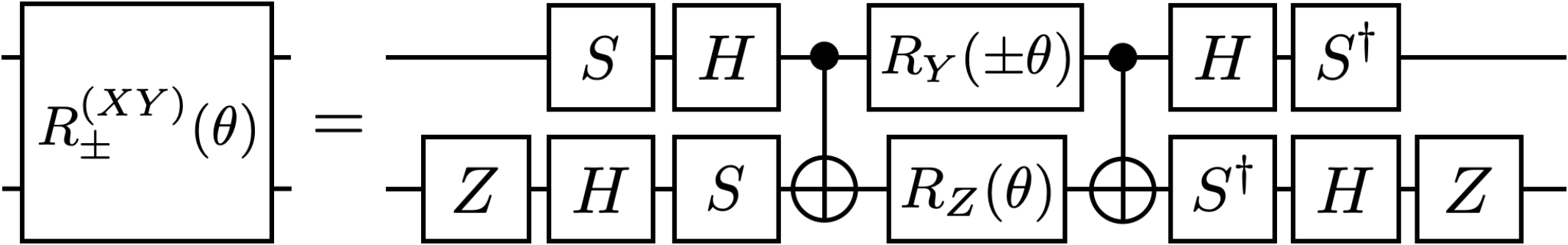}\hfill
    \includegraphics[width=0.45\textwidth]{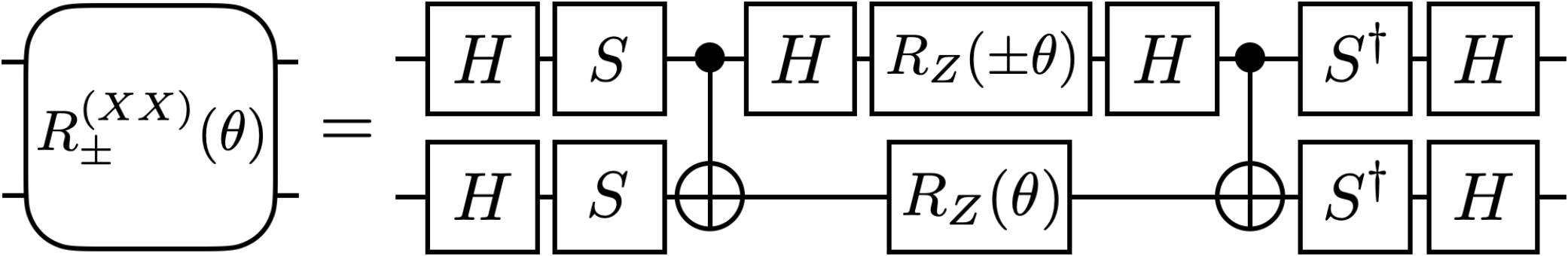}
    \caption{Circuits implementing $R^{(XY)}_{\pm}(\theta)={\rm exp}[-i\tfrac{\theta}{2}(\hat{Y}\hat{X}\pm \hat{X}\hat{Y})]$ (left) and $R^{(XX)}_{\pm}(\theta)={\rm exp}[-i\tfrac{\theta}{2}(\hat{X}\hat{X}\pm \hat{Y}\hat{Y})]$ (right).}
    \label{fig:xy_xx_circuits}
\end{figure}

\begin{figure}[!tb]
    \centering
    \includegraphics[width=0.8\textwidth]{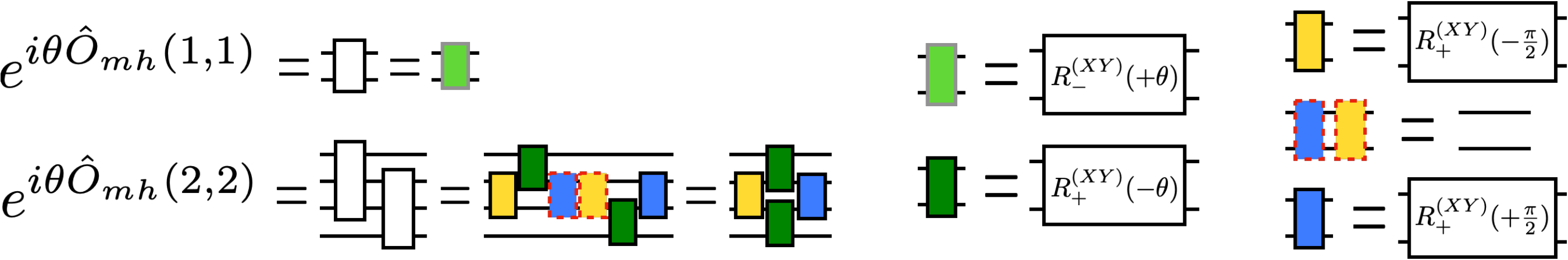}
    \caption{Circuits implementing the unitaries that prepare the 2-step SC-ADAPT-VQE wavepacket. The circuits for the individual blocks $R^{(XY)}_{\pm}(\theta)$ and $R^{(XX)}_{\pm}(\theta)$ are shown in Fig.~\ref{fig:xy_xx_circuits}.}
    \label{fig:2step_wp_adapt}
\end{figure}

\noindent
In order to prepare the SC-ADAPT-VQE vacuum on a quantum computer, the Trotterized exponentials of the pool operators in Eq.~\eqref{eq:poolComm} are converted to sequences of gates, which was treated in detail in previous work~\cite{Farrell:2023fgd}.
The circuit building technique follows the strategy of Ref.~\cite{Algaba:2023enr}, where an ``X''-shaped construction is used to minimize circuit depth and CNOT gate count.
Preparing the SC-ADAPT-VQE hadron wavepacket requires converting the exponential of the pool operators in Eq.~\eqref{eq:PacketPool} to sequences of gates. 
The individual terms in each operator in the wavepacket pool commute, and therefore first-order Trotterization is exact.
The corresponding circuits extend those used for preparing the vacuum, and are shown in Figs.~\ref{fig:xy_xx_circuits} and~\ref{fig:2step_wp_adapt} for the 
2-step SC-ADAPT-VQE wavepacket used in subsequent sections 
(see App.~\ref{app:circuitDetails} for the 10-step SC-ADAPT-VQE circuits). 
These circuits are arranged to maximize cancellations between CNOTs, and minimize the circuit depth.

%

\subsection{Quantum Circuits for Time Evolution }
\noindent
To perform time evolution, a second-order Trotterization of the time-evolution operator with the $\overline{\lambda}=1$ truncated electric interaction will be used,
\begin{equation}
\hat{U}^{\text{(Trot)}}_2(t) \ = \  e^{-i \frac{t}{2} \hat{H}_{kin\text{-1}}} e^{-i \frac{t}{2} \hat{H}_{kin\text{-0}}} e^{-i t  \hat{H}_{m}} e^{-i t \hat{H}_{el}^{(Q=0)}(1)} e^{-i \frac{t}{2} \hat{H}_{kin\text{-0}}}  e^{-i \frac{t}{2} \hat{H}_{kin\text{-1}}} \ ,
\end{equation}
where $\hat{H}_{kin\text{-0}}$ ($\hat{H}_{kin\text{-1}}$) are the hopping terms between even (odd) staggered sites.
This ordering was chosen to maximize the cancellations between neighboring CNOTs. 
A second-order Trotterization is used as it provides a good balance between minimizing both circuit depth and Trotter errors.
In addition, the property of second-order Trotterization $\hat{U}^{(\text{Trot})}_2(t) \hat{U}^{(\text{Trot})}_2(-t) = \hat{1}$ enables a powerful error-mitigation technique~\cite{ARahman:2022tkr}, see Sec.~\ref{sec:Qsim}.

\begin{figure}[!htb]
    \centering
    \includegraphics[width=0.7\columnwidth]{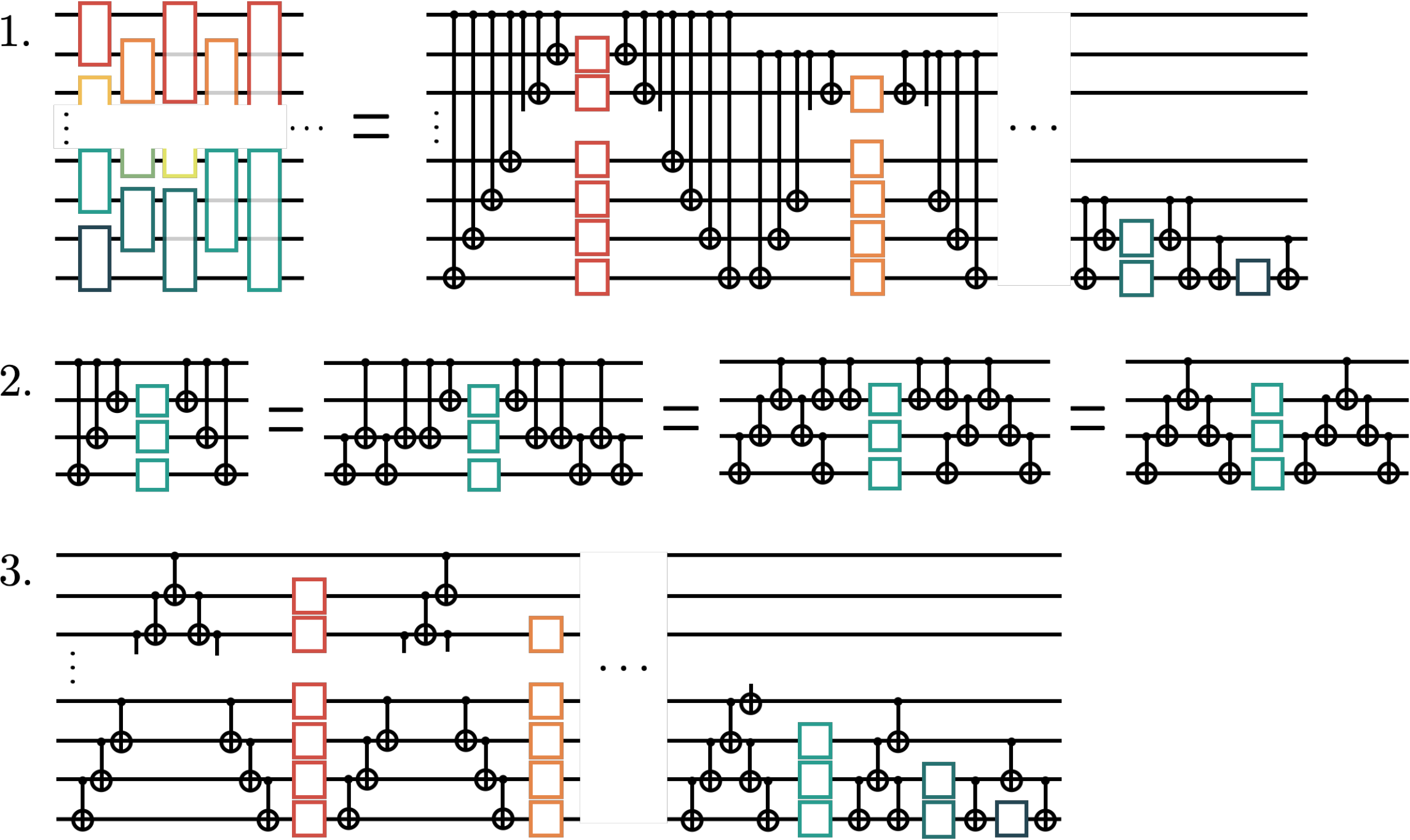}
    \caption{An efficient nearest-neighbor CNOT decomposition for $R_{ZZ}$s between all pairs of $N_q$ qubits.}
    \label{fig:ZZsimplifications}
\end{figure}

\begin{figure}[!htb]
    \centering
    \includegraphics[width=0.85\columnwidth]{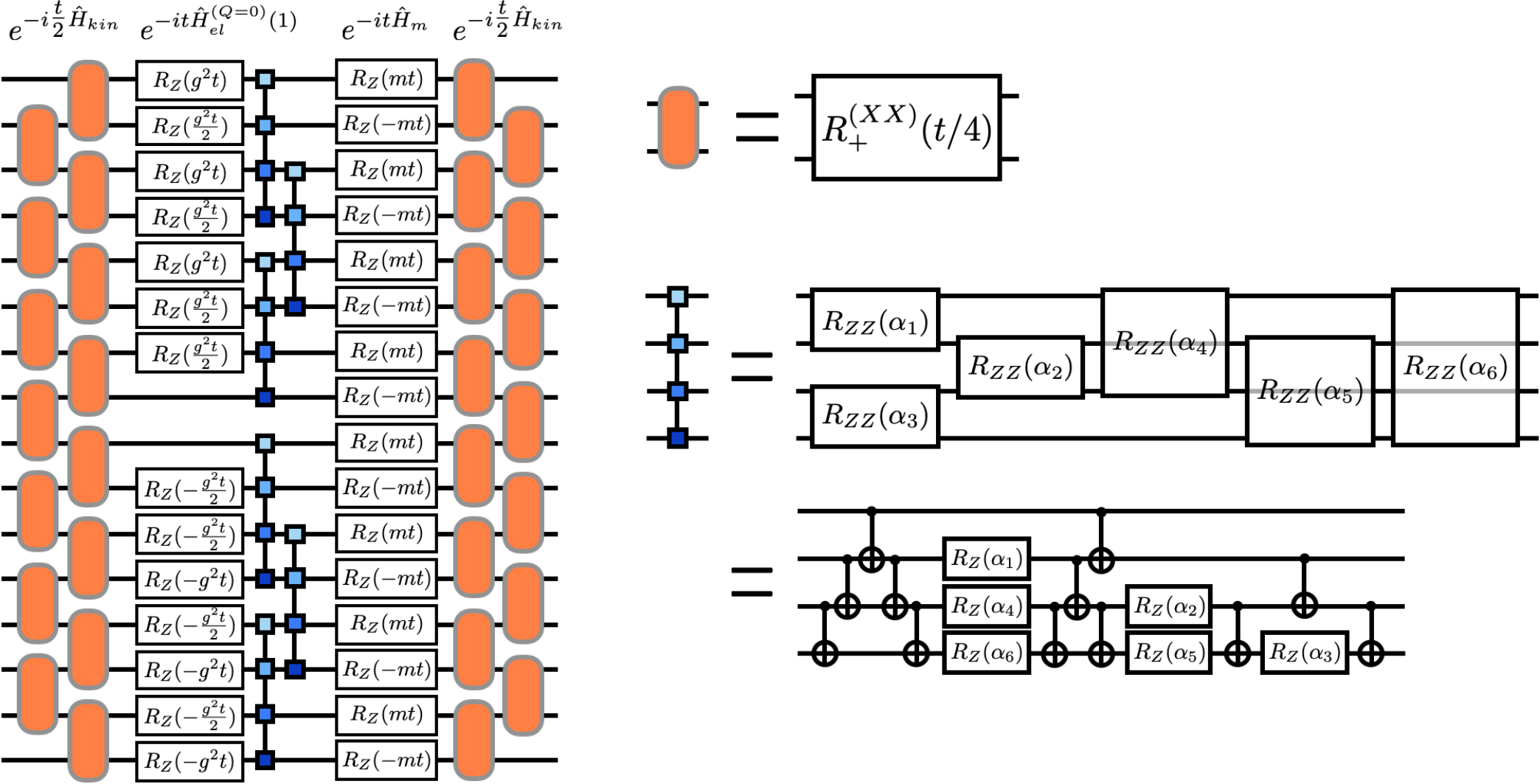}
    \caption{A quantum circuit that implements a single second-order Trotter step associated with the $\overline{\lambda}=1$ truncated Hamiltonian in Eq.~\eqref{eq:spatTrun1} for $L=8$. The orange boxes implement the kinetic term (the right circuit in Fig.~\ref{fig:xy_xx_circuits}) and the blue ``barbells'' are $\hat{Z}\hat{Z}$ rotations. 
    With this ordering, some of the CNOTs in the barbells can be combined with the ones in the kinetic terms.
    The $\alpha_i$ angles can be derived from Eq.~\eqref{eq:spatTrun1} and are given in App.~\ref{app:circuitDetails}.}
    \label{fig:StagCirc}
\end{figure}

The Trotterization of $\hat{H}_m$ only involves single qubit $\hat{Z}$ rotations, which has a straightforward circuit implementation.
The Trotterization of the kinetic terms uses the right circuit in Fig.~\ref{fig:xy_xx_circuits} arranged in a brickwall pattern to minimize circuit depth, and requires $4(2L-1)$ CNOTs per second-order Trotter step.
The Trotterization of $\hat{H}_{el}^{(Q=0)}(1)$ in Eq.~\eqref{eq:spatTrun1} requires nearest-neighbor, next-to-nearest-neighbor and next-to-next-to-nearest-neighbor entangling $R_{ZZ}= e^{-i \frac{\theta}{2} \hat Z \hat Z}$ operations acting between qubits on adjacent spatial sites.
Organizing into blocks of adjacent spatial sites, the problem is to find a nearest-neighbor CNOT decomposition for $R_{ZZ}$s between all pairs of $N_q=4$ qubits.
Generalizing to any $N_q\geq3$, a strategy for constructing these circuits, depicted in Fig.~\ref{fig:ZZsimplifications}, is
\begin{enumerate}
    \item Group all the rotations that share the top qubit.
    \item For each block of grouped rotations, 
    use the bridge decomposition to convert the long-range CNOTs into nearest neighbor ones. Simplify the CNOTs within each block.
    \item Simplify the  CNOTs from neighboring blocks.
\end{enumerate}
These circuits have a total number of CNOTs $N$ and circuit depth $D$ given by
\begin{equation}
    N=2\, \binom{N_q}{2} \ , \ \ D=N_q(N_q-2)+3 \ .
\end{equation}
Compared to the circuits before the nearest-neighbor decomposition (e.g., using the circuits in step 1.), this does not introduce any additional CNOTs, but has a depth that scales as ${\cal O}(N_q^2)$ compared to ${\cal O}(N_q)$.
The $N_q = 4$ circuit used for the $\overline{\lambda}=1$ interaction contributes $12(L-2)$ CNOTs per second-order Trotter step.
Circuits implementing a full second-order Trotter steps are shown in Fig.~\ref{fig:StagCirc}.
Taking into account the CNOT cancellations between the electric and kinetic terms, as well as between adjacent Trotter steps, the total number of CNOTs required is
\begin{equation}
\# \ \text{of CNOTs for }N_T \ 2^\text{nd} \ \text{order Trotter steps with $\overline{\lambda}=1$ :} \ \ 19L-28+(17L-26)(N_T-1) \ .
\end{equation}
For $L=56$, this is 926 CNOTs per additional second-order Trotter step, comparable to the 890 CNOTs required for the 2-step SC-ADAPT-VQE vacuum and hadron wavepacket preparation.
%

\section{Quantifying the Systematic Errors of the Approximations}
\label{sec:Systematics}
\noindent
The systematic errors that are introduced by the approximations we have employed can be analyzed and quantified by performing end-to-end classical simulations using {\tt qiskit}.
The approximations are as follows:
\begin{enumerate}
\item The vacuum is prepared using the 2-step SC-ADAPT-VQE circuits. 
This furnishes an infidelity density of ${\cal I}_L \ = \ {\cal I}/L \ = \ 0.01$ with the exact vacuum.\footnote{The infidelity density ${\cal I}_L$ is a relevant measure for the vacuum as the state is being established across the whole lattice, whereas the infidelity is a relevant figure of merit for the (localized) hadron wavepacket.}
\item A hadron wavepacket is prepared using the 2-step SC-ADAPT-VQE circuits. 
This furnishes an infidelity of ${\cal I}\ = \ 0.05$ with an adiabatically prepared wavepacket.
\item A Hamiltonian with the electric interactions truncated beyond $\overline{\lambda} = 1$ spatial sites is used to evolve the prepared wavepacket forward in time.
\item The time-evolution operator is implemented in quantum circuits using a second-order Trotterization.
\end{enumerate}
This section will focus on a system size of $L=12$, where the classical simulations can be performed exactly. 
The circuit structure and variational parameters for the 
2-step SC-ADAPT-VQE vacuum and wavepacket preparation are given in Table~\ref{tab:AnglesVACWP}.
Note that the (2-step) wavepacket parameters differ slightly from those in Table~\ref{tab:AnglesWP10}, which are for the 10-step SC-ADAPT-VQE ansatz.
\begin{table}[!htb]
\renewcommand{\arraystretch}{1.4}
\begin{tabularx}{\textwidth}{| Y || Y | Y || Y | Y |}
 \hline
   & \multicolumn{2}{c||}{Vacuum}&  \multicolumn{2}{c|}{Wavepacket} \\
   \hline
   \hline
   & $\hat{O}^V_{mh}(1)$ & $\hat{O}^V_{mh}(3)$ & $\hat{O}_{mh}(1,1)$ & $\hat{O}_{mh}(2,2)$ \\
   \hline
    $L=12$ & 0.30738 & -0.04059 & -1.6492 & -0.3281 \\
   \hline
   $L=56$ & 0.30604 & -0.03975 & -1.6494 &  -0.3282 \\
   \hline
\end{tabularx}
\caption{The structure of the SC-ADAPT-VQE preparation circuits for the vacuum and wavepacket. The pool operators in the second row are defined in Eqs.~\eqref{eq:poolComm} and~\eqref{eq:PacketPool}. The parameters for the $L=12$ vacuum were determined in Ref.~\cite{Farrell:2023fgd}, and for the $L=12$ wavepacket in Sec.~\ref{sec:SCADAPTWP}. The wavepacket parameters for $L=56$ are the same as those for $L=14$, and the vacuum parameters are extrapolated via an exponential fit (in $L$).}
\label{tab:AnglesVACWP}
\end{table}

To identify the propagation of hadrons, we choose to measure the local chiral condensate,
\begin{align}
\hat{\chi}_j  = (-1)^j 
\hat{Z}_j + \hat{I} \ ,
\label{eq:localCC}
\end{align}
with eigenvalues of 0 (staggered site $j$ is empty) and $2$ (staggered site $j$ is occupied by a fermion).
It is useful to define the expectation value of the local chiral condensate relative to its vacuum expectation value,
\begin{equation}
{\cal X}_j(t) 
\ = \ \langle \psi_{\text{WP}}\vert
\ \hat{\chi}_j(t)\  \vert \psi_{\text{WP}} \rangle  
\ - \ \langle \psi_{\text{vac}} \vert\  \hat{\chi}_j(t) \  \vert\psi_{\text{vac}} \rangle 
\ .
\label{eq:chij}
\end{equation}
Here, $\hat{\chi}_j(t)$ is the time evolved observable; with exact exponentiation of the full Hamiltonian this would be $\hat{\chi}_j(t) = e^{i t \hat{H}} \hat{\chi}_j e^{-i t \hat{H}}$. 
When using a truncated interaction and/or Trotterization, the time-evolution operator changes.
The states $\vert \psi_{\text{vac}} \rangle $ and $\vert \psi_{\text{WP}} \rangle $ represent the prepared vacuum and wavepacket, either exact or using the SC-ADAPT-VQE approximation.
The subtraction of the vacuum expectation value is also time dependent because, for example, the SC-ADAPT-VQE prepared vacuum is not an eigenstate of the truncated Hamiltonian.
This time-dependent subtraction removes systematic errors that are present in both the wavepacket and vacuum time evolution.
It also proves to be an effective way to mitigate some effects of device errors, see Sec.~\ref{sec:Qsim}.

\begin{figure}[!htb]
    \centering
    \includegraphics[width=0.75\columnwidth]{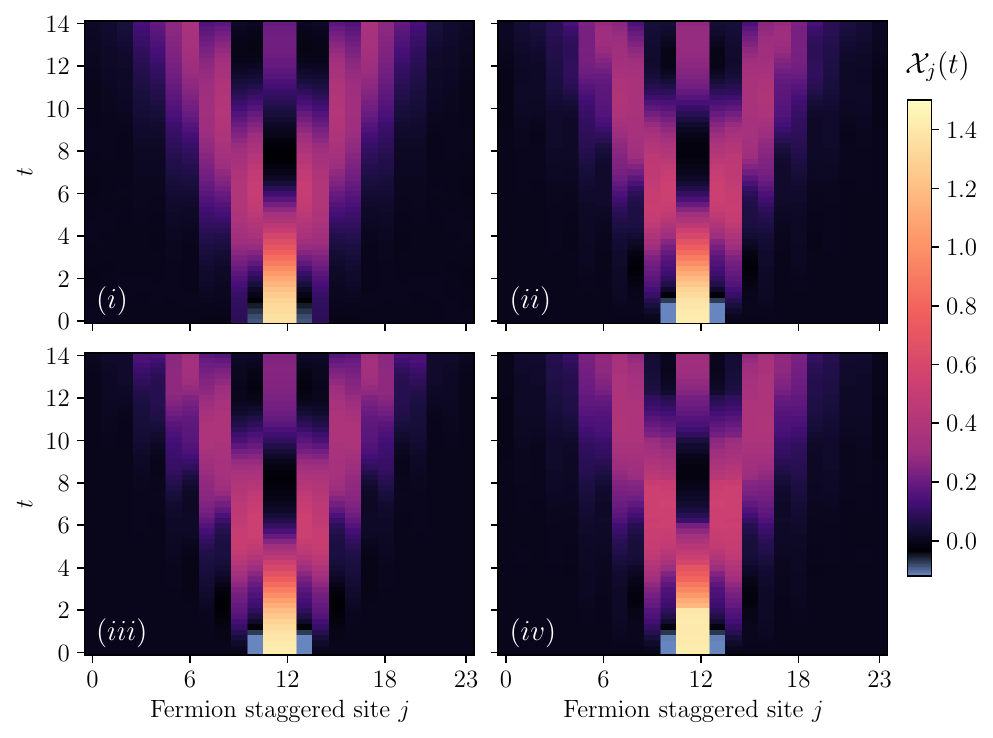}
    \caption{The effects of the approximations introduced in this work on the time evolution of the local chiral condensate ${\cal X}_j(t)$, given in Eq.~\eqref{eq:chij}.
    $(i)$ Without approximations: time evolution of the adiabatically-prepared hadron wavepacket with exact exponentiation of the full Hamiltonian, $\hat{H}$ in Eq.~\eqref{eq:Hgf}.
    $(ii)$ Approximate initial state preparation: time evolution of the 
    2-step SC-ADAPT-VQE hadron wavepacket built on top of the 
    2-step SC-ADAPT-VQE vacuum with exact exponentiation of the full Hamiltonian.
    $(iii)$ The same as $(ii)$, but with the electric interaction replaced with the $\overline{\lambda} = 1$ truncated interaction in Eq.~\eqref{eq:spatTrun1}.
    $(iv)$ The same as $(iii)$, but with time evolution implemented with $2\lceil\frac{t}{2} \rceil$ 
    second-order Trotter steps (maximum step size of $\delta t = 1$).}
    \label{fig:SystematicErr}
\end{figure}
Results obtained for the time evolved chiral condensate are shown in Fig.~\ref{fig:SystematicErr} with four different levels of approximation. 
Small errors are introduced with each approximation, but the results are found to recover expectations
within the uncertainties of the approximations.
Panel $(iv)$ in Fig.~\ref{fig:SystematicErr} shows the time-evolution operator approximated with $2\lceil\frac{t}{2}\rceil$ second-order Trotter steps, giving a maximum step size of $\delta t = 1$.
These step sizes introduce minimal (Trotter) errors, and will be used 
for the time evolution using a digital quantum computer presented in the next section.
The propagation of hadrons outward from an initially localized wavepacket is clearly identified in deviations of the local chiral condensate from its vacuum expectation value.
The oscillations of the condensate at the center of the wavepacket are consistent with expectations, and are discussed further in App.~\ref{app:SFTsrc}.
Due to the symmetry of the initial state, the hadron has equal amplitude to propagate in either direction, with a profile that is bounded by the speed of light ($1$ staggered site per unit time).

The (composite) hadrons that make up the wavepacket are 
(bosonic) vector particles, and some features of the hadron dynamics can be qualitatively understood in the simpler setting of non-interacting 1+1D scalar field theory.
In particular, the light-cone structure of propagating hadrons, the damped oscillations at the origin of the wavepacket and the effects of OBCs in both theories are similar.
This is treated in detail in App.~\ref{app:SFTsrc}, where the (textbook) example of a localized classical source coupled to a scalar field in 1+1D is treated in the continuum and on the lattice, and in App.~\ref{app:OBC}, where OBCs are compared to periodic boundary conditions (PBCs).

\section{Real-time Simulations using IBM's Digital Quantum Computers}
\label{sec:Qsim}
\noindent
The end-to-end simulations performed in the previous section using {\tt qiskit} and classical computers are scaled up to $L=56$ (112 qubits) 
and executed on IBM's 133-qubit {\tt ibm\_torino} Heron processor.
The scalability of the SC-ADAPT-VQE vacuum preparation circuits was demonstrated in Ref.~\cite{Farrell:2023fgd}, where it was shown that the variational parameters are reproduced well by an exponential in $L$.
This enables the extrapolation of the state preparation circuits, determined for $L\leq 14$, to arbitrarily large $L$.
In principle, a similar exponential convergence of parameters for the hadronic wavepacket preparation circuits is expected. 
However, as shown in Sec.~\ref{sec:SCADAPTWP}, the operator ordering and variational parameters of the SC-ADAPT-VQE wavepacket preparation have converged up to the tolerance of the optimizer by $L=14$. 
Therefore, the circuit structure and parameters determined for $L=14$ can be used to initialize the $L=56$ hadron wavepacket.
The operator ordering and parameters used to prepare the 
2-step SC-ADAPT-VQE vacuum and 2-step SC-ADAPT-VQE hadron wavepacket 
for $L=56$ are given in Table~\ref{tab:AnglesVACWP}.

Error mitigation is essential for successful simulations utilizing large quantum volumes~\cite{Kim:2023bwr}.
Here, our error mitigation methods are outlined, and a more detailed discussion can be found in App.~\ref{app:qSimDetails}. 
Through cloud-access, the circuits are sent to {\tt ibm\_torino} using the {\tt qiskit} sampler primitive, which includes both dynamical decoupling~\cite{Viola:1998dsd,2012RSPTA.370.4748S,Ezzell:2022uat} and {\tt M3} measurement mitigation~\cite{Nation:2021kye}.
To mitigate coherent two-qubit gate errors, Pauli twirling~\cite{Wallman:2016nts} is used on the native two-qubit gates, control-$Z$ for {\tt ibm\_torino}.
After twirling, we assume that the coherent two-qubit gate errors are transformed into statistically independent and unbiased incoherent errors, which can be modeled by a Pauli noise channel. 
Observables are then estimated using Operator Decoherence Renormalization (ODR)~\cite{Farrell:2023fgd}, which extends decoherence renormalization~\cite{Urbanek_2021,ARahman:2022tkr,Farrell:2022wyt,Ciavarella:2023mfc} to large systems.\footnote{Instead of setting the single-qubit rotations to zero in the mitigation circuits~\cite{Farrell:2022wyt,Ciavarella:2023mfc,Farrell:2023fgd}, 
they could be replaced by Clifford gates~\cite{Qin:2021jpm,Robbiati:2023eyl}.
}
To implement ODR, two kinds of circuits are run on the device: a ``physics'' circuit, and a ``mitigation'' circuit.
For a simulation of wavepacket dynamics, the physics circuit implements the time evolution of either the wavepacket or the vacuum (to compute ${\cal X}_j(t)$ in Eq.~\eqref{eq:chij}).
The mitigation circuit(s), with a priori known error-free (predicted) results, 
and the physics circuits have similar structures and similar error profiles.
From the mitigation circuits, deviations of measured observables ${\langle \hat{O} \rangle_{\text{meas}}}$ from their predicted values ${\langle \hat{O} \rangle_{\text{pred}}}$ are used to compute the depolarizing noise parameters,
\begin{equation}
\eta_{O} \ = \ 1 \ - \ \frac{\langle \hat{O} \rangle_{\text{meas}}}{\langle \hat{O} \rangle_{\text{pred}}}
\ .
\end{equation}
These $\eta_{O}$ are used to estimate the expectation values from the physics circuits (using the same relation).
For wavepacket (vacuum) time evolution, we choose a mitigation circuit that creates the wavepacket (vacuum), time evolves with half of the Trotter steps until $t/2$ and then evolves for $-t/2$ with the remaining Trotter steps~\cite{ARahman:2022tkr}.
This forwards-backwards time evolution corresponds to the identity operator in the absence of device errors, and restricts our simulations to an even number of Trotter steps.
To determine the $\eta_{O}$, the prediction of a desired observable from the mitigation circuit must be known.
In our case, this requires classically computing $\langle \hat{\chi}_j \rangle$ in both the SC-ADAPT-VQE vacuum and wavepacket.
This can be accomplished even for large systems using the {\tt qiskit} or {\tt cuQuantum} MPS simulator, as was demonstrated in Ref.~\cite{Farrell:2023fgd} for the SC-ADAPT-VQE vacuum up to $L=500$.
Interestingly, our numerical calculations highlight that it is the time evolution, and not the state preparation, that is difficult for classical MPS techniques.

\begin{table}[!tb]
\renewcommand{\arraystretch}{1.4}
\begin{tabularx}{\textwidth}{| c || c | Y | Y | Y | Y | Y || Y | Y |}
 \hline
 $t$ & $N_T$ & \# of CNOTs (per~$t$) & CNOT depth (per~$t$)& \# of distinct circuits (per~$t$)
 &   \# of twirls (per circuit) & \# of shots (per twirl) 
 &  Executed CNOTs  ($\scalemath{0.9}{\times 10^9}$) & Total \# of shots ($\scalemath{0.9}{\times 10^6}$) \\
   \hline
   \hline
    1 \& 2 & 2 &  2,746    & 70     & 4 & 480 & 8,000  & $4\times 2\times 10.5$ & $4\times 2\times 3.8$\\
    \hline
    3 \& 4 & 4 & 4,598 & 120 & 4 & 480 & 8,000 & $4\times 2\times 17.7$ & $4\times 2\times 3.8$\\
    \hline
    5 \& 6 & 6 & 6,450 & 170  & 4 & 480 & 8,000 & $4\times 2\times 24.8$ & $4\times 2\times 3.8$\\
    \hline
    7 \& 8 & 8 & 8,302 & 220 & 4  & 480 & 8,000 & $4\times 2\times 31.9$ & $4\times 2\times 3.8$\\
    \hline
    9 \& 10 & 10 & 10,154 & 270 & 4  & 160 & 8,000 & $4\times 2\times 13.0$  & $4\times 2\times 1.3$\\
    \hline
    11 \& 12 & 12 & 12,006 & 320 & 4  &  160 & 8,000 & $4\times 2\times 15.4$ & $4\times 2\times 1.3$\\
    \hline
    13 \& 14 & 14 & 13,858 & 370 & 4  & 160 & 8,000 & $4\times 2\times 17.7$ & $4\times 2\times 1.3$ \\
    \hline \hline
    {\bf Totals} & & & & & & & $1.05 \times 10^{12}$ & $1.54\times 10^8$ \\
    \hline
\end{tabularx}
\caption{Details of our quantum simulations performed using 112 qubits of IBM's {\tt ibm\_torino} Heron processor. 
For a given simulation time, $t$ (first column), the second column gives the number of employed Trotter steps $N_T$.
The third and fourth columns give the number of CNOTs and corresponding CNOT depth.
The CNOT totals given in the third column include the cancellations that occur during transpilation, and the CNOT depth should be compared to the minimum depth that is equal to twice the number of CNOTs/qubit (49, 82, 115, 148, 181, 214, 247 for increasing $N_T$) to assess the sparsity of the circuits. 
The fifth column gives the number of distinct circuits per $t$ (this number does not include the circuits needed for readout mitigation) and the sixth column gives the number of Pauli-twirls executed per distinct circuit.
For each twirl, 8,000 shots are performed (seventh column).
The total number of executed CNOT gates are given in the eighth column, and the total number of shots are given in the ninth column.
The total number of CNOT gates applied in this production is one trillion, and the total number of shots is 154 million.} 
 \label{tab:QsimCNOT}
\end{table}
\begin{figure}[!t]
    \centering
    \includegraphics[width=\columnwidth]{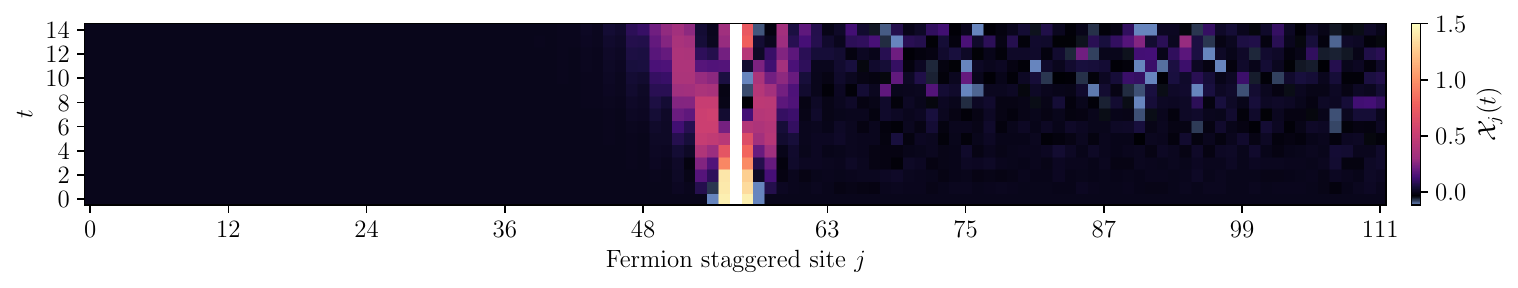}
    \caption{The time evolution of the vacuum subtracted chiral condensate $\mathcal{X}_j(t)$, defined in Eq.~\eqref{eq:chij}, for a $L=56$ (112 qubits) spatial-site lattice. 
    The initial state is prepared using the 
    2-step SC-ADAPT-VQE vacuum and wavepacket preparation circuits.
    Time evolution is implemented using a second-order Trotterization of the Hamiltonian with the $\overline{\lambda}=1$ truncated electric interaction.
    The left side shows the results of error-free classical simulations from the {\tt cuQuantum} MPS simulator, while the right side shows the CP-averaged results obtained using IBM's superconducting-qubit digital quantum computer {\tt ibm\_torino} (both sides show the MPS result for $t=0$).
    Due to CP symmetry, the right and left halves would be mirror images of each other in the absence of device errors.
    A more detailed view for each time slice is given in Fig.~\ref{fig:srcmvac_evol}, and discussions of the error-mitigation techniques are presented in the main text and App.~\ref{app:qSimDetails}.}
    \label{fig:IBMresultsMPS}
\end{figure}
We implement time evolution for $t=\{1,2,\ldots,14\}$ with $2\lceil\frac{t}{2} \rceil $ second-order Trotter steps (a maximum step size of $\delta t = 1$).
As shown in the previous section, this step size does not introduce significant Trotter errors.
The number of CNOTs and corresponding CNOT depth for each simulation time are given in Table~\ref{tab:QsimCNOT}, and range from 2,746 CNOTs (depth 70) for 2 Trotter steps to 13,858 CNOTs (depth 370) for 14 Trotter steps.
\begin{figure}[!htb]
    \centering
    \includegraphics[width=\columnwidth]{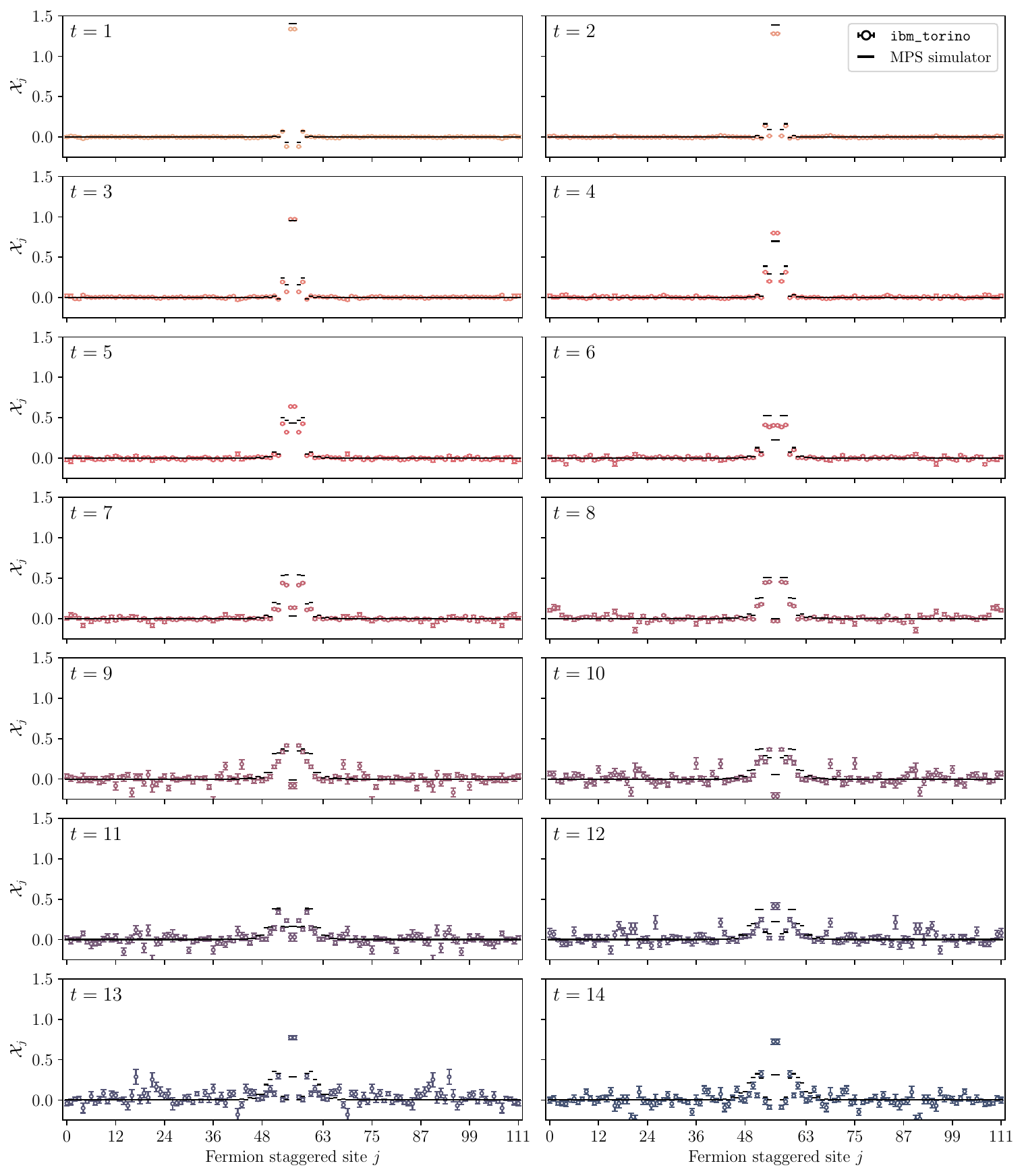}
    \caption{A detailed view of the time evolution of the vacuum subtracted chiral condensate shown in Fig.~\ref{fig:IBMresultsMPS} for each simulation time.
    The open circles are CP averaged results obtained using IBM's superconducting-qubit digital quantum computer {\tt ibm\_torino}.
    The black dashes are the error-free expectations obtained from the {\tt cuQuantum} MPS classical simulator.
    A complete discussion of the error-mitigation techniques, post processing and statistical uncertainties is presented in the main text and App.~\ref{app:qSimDetails}.
    The results are tabulated in App.~\ref{app:tables}.}
    \label{fig:srcmvac_evol}
\end{figure}
The results for ${\cal X}_j(t)$ obtained from {\tt ibm\_torino} and the MPS simulator are shown in Fig.~\ref{fig:IBMresultsMPS}, with a breakdown of each $t$ given in Fig.~\ref{fig:srcmvac_evol} (the separate evolutions of the wavepacket and vacuum are shown in Fig.~\ref{fig:src_vac_evol}).
For each time, four circuits are run; time evolution of the wavepacket, time evolution of the vacuum, forward-backward time evolution of the wavepacket and forward-backward time evolution of the vacuum.
For $t=1-8$, 480 twirled instances of each circuit are run, and for $t=9-14$, 160 twirled instances are run. 
Each twirled instance has 8,000 shots, using a total of $\sim 1.5 \times 10^8$ shots for the complete production.
We have estimated the uncertainties in the results from the quantum computer using bootstrap-mean resampling.\footnote{Due to the noisy nature of the device, the utility of the Hodges-Lehmann (HL) estimator was studied, and consistent results were obtained.
The HL estimator has been considered in lattice QCD studies to mitigate the impact of outliers in nuclear correlation functions~\cite{Beane:2014oea,Orginos:2015aya,NPLQCD:2020lxg}.}
The expected results are determined by using the {\tt cuQuantum} MPS simulator with maximum bond dimension 200.
The run time and convergence of the MPS simulations are discussed in App.~\ref{app:MPSSim}.

The individual time evolutions of the wavepacket and vacuum, used to compute ${\cal X}_j(t)$, are shown in Fig.~\ref{fig:src_vac_evol} of App.~\ref{app:qSimDetails}.
A systematic error in the chiral condensate away from the center of the lattice is seen to increase with simulation time.
Fortunately, it is similar for the wavepacket and vacuum evolution, and largely cancels in the subtraction to form ${\cal X}_j(t)$, as shown in Fig.~\ref{fig:srcmvac_evol}.
The origin of this systematic error is currently unknown to us, and either stems from a deficiency in our error-mitigation techniques, or from insufficient convergence in the MPS simulations.
Without the approximations in the state preparation and time evolution, the chiral condensate would not evolve in regions that are locally the vacuum.
This qualitatively holds for smaller systems with $L\leq 14$ that can be simulated exactly.
The results from the quantum computer agree with these expectations, showing little evolution of the chiral condensate in the vacuum (right column of Fig.~\ref{fig:src_vac_evol}).
The MPS simulations, on the other hand, show significant evolution of the vacuum chiral condensate.
For the range of maximum bond dimensions we have been able to explore, it appears that the chiral condensate has converged at the level of $10^{-2}$ for late times.
However, these results are not exact, and at this point we cannot rule out systematic errors being present in the MPS simulations.
From preliminary investigations, it appears that the vacuum evolution is due to $\overline{\lambda}=1$ being too small for exponential convergence.
This is not surprising since the relevant ratio for exponential convergence is $\propto \overline{\lambda}/\overline{\xi}$, with possibly a prefactor proportional to, for example, $2\pi$.
However, the maximum bond dimension required for convergence becomes significantly larger with increasing $\overline{\lambda}$, and it is unclear if this conclusion is consistent.
A future detailed study of the effects of increasing the precision of the state preparation, increasing $\overline{\lambda}$, and increasing the number of Trotter steps will be needed to determine if this discrepancy is due to errors in the MPS simulation or from imperfect error mitigation.

The results shown in Figs.~\ref{fig:IBMresultsMPS} and \ref{fig:srcmvac_evol} demonstrate that, by implementing a series of exponentially convergent approximations (beyond Trotterization), wavepackets of hadrons can be prepared and evolved forward in time with available quantum computers.
Propagating hadrons are clearly identified as a disturbance in the chiral condensate, with random fluctuations due to device errors outside of the hadron's light-cone.
It should be emphasized that obtaining ${\cal X}_n(t) = 0$ outside of the light-cone using IBM's device is a non-trivial result, as it requires cancellations between the wavepacket and vacuum evolutions.
The simulations performed using {\tt ibm\_torino} show qualitative agreement with classical MPS results, but degrade with increasing number of Trotter steps (circuit depth).
The simulations highlight that device errors dominate over the systematic errors due to approximate state preparation and time evolution.
The results qualitatively recover expectations, but often differ by many standard deviations from classical expectations, indicating that we do not have a complete quantification of uncertainties.
This is not surprising given the simplicity and limitations of the assumed error model.
Despite the device errors, it is clear that current hardware is capable of creating and possibly colliding (composite) hadrons over a meaningful time interval.
Such simulations could provide first glimpses of inelastic hadron scattering and fragmentation in the Schwinger model that are beyond present capabilities of classical computing.

\section{Summary and Outlook}
\label{sec:Summary}
\noindent
Quantum computing offers the potential of reliably simulating the collisions of high-energy hadrons and nuclei directly from quantum chromodynamics, the quantum field theory describing the strong interactions.
First steps are being taken to develop scalable techniques and algorithms for QCD simulations by working with the Schwinger model defined in 1+1D. 
Towards these goals, this work develops
protocols for quantum simulations of hadron dynamics that are demonstrated on a $L=56$ (112 qubit) lattice using IBM's superconducting-qubit digital quantum computer, {\tt ibm\_torino}.
These simulations start with establishing a wavepacket of hadrons in the center of the lattice on top of the vacuum.
The necessary quantum circuits for the creation of this wavepacket are determined using the SC-ADAPT-VQE algorithm that was recently introduced by the authors in Ref.~\cite{Farrell:2023fgd}.
In SC-ADAPT-VQE, low-depth circuits for state preparation are determined on a series of small lattices using {\it classical computers}, and then systematically scaled up to prepare states on a {\it quantum computer}.
For the present purposes, the SC-ADAPT-VQE circuits are variationally optimized to have maximal overlap with an adiabatically prepared hadron wavepacket.
The vacuum and hadronic wavepacket that are initialized on the quantum computer are then time evolved using a second-order Trotterization of the time evolution operator.
Naively, the electric interaction between fermions is all-to-all, giving rise to a prohibitive ${\cal O}(L^2)$ scaling in the number of two-qubit gates needed for time evolution.
Motivated by confinement, an approximation that truncates the electric interaction between distant charges is introduced.
This interaction converges exponentially with increasing interaction distance, and improves the scaling of the number of two-qubit gates required for time evolution to ${\cal O}(\overline{\lambda} L)$, where $\overline{\lambda}$ is proportional to the confinement length scale.
These new methods for state preparation are verified on small systems using a classical simulator, and then applied to time evolve hadron wavepackets on a $L=56$ (112 qubit) lattice using {\tt ibm\_torino}.
Our digital quantum simulations utilize some of the largest quantum volumes to date~\cite{Yu:2022ivm,Kim:2023bwr,Shtanko:2023tjn,Farrell:2023fgd,Baumer:2023vrf,Chen:2023tfg,Liao:2023eug,Chowdhury:2023vbk}, with up to 13,858 two-qubit entangling gates applied (CNOT depth of 370).
A large number of shots with which to implement the error mitigation techniques is found to be essential to the success of our simulations. Our results show clear signatures of hadron propagation through modifications of the local chiral condensate.

Real-time dynamics typically explore highly-entangled regions of Hilbert space and, as a result, classical methods scale unfavorably with simulation time $t$, lattice volume $L$, and energy.
To explore this in more detail, our quantum simulations have been compared to classical MPS circuit simulations using {\tt qiskit} and {\tt cuQuantum}.
We have found that our initial state preparation circuits can be simulated relatively easily with these simulators.
However, the bond dimension needed for proper convergence grows rapidly as more steps of Trotterized time evolution are added to the quantum circuit.
All of this points to a potential near-term quantum advantage for the simulation of hadronic dynamics.
In particular, it is likely that the simulation of high-energy hadronic collisions will exceed the capabilities of classical computing for simulation times and volumes that are not excessively large.
Exactly where such a quantum advantage can be realized remains to be established.

On this path, future work will use the hadron wavepacket preparation and time evolution circuits that we have presented here to simulate hadron scattering. 
Evolving out to later times will require time-evolution methods that improve upon Trotterization.
A promising direction is to use SC-ADAPT-VQE to find low-depth circuits for simulating over the early times.
The light-cone restricts early-time dynamics to only a modest number of qubits, and scalable low-depth circuits can likely be found with classical computing.
Another direction worth pursuing  is to approach the continuum by taking $m$ and $g$ smaller, increasing the correlation length.
These longer correlation lengths will require deeper state-preparation circuits and larger truncations of the electric interaction to reach a target simulation quality.
Further into the future, improved methods for hadron detection will also be needed. 
Finally, it will be necessary to extend these techniques to non-Abelian gauge theories and higher dimensions to perform more realistic simulations of QCD.

\begin{acknowledgements}
\noindent
We would like to thank Silas Beane, Ivan Chernyshev, Niklas Mueller and Nikita Zemlevskiy for insightful comments. 
We would also like to thank the participants of the {\it Quantum Error Mitigation for Particle and Nuclear Physics} IQuS-INT workshop held between May 9-13, 2022 for helpful discussions. We would like to thank Yang Gao (NVIDIA) for help in developing our {\tt cuQuantum} MPS simulation code.
This work was supported, in part, by the U.S. Department of Energy grant DE-FG02-97ER-41014 (Farrell), by U.S. Department of Energy, Office of Science, Office of Nuclear Physics, InQubator for Quantum Simulation (IQuS)\footnote{\url{https://iqus.uw.edu/}} under Award Number DOE (NP) Award DE-SC0020970 via the program on Quantum Horizons: QIS Research and Innovation for Nuclear Science\footnote{\url{https://science.osti.gov/np/Research/Quantum-Information-Science}} (Ciavarella, Farrell, Savage), the Quantum Science Center (QSC)\footnote{\url{https://qscience.org}} which is a National Quantum Information Science Research Center of the U.S.\ Department of Energy (DOE) (Illa), and by the U.S. Department of Energy (DOE), Office of Science under contract DE-AC02-05CH11231, through Quantum Information Science Enabled Discovery (QuantISED) for High Energy Physics (KA2401032) (Ciavarella).
This work is also supported, in part, through the Department of Physics\footnote{\url{https://phys.washington.edu}} and the College of Arts and Sciences\footnote{\url{https://www.artsci.washington.edu}} at the University of Washington.
This research used resources of the Oak Ridge Leadership Computing Facility (OLCF), which is a DOE Office of Science User Facility supported under Contract DE-AC05-00OR22725.
We acknowledge the use of IBM Quantum services for this work. The views expressed are those of the authors, and do not reflect the official policy or position of IBM or the IBM Quantum team. This work was enabled, in part, by the use of advanced computational, storage and networking infrastructure provided by the Hyak supercomputer system at the University of Washington.\footnote{\url{https://itconnect.uw.edu/research/hpc}}
This research was done using services provided by the OSG Consortium~\cite{osg07,osg09,osgweb,osgweb2}, which is supported by the National Science Foundation awards \#2030508 and \#1836650.
We have made extensive use of Wolfram {\tt Mathematica}~\cite{Mathematica}, {\tt python}~\cite{python3,Hunter:2007}, {\tt julia}~\cite{Julia-2017}, {\tt jupyter} notebooks~\cite{PER-GRA:2007} in the {\tt Conda} environment~\cite{anaconda}, NVIDIA {\tt cuQuantum} Appliance~\cite{cuquantum}, and IBM's quantum programming environment {\tt qiskit}~\cite{qiskit}. The DMRG calculations were performed using the C++ {\tt iTensor} library~\cite{fishman2022itensor}.
\end{acknowledgements}

\clearpage
\appendix

\FloatBarrier
\section{The Classical Dynamics of a Sourced Non-Interacting Scalar Field}
\label{app:SFTsrc}
\noindent
The spectrum of the Schwinger model consists of composite hadrons due to confinement.
Unlike the underlying electron and positron degrees of freedom, which are fermions, the hadrons are bosonic scalar and vector particles.
Important features of the hadronic dynamics simulated in this work can be understood in the simpler setting of a non-interacting scalar field evolving from a localized source.
The framework for the latter is well known, and can be found in quantum field theory textbooks, for example, Ref.~\cite{Peskin:1995ev}.
The spatial and temporal extents of the hadron wavepacket (in the Schwinger model) that we work with are approximately determined by the correlation length, $\xi$, and we model this by a Gaussian source for the scalar field (describing the Schwinger model vector hadron). The Klein-Gordon equation in the presence of a classical source,
\begin{equation}
    \left( \partial^\mu\partial_\mu + m^2 \right)\phi(x,t) = j(x,t)
    \ ,
    \label{eq:KG1}
\end{equation}
is solved in 1+1D in infinite volume and with vanishing lattice spacing by
\begin{equation}
    \phi_j(x,t) = \phi_{j=0}(x,t) 
    \ +\ 
    i \int dy\ dt^\prime \ G_R(x-y, t-t^\prime) j(y,t^\prime)
    \ ,
    \label{eq:KG2}
\end{equation}
where $G_R(a,b)$ is the retarded Green's function and $\phi_{j=0}(x,t)$ is the field in the absence of the source.
The effective source we consider is 
\begin{equation}
    j(x,t) = J_0\ \sqrt{\frac{\alpha}{\pi}}\ e^{-\alpha x^2}\ \delta(t)
    \ \ ,\ \ 
    \int dx\ dt\ j(x,t) = J_0
    \ .
    \label{eq:KG3}
\end{equation}
After writing the propagator in momentum space, and using the spatial symmetry of the source, the field in the presence of the source is given by
\begin{equation}
    \phi_j(x,t) = \phi_{j=0}(x,t) 
    \ +\ 
2 J_0 \int_0^\infty \frac{dp}{2\pi} \frac{e^{-p^2/(4 \alpha)}}{\omega_p}\ 
\cos p x\ \sin \omega_p t \ ,
    \label{eq:KG4}
\end{equation}
where $\omega_p = \sqrt{p^2+m^2}$.
In a finite volume with a discrete set of uniformly spaced lattice points, it is straightforward to derive the appropriate analogous relation.
Spatial integrals are replaced by a discrete sum over the finite number lattice sites, and momentum integrals are replaced by sums over momentum modes within the first Brillouin zone (the exact set of modes are determined by the selected boundary conditions imposed on the field).
\begin{figure}[!b]
    \centering
    \includegraphics[width=0.95\columnwidth]{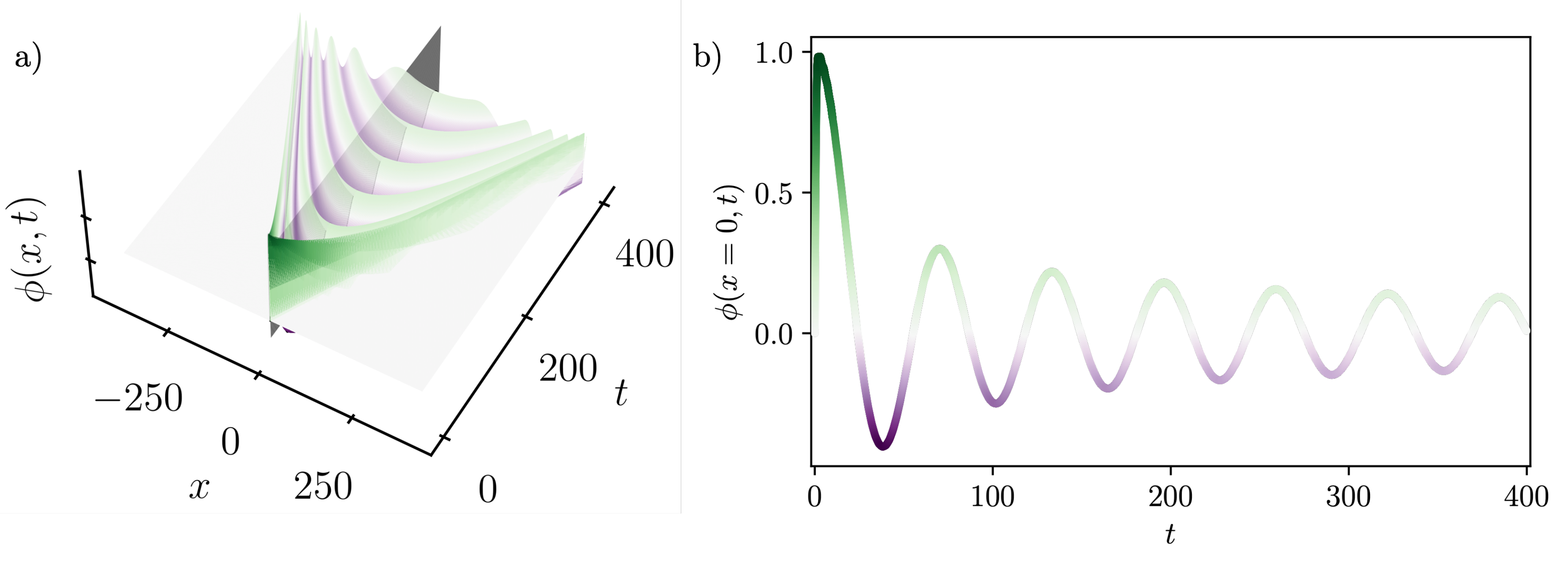}
    \caption{a) The free scalar field downstream from a Gaussian source given in Eq.~\eqref{eq:KG3}
    with $m=0.1$ and $\alpha=J_0=1$, determined by Eq.~\eqref{eq:KG4}. b) Profile of the scalar field at $x=0$.}
    \label{fig:KGdynamics}
\end{figure}
Figure~\ref{fig:KGdynamics} shows the downstream field in spacetime from the source given in Eq.~\eqref{eq:KG3}, with parameters $m=0.1$ and $\alpha=J_0=1$.
The light cone at $x=t$ is clear, with the field decaying exponentially beyond these lines. 
Importantly, the field near the origin is seen to ``ring down'', continuing to emit particles until the initially localized energy density is dispersed via particle production.

The total energy injected into the field by the source is
\begin{equation}
\langle \hat H \rangle = \sqrt{\frac{\alpha}{8 \pi}}\ J_0^2 \ ,
    \label{eq:KG5}
\end{equation}
where $\hat{H}$ is the free Hamiltonian without the source, and the energy of the vacuum has been set to zero.
The probability of creating a particle in the $|p\rangle$ momentum state, ${\rm Prob}(|p\rangle)$, and the expectation value of the total number of particles produced in such an event, $N_\phi$, are
\begin{equation}
{\rm Prob}(|p\rangle) = 
\frac{J_0^2}{2 \omega_p} e^{-\frac{p^2}{2 \alpha}}
\ \ ,\ \ 
N_\phi  =  
\frac{J_0^2}{4\pi} \ e^{\frac{m^2}{4\alpha}}\ K_0\left(\frac{m^2}{4\alpha}\right) \ ,
    \label{eq:KG6}
\end{equation}
with $K_0$ being the modified Bessel function of the second kind of order zero.

\FloatBarrier
\section{Aspects of Open Boundary Conditions}
\label{app:OBC}
\noindent
Ideally, quantum simulations of lattice field theories would utilize periodic boundary conditions (PBCs) in order to maintain the translation invariance of free space (in the continuum limit).
However, without connectivity between the initial and final lattice sites, as is the case in some quantum computers, simulations can be performed with OBCs.
In this appendix, we demonstrate some key features of OBCs in the context of scalar field theory, and make connections to the Schwinger model.

The Hamiltonian describing non-interacting lattice scalar field theory with continuous fields at each lattice site and with OBCs is given by 
\begin{equation}
\hat H_{\rm lsft} = 
\frac{1}{2} \sum_{j=0}^{L-1} \hat \Pi_j^2
+ \frac{1}{2} \sum_{j=0}^{L-1} m_0^2 \hat \phi_j^2
-\frac{1}{2} \sum_{\substack{j=0\\ j-1\ge 0\\ j+1\le L-1}}^{L-1} 
\hat\phi_j (\hat\phi_{j+1}+\hat\phi_{j-1}-2\hat\phi_j)
= 
\frac{1}{2}
\hat \Pi^2 
\ +\ 
\frac{1}{2}
\Phi^T \left[\ 
m_0^2 \hat I
\ +\ 
{\cal G}
\right] \Phi \ ,
    \label{eq:OBC1}
\end{equation}
where 
\begin{equation}
{\cal G} = 
\left(
\begin{array}{cccccc}
2 & -1 & 0 & 0 & \cdots & 0 \\
-1 & 2 & -1 & 0 & \cdots & 0 \\
0 & -1 & 2 & -1 & \cdots & 0 \\
\vdots & & & & & \vdots \\
0 & 0 & 0 & 0 & \cdots & -1 \\
0 & 0 & 0 & 0 & \cdots & 2 \\
\end{array}
\right)
\ \ ,\ \ 
\Phi^T = \left(\phi_0, \phi_1, \cdots ,\phi_{L-1} \right) \ ,
    \label{eq:OBC2}
\end{equation}
and where $\hat\Pi$ is the conjugate-momentum operator.
The only difference between this expression and that for PBCs is the absence of terms in the extreme anti-diagonal entries in ${\cal G}$, which renders the matrix non-circulant, reflecting the lack of discrete translational invariance.
An orthogonal transformation can be applied to the fields to diagonalize the Hamiltonian matrix,
\begin{equation}
\Phi = V \Psi
\ \ ,\ \
\hat H \ = \ 
\frac{1}{2}
\hat \Pi^2
\ +\ 
\frac{1}{2}
\Psi^T  
\ 
\Omega^2 \ \Psi 
\ ,
    \label{eq:OBC3}
\end{equation}
where $\Omega$ is a $L\times L$ diagonal matrix with eigenvalues $\omega_i$.
Therefore, the $L$ towers of single-particle energy eigenvalues of these systems are 
\begin{equation}
E_i = \left(n_i + \frac{1}{2} \right) \omega_i \ ,
    \label{eq:OBC4}
\end{equation}
where $n_i$ are the number of bosons with energy $\omega_i$, with a vacuum energy that is the sum of zero-point energies,
\begin{equation}
E_{\rm vac} = \frac{1}{2} \sum_i \omega_i \ .
    \label{eq:OBC5}
\end{equation}
%

\subsection{OBCs and PBCs for \texorpdfstring{$L=4$}{L=4}}
\label{app:OBCL4}
\noindent
It is instructive to consider the similarities and differences 
between OBCs and PBCs for non-interacting scalar field theory on $L=4$ lattice sites.
It is well known that the structure of the Hamiltonian in Eq.~\eqref{eq:OBC1} indicates that this (and other such systems) can be diagonalized by the eigenvectors of ${\cal G}$, and are hence independent of the mass and conjugate momentum (as these are both local operators).

For OBCs, the $\omega_i$ are
\begin{align}
\omega_i & =
\left\{ \sqrt{m_0^2 + \frac{1}{2}(3-\sqrt{5})} , 
\sqrt{m_0^2 + \frac{1}{2}(5-\sqrt{5})} ,
\sqrt{m_0^2 + \frac{1}{2}(3+\sqrt{5})} , 
\sqrt{m_0^2 + \frac{1}{2}(5+\sqrt{5})} \right\}
\nonumber\\
& = 
\left\{ \sqrt{m_0^2 + \frac{1}{2}(3-\sqrt{5})} , 
\sqrt{m_0^2 + 2 + \frac{1}{2}(1-\sqrt{5})} , 
\sqrt{m_0^2 + 2 - \frac{1}{2}(1-\sqrt{5})} , 
\sqrt{m_0^2 + 4 - \frac{1}{2}(3-\sqrt{5})} \right\}
\nonumber\\
& = 
\left\{ \sqrt{m_0^2 + 0.3819} , 
\sqrt{m_0^2 + 1.3819} , 
\sqrt{m_0^2 + 2.6180} , 
\sqrt{m_0^2 + 3.6180} \right\}
\ ,
\label{eq:OBCL4}
\end{align}
which are to be compared with those from PBCs,
\begin{align}
\omega_i & =
\left\{ m_0 , 
\sqrt{m_0^2 + 4\sin^2 \frac{\pi}{4}} ,
\sqrt{m_0^2 + 4\sin^2 \frac{\pi}{4}} , 
\sqrt{m_0^2 + 4\sin^2 \frac{\pi}{2}} \right\}
\nonumber\\
& =
\left\{ m_0 ,
\sqrt{m_0^2 + 2} , 
\sqrt{m_0^2 + 2} , 
\sqrt{m_0^2 + 4} \right\}
\ .
\label{eq:PBCL4}
\end{align}
The kinetic contributions to the energies in Eq.~\eqref{eq:OBCL4} correspond to ``momentum modes'' with $k=n \pi/5$ with $n=\{1,2,3,4\}$, and generalizes to $k=n \pi/(L+1)$ with $n=\{1,\ldots, L\}$.\footnote{A more direct comparison between  
Eq.~(\ref{eq:OBCL4}) and Eq.~(\ref{eq:PBCL4}) can be made using relations such as $4 \sin^2 \frac{\pi}{10} = (3-\sqrt{5})/2$.}
The energies of the OBC states are split around the energies of the PBC states, 
with the lowest is raised, and highest lowered.
This splits the degeneracies of the left- and right-moving momentum eigenstates associated with PBCs.
These features extend to larger values of $L$, with the splittings reducing with increasing $L$.

\begin{figure}[!ht]
	\centering
	\includegraphics[width=\columnwidth]{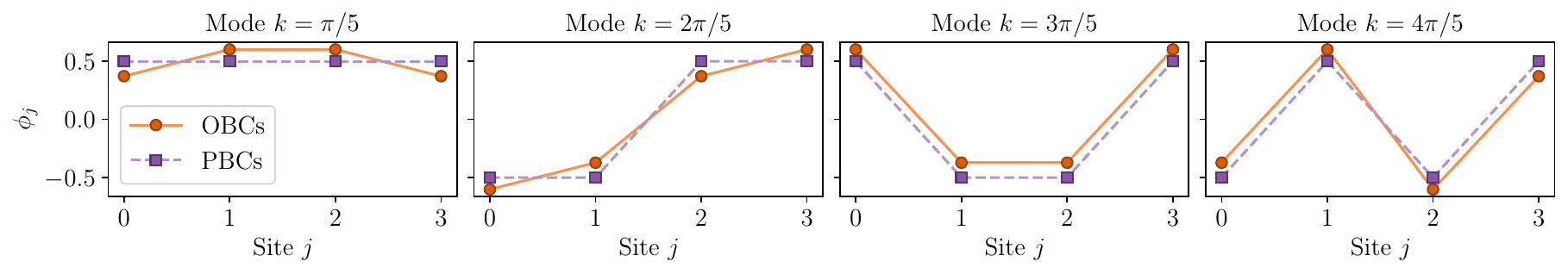}
	\caption{The eigenvectors of the $L=4$ lattice scalar field theory with PBCs (purple) and OBCs (orange).}
	\label{fig:OPBCevecsL4}
\end{figure}

The eigenstates can all be made real by global phase rotations, and identification of these states with the associated states with PBCs can be made by forming linear combinations of the degenerate PBC states.
Figure~\ref{fig:OPBCevecsL4} shows the eigenstates for PBCs and OBCs.
Even for $L=4$, the difference between the eigenstates is not large, and diminishes with increasing $L$.
%

\subsection{Matching the Schwinger Model to Non-Interacting Scalar Field Theory for \texorpdfstring{$L=8$}{L=8} and \texorpdfstring{$L=14$}{L=14}
with OBCs}
\label{app:OBCSMmatch}
\noindent
In large enough spatial volumes, 
it is expected that the low-lying continuum states of the Schwinger model will be approximately recovered by an effective field theory (EFT) of scalar and vector particles~\cite{Luscher:1986pf,Gasser:1987zq,Luscher:1990ux,2001afpp.book..683V,Beane:2003da,PhysRevD.70.074029,Colangelo_2005}. 
To explore this more with OBC simulations, the mass of the scalar particle needs to be determined from the spectrum of the Schwinger model.
As the energies of the states of the scalar field depend in a non-trivial way on the mass of the scalar particle, this is accomplished numerically.

In the Schwinger model, fermions are discretized on a lattice with $2L$ staggered sites, corresponding to $L$ spatial sites.
To match to the spectrum of lattice scalar field theory, a conversion must be performed to switch from units of staggered lattice spacing, $a_{st}$, to units of spatial lattice spacing, $a_{sp}$.
A dimensionless energy, $\Delta_{st}$, in the Schwinger model is related to a physical energy by $\Delta_{st} = a_{st} E /(\hbar c)$, where $\hbar c=197.32$ MeV fm, and $E$ is an energy in units of MeV.
The corresponding quantity in terms of the spatial lattice spacing is $\Delta_{sp} = a_{sp} E /(\hbar c) = 2 a_{st} E /(\hbar c) = 2 \Delta_{st}$.

Exact diagonalization of the Schwinger model Hamiltonian with parameters $m=0.5,g=0.3,L=8$ gives a gap to the first excited state (vector hadron mass) of $a_{st} E_1 = a_{st} m_{\text{hadron}} = 1.15334$.
In the $L=8$ non-interacting scalar field theory, this corresponds to an excitation of $a_{sp}\  \omega_1= 2 m_{\text{hadron}} = 2.30668$.
Fitting the bare scalar field mass $m_0$ to this value gives $m_{0}^{({\rm fit})} = 2.28039$ in spatial lattice units, which can be then used to predict higher-lying states in the Schwinger model spectrum.
Converting back to the staggered lattice spacing gives the values of $a_{st}\omega_i$ to be compared with the exact results from the Schwinger model, $a_{st}E_i$, shown in Table~\ref{tab:L8comp}.
\begin{table}[!t]
\renewcommand{\arraystretch}{1.4}
\begin{tabularx}{\textwidth}{|c || Y | Y | Y | Y | Y | Y | Y | Y | Y | Y | }
 \hline
Quantity & State 1 & State 2 & State 3 & State 4 & State 5 & State 6 & State 7 & State 8 & State 9 & State 10  \\
 \hline\hline
$a_{st} \ E_i$ &1.15334& 1.19133& 1.25209& 1.33035& 1.33728& 1.38401& 1.41968& 1.44693& 1.47535& 1.51249\\
\hline
$a_{st} \ \omega_i$ &1.15334& 1.19039& 1.24501& 1.30890 & - & 1.37363& 1.43180 & - & 1.47752& 1.50662\\
 \hline
\end{tabularx}
\caption{The lowest-lying energies, $a_{st} \ E_i$,  of the Schwinger model with Hamiltonian parameters $m=0.5, g=0.3$ and  $L=8$. These are compared with the lowest-lying eigenvalues of a non-interacting scalar field theory with OBCs, $a_{st} \ \omega_i$, with a scalar mass parameter fit to reproduce $a_{st} \ E_1$.}
 \label{tab:L8comp}
\end{table}
Each of the energies $a_{st} \ \omega_i$ can be identified with an energy in the Schwinger model, within $\sim 2\%$, indicating that the low-lying spectrum is largely from the motion of a single hadron on the lattice.
We assume that the two states that do not correspond to states in the scalar theory result from internal excitations of the single particle state in the Schwinger model.
This analysis can be repeated for $L=14$ where it is found that $a_{st} E_1 = a_{st} m_{\text{hadron}} =  1.1452$ and $m_{0}^{({\rm fit})} = 2.28096$ (spatial lattice units). 
These quantities are very similar to the $L=8$ ones, as expected since $m_{\text{hadron}} \ll L$ and finite-size effects are small.
Table~\ref{tab:L14comp} shows the energy levels in the Schwinger model compared with those predicted from non-interacting scalar field theory fit to the lowest level.
Good agreement is again found, supporting the identification of the excited states in the Schwinger model with OBC momentum modes.
\begin{table}[!t]
\renewcommand{\arraystretch}{1.4}
\begin{tabularx}{\textwidth}{|c || Y | Y | Y | Y | Y | }
 \hline
Quantity & State 1 & State 2 & State 3 & State 4 & State 5 \\
 \hline\hline
$a_{st} \ E_i$ &1.1452 &1.1588&1.1812&1.2118&1.2496\\
\hline
$a_{st} \ \omega_i$ &1.1452&1.1592&1.1816&1.2108&1.2452\\
 \hline
\end{tabularx}
\caption{The same as Table~\ref{tab:L8comp} but for $L=14$.}
 \label{tab:L14comp}
\end{table}
%

\subsection{Sources with OBCs}
\label{app:OBCSMsrcs}
\noindent
The analysis in App.~\ref{app:SFTsrc} related to source dynamics in non-interacting scalar field theory is performed in infinite volume and in the continuum limit.
To better understand the impact of finite-volume and OBCs, it is helpful to consider the retarded-Green's function on such lattices.
The Green's function in Eq.~\eqref{eq:KG2} in 3+1D is given by 
\begin{align}
D_R({\bf x},{\bf y},t,0) & =
\theta(t)\ 
\int\ \frac{d^3{\bf k}}{(2\pi)^3}\ \frac{1}{2\omega_k}\ 
\left(
e^{-i \left( \omega_k t - {\bf k}\cdot ({\bf x}-{\bf y}) \right)}
\ -\ 
e^{+i \left( \omega_k t - {\bf k}\cdot ({\bf x}-{\bf y}) \right)}
\right)
\nonumber\\
& = - i \theta(t)\ 
\int\ \frac{d^3{\bf k}}{(2\pi)^3}\ \frac{1}{\omega_k}\ 
\sin \omega_k t\ 
e^{i  {\bf k}\cdot ({\bf x}-{\bf y}) } \ .
    \label{eq:GROBC1}
\end{align}
In a 3+1D finite volume with PBCs, this becomes
\begin{equation}
D_R({\bf x},{\bf y},t,0) \rightarrow 
- i \theta(t)\ 
\frac{1}{L^3} \sum_{\bf k}
\ \frac{1}{\omega_{\bf k}}\ 
\sin \omega_{\bf k} t\ 
e^{i  {\bf k}\cdot ({\bf x}-{\bf y}) }
 =
- i \theta(t)\ 
\sum_{\bf n}
\ \frac{1}{\omega_{\bf n}}\ 
\sin \omega_{\bf n} t\ 
\psi_{\bf n}^\dagger({\bf y})
\psi_{\bf n}({\bf x}) \ ,
    \label{eq:GROBC2PBC}
\end{equation}
where $\psi_{\bf n}({\bf x})$ is an appropriately normalized lattice eigenstate subject to PBCs, defined by a triplet of integers ${\bf n}$,
\begin{equation}
\psi_{\bf n}({\bf x}) = \frac{1}{L^{3/2}}\ 
e^{i  {\bf k}\cdot {\bf x} }
\ \ ,\ \ 
\sum_{\bf x}
\psi_{\bf n}^\dagger({\bf x}) \psi_{\bf m}({\bf x}) = \delta^{(3)}_{{\bf n},{\bf m}}
\ \ ,\ \ 
\sum_{\bf n}
\psi_{\bf n}^\dagger({\bf y}) \psi_{\bf n}({\bf x}) = \delta^{(3)}_{{\bf x},{\bf y}} \ ,
    \label{eq:GROBC2PBCb}
\end{equation}
with ${\bf k}=2 \pi {\bf n}/L$.
To transition to OBCs, the OBC eigenstates $\psi_{\bf m}({\bf x})$ are used.
For simulations in 1+1D with OBCs, the relevant retarded Green's function is
\begin{equation}
D_R(x,y,t,0) = 
- i \theta(t)\ 
\sum_n
\ \frac{\sin \omega_n t}{\omega_n}\ 
\psi_n^\dagger(y)
\psi_n(x) \ ,
    \label{eq:GROBC2OBCa}
\end{equation}
with appropriately orthonormalized wavefunctions, such as those shown in Fig.~\ref{fig:OPBCevecsL4}.

Consider a source with a Gaussian profile, as was considered earlier, 
on a lattice of length $L$,
\begin{equation}
j_L(y) = \eta\ \sum_{n=0}^{L-1}\ \delta (y-n)\ e^{-\alpha (y - \frac{L-1}{2})^2} \ .
    \label{eq:jy}
\end{equation}
where $\eta$ is the appropriate normalization factor determined by requiring,
\begin{equation}
\int_{-\infty}^{+\infty}\ dy\ j_L(y) = J_0
 \ = \ \
 \eta \sum_{n=0}^{L-1} e^{-\alpha \left (\frac{L-1-2n}{2}\right )^2 } 
 \ \approx \ \eta \sqrt{\frac{\pi}{\alpha}}\ \left[
1\ +\ 2\sum_{p=1}^\infty\ (-)^p e^{-\pi^2 p^2/\alpha} \right]
 \equiv 
\eta \sqrt{\frac{\pi}{\alpha}}\  S(\alpha) \ .
    \label{eq:jyb}
\end{equation}
The approximate equality holds for a well-localized source with large $L$ and small $\alpha$ (in which case the bounds of the sum can extended to $\pm\infty$ with exponentially-suppressed errors, and the Poisson resummation formula can be used).
The function $S(\alpha)$ rapidly approaches the continuum result of unity, for decreasing $\alpha$.
Therefore, the sources can be written as
\begin{equation}
j_L(y) = J_0\ \sqrt{\frac{\alpha}{\pi}}\ \frac{1}{S(\alpha)}\ 
\sum_{n=0}^{L-1}\ \delta (y-n)\ e^{-\alpha (y - \frac{L-1}{2})^2} \ ,
    \label{eq:jyc}
\end{equation}
which is the discrete version of Eq.~\eqref{eq:KG3}.
The expression for the downstream field from the source is given by Eq.~\eqref{eq:KG2}, and can be written as
\begin{equation}
    \phi_j(x,t) = \phi_{j=0}(x,t) 
    \ +\ 
    J_0\ \sqrt{\frac{\alpha}{\pi}}\ \frac{1}{S(\alpha)}\ 
    \sum_{n=1}^L\ 
    \left(
    \sum_{y=0}^{L-1} \psi_n^\dagger(y)
    e^{-\alpha (y - \frac{L-1}{2})^2}
    \right)
    \ \frac{\sin \omega_n t}{\omega_n}\ 
    \psi_n(x) \ .
    \label{eq:phiDS}
\end{equation}
The expression in Eq.~(\ref{eq:phiDS}) is the corresponding result to Eq.~(\ref{eq:KG4}) but in a finite volume with OBCs.
Numerically, evaluating the field evolution from the source are the same until boundary effects become important.

\FloatBarrier
\section{Truncated Electric Interactions for Odd \texorpdfstring{$L$}{L}}
\label{app:truncHam}
\noindent
The Hamiltonian corresponding to Eq.~\eqref{eq:spatTrunce} for odd $L$ is
\begin{align}
\hat{H}_{el}^{(Q=0)}(\bar{\lambda}) \ = & \ \frac{g^2}{2}\left\{ \sum_{n=0}^{\frac{L-3}{2}} \left[ \left( L - \frac{5}{4} - 2n \right) \hat{\overline{Q}}^2_n + \frac{1}{2} \hat{\overline{Q}}_n \hat{\delta}_n + \frac{1}{4} \hat{\delta}^2_n + \left( \frac{7}{4} + 2n \right) \hat{\overline{Q}}^2_{\frac{L+1}{2}+n} - \frac{1}{2} \hat{\overline{Q}}_{\frac{L+1}{2}+n} \hat{\delta}_{\frac{L+1}{2}+n} + \frac{1}{4} \hat{\delta}^2_{\frac{L+1}{2}+n} \right] \right. \nonumber \\
& + \frac{1}{4}(\hat{\overline{Q}}^2_{\frac{L-1}{2}}+\hat{\delta}^2_{\frac{L-1}{2}}) \nonumber \\
&+ 2\sum_{n=0}^{\frac{L-5}{2}}\ \ \sum_{m=n+1}^{\min(\frac{L-3}{2},n+\bar{\lambda})} \left[ \left( L-1 - 2m \right) \hat{\overline{Q}}_n\hat{\overline{Q}}_m + \frac{1}{2} \hat{\overline{Q}}_{n} \hat{\delta}_{m} + \left( 2 + 2n \right) \hat{\overline{Q}}_{\frac{L+1}{2}+n}\hat{\overline{Q}}_{\frac{L+1}{2}+m} - \frac{1}{2} \hat{\overline{Q}}_{\frac{L+1}{2}+m} \hat{\delta}_{\frac{L+1}{2}+n} \right]  \nonumber \\
&+ \left. \frac{1}{2} \sum_{n=1}^{\min(\frac{L-1}{2},\bar{\lambda})} \left[ \hat{\overline{Q}}_{\frac{L-1}{2}-n} \hat{\overline{Q}}_{\frac{L-1}{2}} + \hat{\overline{Q}}_{\frac{L-1}{2}-n} \hat{\delta}_{\frac{L-1}{2}} +  \hat{\overline{Q}}_{\frac{L-1}{2}+n} \hat{\overline{Q}}_{\frac{L-1}{2}} - \hat{\overline{Q}}_{\frac{L-1}{2}+n} \hat{\delta}_{\frac{L-1}{2}} \right] \right\} \ .
\end{align}
%

\clearpage
\section{Further Details on Circuit Construction}
\label{app:circuitDetails}
\noindent
\begin{figure}[!ht]
    \centering
    \includegraphics[width=\textwidth]{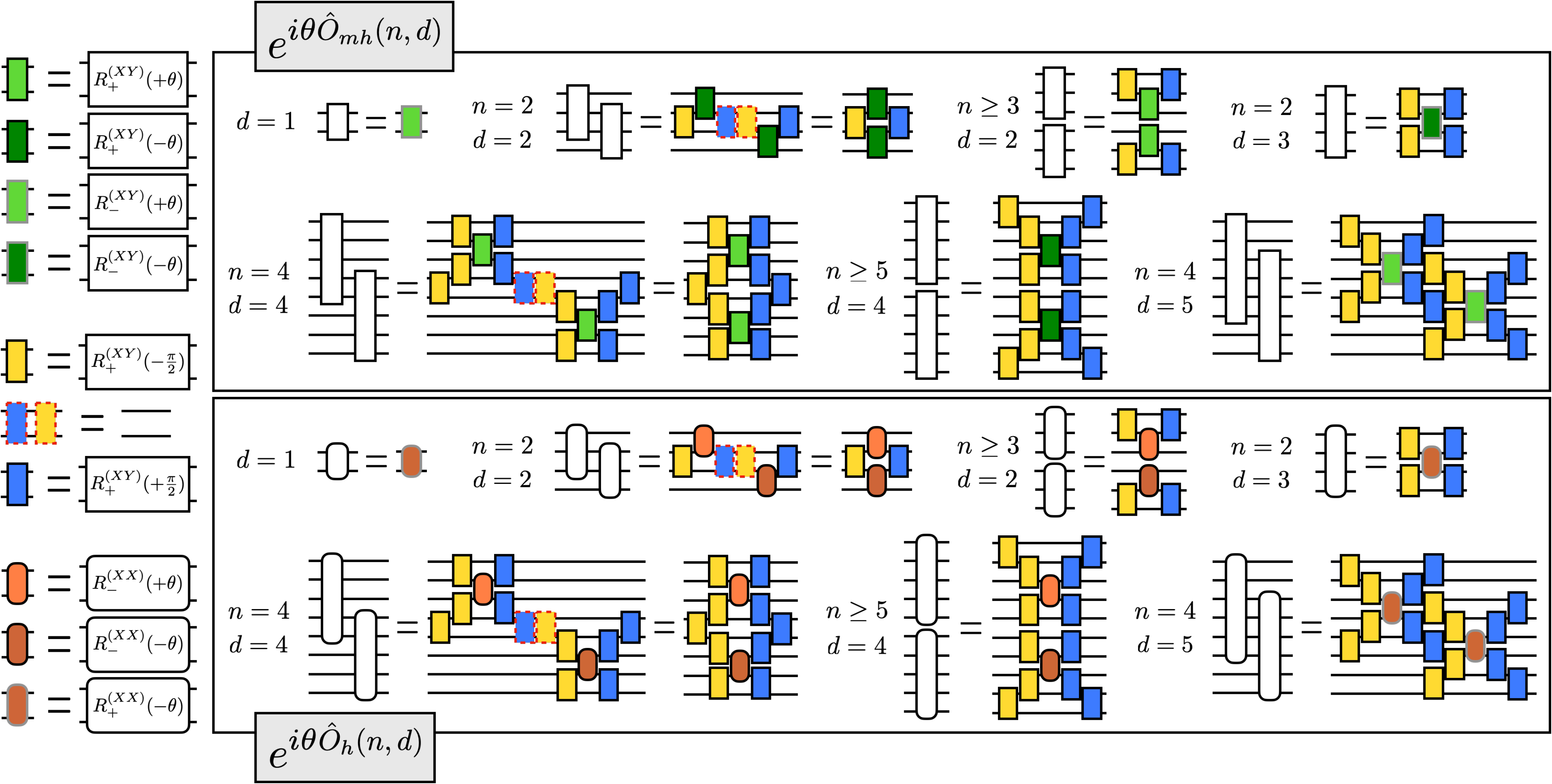}
    \caption{Efficient circuits implementing the unitaries corresponding to the wavepacket pool operators in Eq.~\eqref{eq:PacketPool}. 
    The circuits for the individual blocks $R^{(XY)}_{\pm}(\theta)$ and $R^{(XX)}_{\pm}(\theta)$ are shown in Fig.~\ref{fig:xy_xx_circuits}.}
    \label{fig:blocks_wp_adapt}
\end{figure}
The circuit implementation of the operators from the wavepacket pool in Eq.~\eqref{eq:PacketPool} for $d\leq 5$ is shown in Fig.~\ref{fig:blocks_wp_adapt}. 

What follows next is a code snippet for constructing the Trotterized time evolution of the electric interaction.
Specifically, this code computes the angles $\alpha_i$ in the barbells used to implement the two-qubit $R_{ZZ}$ rotations in Fig.~\ref{fig:StagCirc}.
For a single Trotter step of size $t$, the circuit construction code is:
\lstset{language=Python}
\begin{lstlisting}
circ = QuantumCircuit(2*L)
if np.floor(L/4) == np.floor((L-2)/4):
    for n in range(0,int((L-2)/2),2):
        if n==0:
            a1=(g**2*t)*(1+n*4)/4
            a2=(g**2*t)*(3+n*4)/4
            a3=0
            a4=a1
            a5=a2
            a6=a1
            circ.append(barbell(a1,a2,a3,a4,a5,a6), [L+2*n,L+1+2*n,L+2+2*n,L+3+2*n])
            circ.append(barbell(a3,a2,a1,a5,a4,a6), [L-4-2*n,L-3-2*n,L-2-2*n,L-1-2*n])
        else:
            a1=0
            a2=(g**2*t)*(3+n*4)/4
            a3=0
            a4=(g**2*t)*(1+n*4)/4
            a5=a2
            a6=a4
            circ.append(barbell(a1,a2,a3,a4,a5,a6), [L+2*n,L+1+2*n,L+2+2*n,L+3+2*n])
            circ.append(barbell(a3,a2,a1,a5,a4,a6), [L-4-2*n,L-3-2*n,L-2-2*n,L-1-2*n])
    for n in range(1,int((L-2)/2),2):
        a1=(g**2*t)*(1+n*4)/4
        a2=(g**2*t)*(3+n*4)/4
        a3=(g**2*t)*(5+n*4)/4
        a4=a1
        a5=a2
        a6=a1
        circ.append(barbell(a1,a2,a3,a4,a5,a6), [L+2*n,L+1+2*n,L+2+2*n,L+3+2*n])
        circ.append(barbell(a3,a2,a1,a5,a4,a6), [L-4-2*n,L-3-2*n,L-2-2*n,L-1-2*n])
else:
    for n in range(0,int((L-2)/2),2):
        a1=(g**2*t)*(1+n*4)/4
        a2=(g**2*t)*(3+n*4)/4
        a3=(g**2*t)*(5+n*4)/4
        a4=a1
        a5=a2
        a6=a1
        circ.append(barbell(a1,a2,a3,a4,a5,a6), [L+2*n,L+1+2*n,L+2+2*n,L+3+2*n])
        circ.append(barbell(a3,a2,a1,a5,a4,a6), [L-4-2*n,L-3-2*n,L-2-2*n,L-1-2*n])
    for n in range(1,int((L-2)/2),2):
        a1=0
        a2=(g**2*t)*(3+n*4)/4
        a3=0
        a4=(g**2*t)*(1+n*4)/4
        a5=a2
        a6=a4
        circ.append(barbell(a1,a2,a3,a4,a5,a6), [L+2*n,L+1+2*n,L+2+2*n,L+3+2*n])
        circ.append(barbell(a3,a2,a1,a5,a4,a6), [L-4-2*n,L-3-2*n,L-2-2*n,L-1-2*n])
\end{lstlisting}
In this code, {\tt barbell(a1,a2,a3,a4,a5,a6)} is the circuit block shown in Fig.~\ref{fig:StagCirc}, with {\tt ai} being the angles $\alpha_i$, and each block is being appended to the circuit {\tt circ}, starting from the center  of the lattice 
and progressing outwards. 
CP symmetry is used to relate the angles from the first half of the lattice (labeled as $1{\rm st}$) to the second half (labeled as $2{\rm nd}$) through   the following relations: $\left. \alpha_1 \right|_{1{\rm st}} = \left. \alpha_3 \right|_{2{\rm nd}}$, $\left. \alpha_3 \right|_{1{\rm st}} = \left. \alpha_1 \right|_{2{\rm nd}}$, $\left. \alpha_4 \right|_{1{\rm st}} = \left. \alpha_5 \right|_{2{\rm nd}}$, and $\left. \alpha_5 \right|_{1{\rm st}} = \left. \alpha_4 \right|_{2{\rm nd}}$. 
Also, the angles within a block are not all independent:  $\left. \alpha_1 \right|_{2{\rm nd}} = \left. \alpha_4 \right|_{2{\rm nd}} = \left. \alpha_6 \right|_{2{\rm nd}}$ and $\left. \alpha_2 \right|_{2{\rm nd}} = \left. \alpha_5 \right|_{2{\rm nd}}$. 
The blocks that start at qubit  
$2+4n$ with $n\in\{0,1,\ldots\}$ (or end at $2L-2-4n$) have  
$\alpha_1=\alpha_3=0$ to avoid repeating rotations from the blocks starting at $4n$ (or ending at $2L-4n$). 
There is an exception when $\lfloor\frac{L}{4}\rfloor=\lfloor\frac{L-2}{4}\rfloor$, and only $\left. \alpha_3 \right|_{2{\rm nd}}=\left. \alpha_1 \right|_{1{\rm st}}=0$.

\section{Another Way to Create Hadron Wavepackets}
\label{app:chempot}
\noindent
In the main text, circuits are constructed that optimize the overlap with an adiabatically prepared hadron wavepacket.
Here, an alternative method for preparing hadron wavepackets is presented based on minimizing the energy in the single-hadron sector.
Desirable features of a hadronic wavepacket are that it is localized (i.e., outside of the wavepacket profile, the system is locally in the vacuum), and that it is composed of single hadrons.
When establishing a wavepacket on top of the interacting vacuum, as is done in the main text, localizability can be implemented at the level of the operator pool.
For example, by only including operators in the pool that have support over a predefined spatial interval, $l$, it is guaranteed that outside of $l$ is vacuum.
To ensure that the wavepacket is composed of single hadron states, consider adding a vacuum chemical potential, $\mu$, to the Hamiltonian,
\begin{equation}
\hat{H}_{\text{1-hadron}} \ = \ \hat{H} \ + \ \mu \vert \psi_{\text{vac}} \rangle \langle \psi_{\text{vac}}\vert \ .
\label{eq:H1chem}
\end{equation}
For $\mu > m_{\text{hadron}}$, the ground state of $\hat{H}_{\text{1-hadron}}$ 
in the $Q=0$ sector is the lowest-energy state of a single hadron.
The strategy for building a wavepacket is to minimize $\langle \psi_{\text{ansatz}} \vert \hat{H}_{\text{1-hadron}}\vert  \psi_{\text{ansatz}} \rangle$, where $\vert  \psi_{\text{ansatz}} \rangle$ is adaptively built using a localized operator pool. 
The resulting state will be the lowest energy configuration orthogonal to the vacuum that is localized within the interval $l$. 
The prepared state will primarily be a superposition of single hadrons, with multi-hadron contributions decreasing as $l$ increases.

As an example, consider using this procedure to construct a single-hadron wavepacket with an operator pool localized to $l=2$ sites on either side of the midpoint of the lattice.
Starting from the operator pool in Eq.~\eqref{eq:PacketPool}, the $l=2$ pool consists of $\hat{O}_m(1)$, $\hat{O}_m(2)$, $\hat{O}_{mh}(1,1)$, $\hat{O}_{h}(1,1)$, $\hat{O}_{mh}(2,d)$ and $\hat{O}_{h}(2,d)$ with $d=\{1,2,3\}$ .
Choosing $\mu = 2.5 \, m_{\text{hadron}}$ pushes the energy of the vacuum above two-particle threshold (which is slightly below $2 \, m_{\text{hadron}}$ due to the presence of a two-hadron bound state), and is found to be effective for our purposes.
To update the SC-ADAPT-VQE ansatz, the gradient can be computed with 
\begin{equation}
\frac{\partial}{\partial\theta_i}\left. \langle \psi_{\text{ansatz}}\vert e^{-i \theta_i \hat{O}_i}\hat{H}_{\text{1-hadron}} e^{i \theta_i \hat{O}_i}\vert \psi_{\text{ansatz}}\rangle \right|_{\theta_i=0}
    = - \text{Im} \left [\langle \psi_{\text{ansatz}}\vert \left ( [ \hat{H},\hat{O}_i ] \ + \ 2 \mu \, \vert \psi_{\text{vac}} \rangle \langle \psi_{\text{vac}} \vert \hat{O}_i \right ) \vert \psi_{\text{ansatz}} \rangle \right ] 
    \ .
\end{equation}
Note that it can be necessary to bias the initial parameters to avoid the optimizer choosing $\theta_i=0$ because the initial state is a local maxima of energy (and second derivatives are then required).
Due to the limited size of the operator pool, the SC-ADAPT-VQE algorithm converges relatively well after 4 steps, with the optimal operators and associated variational parameters shown in Table~\ref{tab:AnglesWPchem}.
\begin{table}[!tb]
\renewcommand{\arraystretch}{1.4}
\begin{tabularx}{\textwidth}{|c || Y | Y | Y | Y ||}
 \hline
 \diagbox[height=23pt]{$L$}{$\theta_i$} & $\hat O_{mh}(1,1)$ & $\hat O_{mh}(2,3)$ & $\hat O_{mh}(2,2)$ & $\hat O_{mh}(2,1)$  \\
 \hline\hline
 7  &  -2.4342  & -0.9785  & 0.0819  & 0.2599  \\
 \hline
8  &  2.4343  & 0.9778  & 0.0808  & -0.2591  \\
\hline
9  &  -2.4342  & -0.9780  & 0.0812  & 0.2594  \\
\hline
10  &  2.4340  & 0.9778  & 0.0811  & -0.2595  \\
\hline
11  &  -2.4339  & -0.9776  & 0.0810  & 0.2593  \\
\hline
12  &  2.4321  & 0.9781  & 0.0837  & -0.2605  \\
\hline
13  &  -2.4343  & -0.9780  & 0.0810  & 0.2593  \\
 \hline
\end{tabularx}
\caption{The operator ordering and variational parameters that minimize $\hat{H}_{\text{1-hadron}}$, and prepare a hadron wavepacket for $L=7-13$.}
 \label{tab:AnglesWPchem}
\end{table}
The resulting state has an $L$-independent energy expectation value of $\langle \psi_{\text{ansatz}} \vert \hat{H} \vert \psi_{\text{ansatz}} \rangle = 1.18 \, m_{\text{hadron}}$, and overlap onto the vacuum state of $\vert\langle \psi_{\text{vac}} \vert \psi_{\text{ansatz}} \rangle \vert^2 = 8.5\times 10^{-5}$.
These results show that the prepared wavepacket is primarily composed of single-hadron states, and both $\langle \psi_{\text{ansatz}} \vert \hat{H} \vert \psi_{\text{ansatz}} \rangle$ and $\vert\langle \psi_{\text{vac}} \vert \psi_{\text{ansatz}} \rangle \vert^2$ can be further reduced by increasing $l$, i.e., de-localizing the prepared wavepacket.

\FloatBarrier
\section{Details on the 112-qubit MPS Simulations}
\label{app:MPSSim}
\noindent
The 112-qubit quantum simulations in Sec.~\ref{sec:Qsim} are compared to the expected, error-free, results determined using the {\tt qiskit} and {\tt cuQuantum} MPS circuit simulators.
MPS techniques are approximations that can be improved by increasing the bond dimension in the MPS ansatz. 
A higher bond dimension increases the maximum amount of entanglement in the ansatz state, at the cost of longer run-time on a classical computer.
As a result, simulations that explore highly-entangled states are promising candidates for a near-term quantum advantage.
Our numerical investigations have found a large contrast between the bond dimension needed for state preparation and time evolution.
The initial hadron wavepacket coincides with the vacuum state outside of the few sites where the wavepacket has support.
This state has a low amount of entanglement as the ground states of gapped 1D systems have area-law entanglement~\cite{Hastings_2007,arad2013area,Brand_o_2014}. 
Therefore, a relatively small bond dimension can be used in the MPS simulations to faithfully reproduce the preparation of the vacuum and initial hadron wavepacket.
Time evolution, on the other hand, involves a superposition of many single-hadron states, which disturb the vacuum as they propagate.
This produces a significant amount of entanglement, and subsequently requires a larger bond dimension.

\begin{figure}[!ht]
    \centering
    \includegraphics[width=0.9\columnwidth]{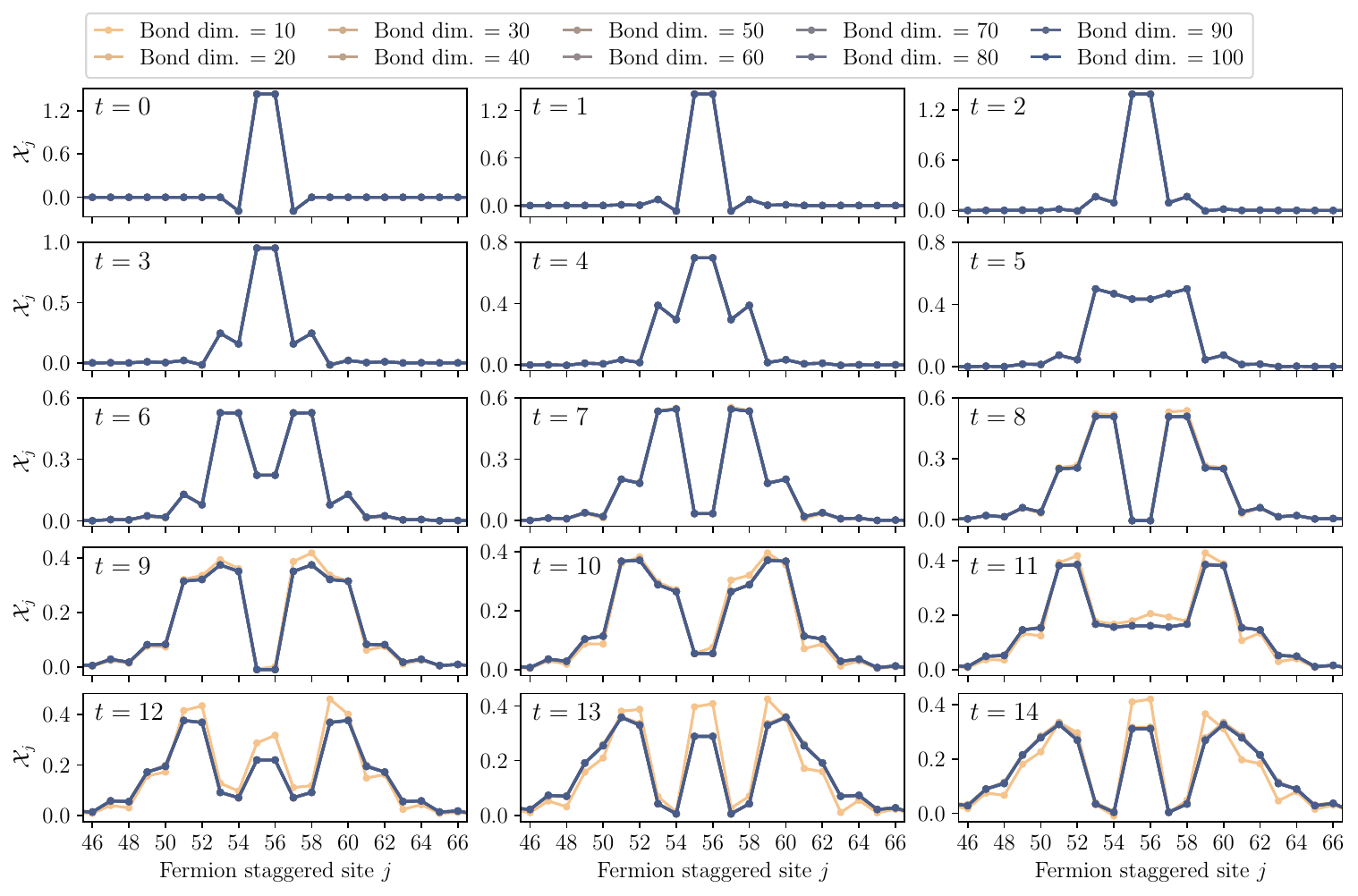}
    \caption{Convergence of the vacuum subtracted local chiral condensate, ${\cal X}_j(t)$, with different maximum bond dimension in the {\tt cuQuantum} MPS simulator.
    Results are shown for $L=56$, and are focused around the center of the lattice where the convergence is the slowest.}
    \label{fig:MPS_Convergence}
\end{figure}

\begin{figure}[!ht]
    \centering
    \includegraphics[width=0.9\columnwidth]{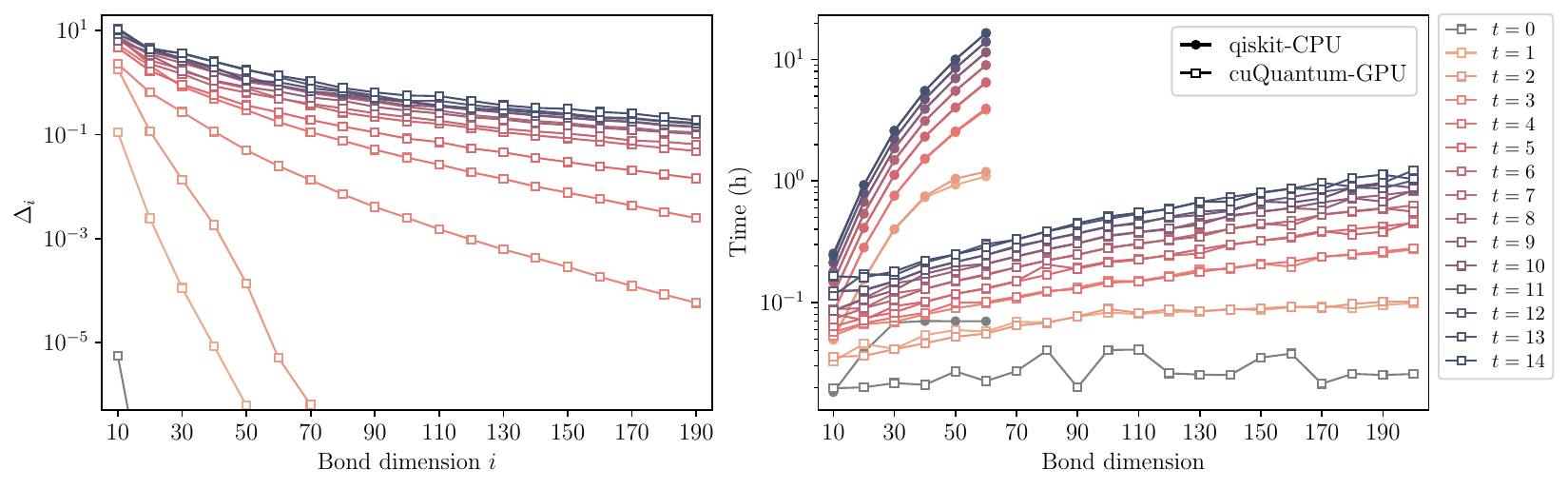}
    \caption{The left panel shows the relative convergence of the chiral condensate, $\Delta_i$, defined in Eq.~\eqref{eq:deltai}, for a selection of times as a function of maximum bond dimension, while the right panel shows the computational run time using the {\tt qiskit} MPS simulator on a single 40-core CPU-node and the {\tt cuQuantum} MPS simulator on a single NVIDIA RTX A5000, for $t=\{0,1,\ldots,14\}$ with different maximum bond dimension.}
    \label{fig:MPS_Runtime}
\end{figure}

The bond dimension needed for convergence of the chiral condensate for different simulation times is shown in Fig.~\ref{fig:MPS_Convergence}.
It is seen that a relatively small bond dimension is sufficient for convergence, even out to late times. 
This should be compared to the convergence of $\langle \psi_{\text{WP}} \vert \hat{\chi}_j \vert\psi_{\text{WP}} \rangle$
in the left panel of Fig.~\ref{fig:MPS_Runtime}, where the quantity
\begin{equation}
    \Delta_i \ = \ \sum_j \ | \ \langle \psi^{MPS_i}_{\rm WP} \vert \ \hat{\chi}_j(t) \ \vert \psi^{MPS_i}_{\rm WP} \rangle \ - \ \langle \psi^{MPS_{i+10}}_{\rm WP} \vert \ \hat{\chi}_j(t) \ \vert \psi^{MPS_{i+10}}_{\rm WP} \rangle \ | \ ,
\label{eq:deltai}
\end{equation}
is computed for different bond dimensions.
This quantity determines how much the local chiral condensate of the evolved wavepacket changes as the maximum bond dimension is increased from $i$ to $i+10$.
This reveals that MPS calculation of the chiral condensate of 
(a) the initial state can be done very efficiently (results with a maximum bond dimension of 10 have already converged below a $10^{-5}$ precision), and 
(b) the evolved wavepacket converges  slowly, especially at late times. 
Indeed, the quick convergence of ${\cal X}_n(t)$ in Fig.~\ref{fig:MPS_Convergence} is due to the cancellations of errors between the MPS simulation of the wavepacket and vacuum evolution.
These MPS simulations take increasingly more compute run-time as the bond dimension increases.
This is illustrated in the right panel of Fig.~\ref{fig:MPS_Runtime}, where the run-time for a selection of times  and various bond dimensions are shown. In this panel, we compare the performance of the CPU-based {\tt qiskit} MPS simulator, run on a single 40-core CPU-node on Hyak, and the GPU-based {\tt cuQuantum} MPS simulator, run on a single NVIDIA RTX A5000 through the OSG Pool.
%

\FloatBarrier
\section{Further Details about the Error Mitigation and Analysis}
\label{app:qSimDetails}
\begin{figure}[!b]
    \centering
    \includegraphics[width=0.5\columnwidth]{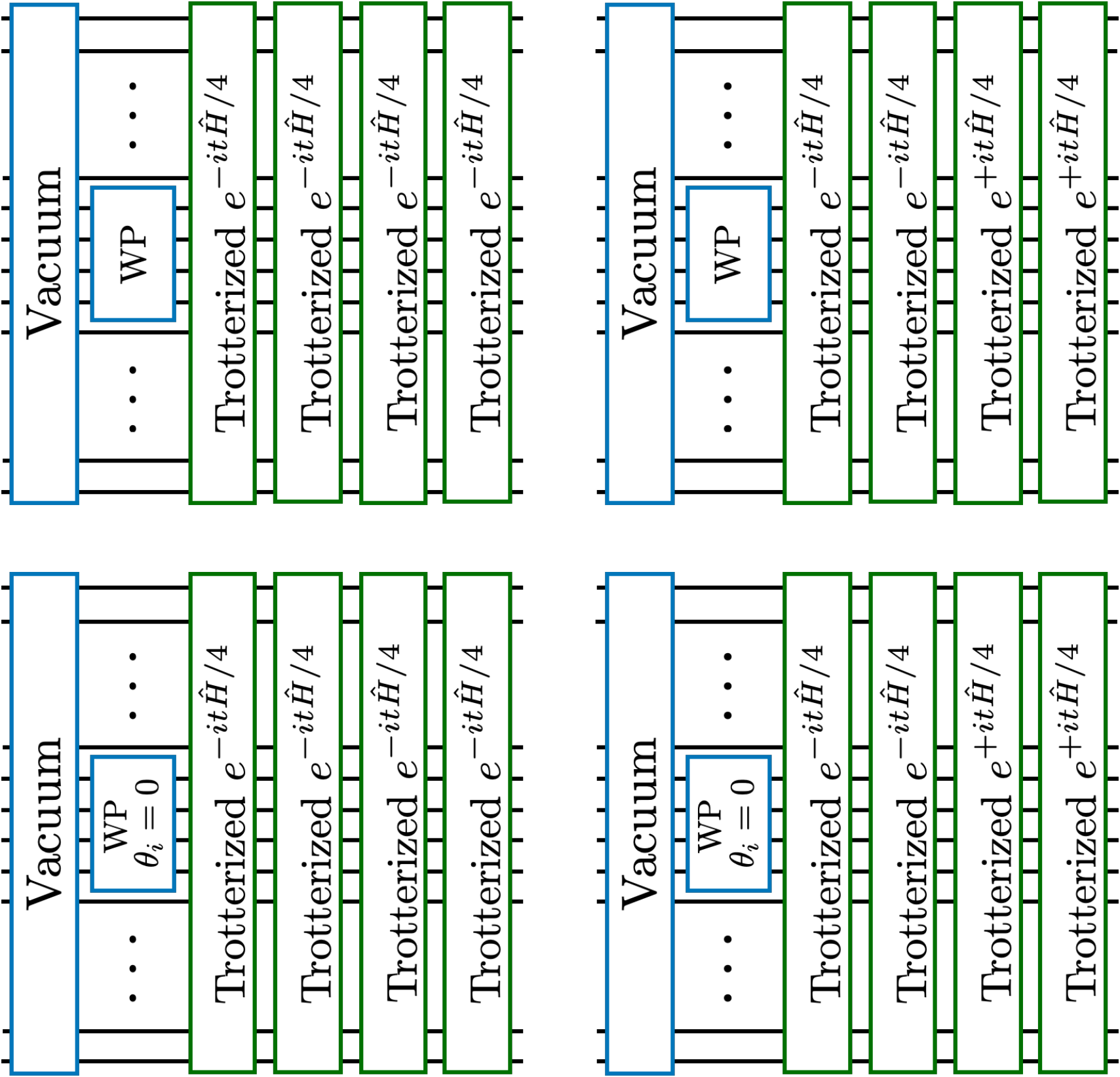}
    \caption{The four types of circuits run on {\tt ibm\_torino}.
    Blue boxes denote SC-ADAPT-VQE circuits, and green boxes denote Trotterized 
    time-evolution circuits.
    Shown are examples for 4 Trotter steps of time evolution, with straightforward extension to other even number of Trotter steps.
    (Lower) upper-left show the circuits used for the time evolution of the (vacuum) wavepacket. 
    (Lower) upper-right are the forwards-backwards time-evolution circuits used for ODR error-mitigation of the wavepacket (vacuum).
    The $\theta_i=0$ in the wavepacket circuit box denotes that the SC-ADAPT-VQE parameters are set to zero, i.e., it is the identity operator in the absence of device errors.}
    \label{fig:FourCircuits}
\end{figure}

\noindent
For each time $t=\{1,2,\ldots,14\}$, four kinds of circuits are run on the quantum computer: time evolution of the vacuum, time evolution of the wavepacket, and the corresponding forward-backward evolution for ODR error mitigation, see Fig.~\ref{fig:FourCircuits}.
Each circuit for $t=1-8$ is run with 480 twirls and each circuit for $t=9-14$ is run with 160 twirls; each twirl with 8,000 shots, as displayed in Table~\ref{tab:QsimCNOT}.
The longest continuous one-dimensional chain on {\tt ibm\_torino} that we utilize is 112 qubits, corresponding to a $L=56$ lattice (see layout in Fig.~\ref{fig:SimTricks}). 
We use two lattice-to-qubit mappings to minimize the effects of poorly performing qubits.
Half of the twirls assign staggered site 0 to the top-right device qubit, and the other half assign staggered site 0 to the bottom left device qubit.
Averaging over multiple layouts mitigates some of the effects of qubit-specific noise.
Indeed, in our simulations there are twirled instances where qubits perform poorly, either due to decoherence or to readout errors.
Such errors can be identified and removed from analysis by filtering out measurements where $\langle \hat{Z}_j \rangle_{\text{meas}}/\langle \hat{Z}_j \rangle_{\text{pred}} < \epsilon$ in the mitigation runs, with $\epsilon$ some predetermined threshold.\footnote{This type of event post-selection, requiring device performance to exceed a specified level in interleaved calibration circuits, has been employed previously, for example, Ref.~\cite{Klco:2019xro}.}
If this ratio is negative, then the qubit has flipped, and if it is 0 then the qubit has completely decohered, i.e., it has become a maximally mixed state.
We choose $\epsilon = 0.01$, and do not see much difference varying up to $\epsilon = 0.05$.
Our scheduling of  jobs interleaves physics and mitigation circuits with the same twirl.
Poorly performing qubits, identified from measuring the mitigation circuit, are cut from both the ensemble of mitigation and associated physics measurements.\footnote{
Note that 160 of the 480 twirls for $t=1-7$ do not interleave physics and mitigation. Instead, they are sent in batches of 40 circuits with uncorrelated twirls between mitigation and physics circuits. 
In this case a qubit measurement of physics circuit $n$ in the batch is cut if the corresponding qubit measurement in mitigation circuit $n$ is cut.
Surprisingly, no improvement is found when correlating the twirls and interleaving mitigation and physics circuits.}

\begin{figure}[!htb]
    \centering
    \includegraphics[width=\columnwidth]{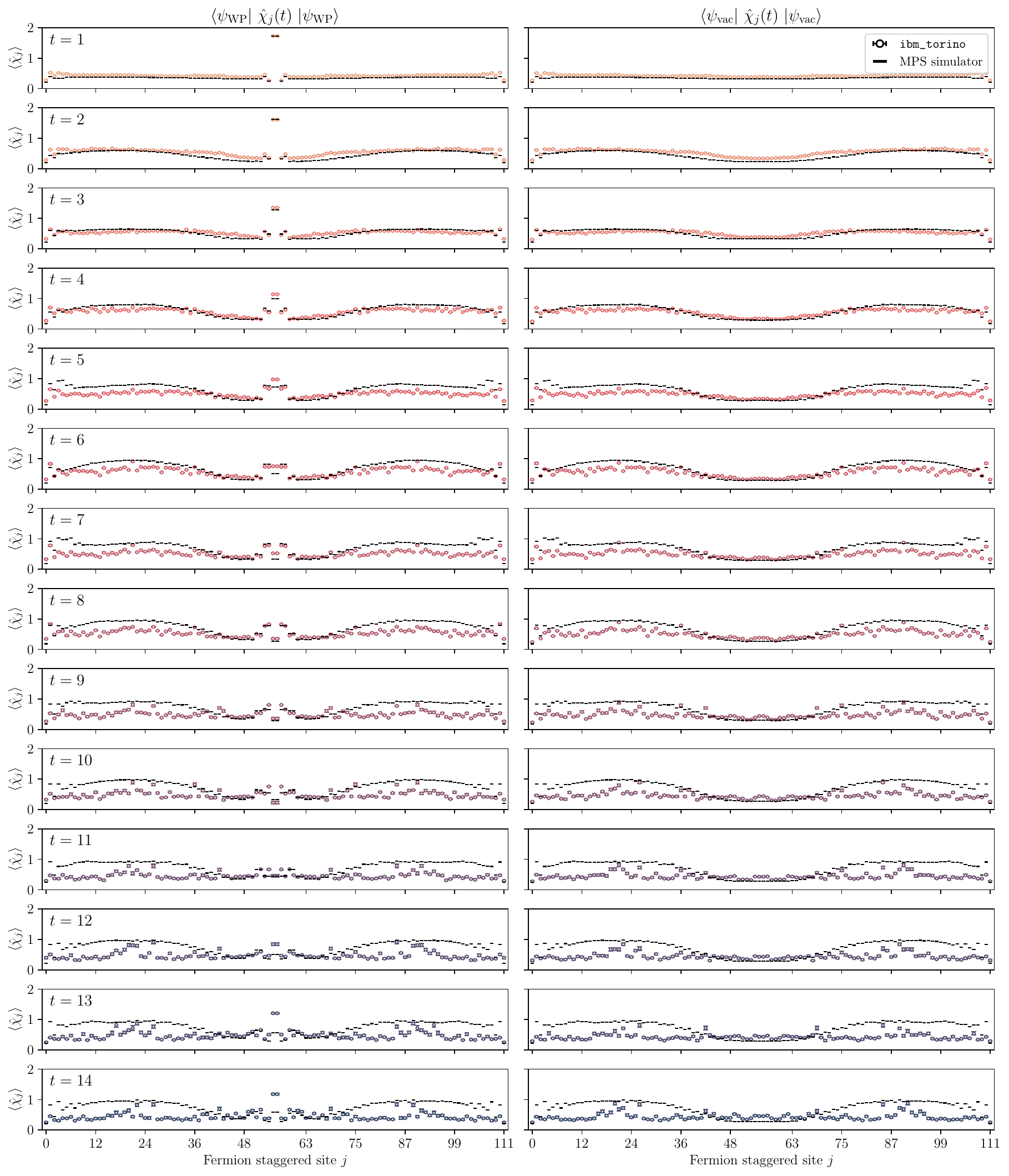}
    \caption{CP-averaged local chiral condensate for the time-evolved wavepacket (left subpanels) and vacuum (right subpanels). The points are obtained from {\tt ibm\_torino}, and the black lines from the {\tt cuQuantum} MPS circuit simulator.}
    \label{fig:src_vac_evol}
\end{figure}

The results of measurements related by CP symmetry are combined.
For $\hat{Z}_j$, this means combining $\langle \hat{Z}_j \rangle$ and $-\langle \hat{Z}_{2L-1-j} \rangle$ (for runs with 480 twirls, this can lead to up to 960 independent measurements for $\langle \hat{Z}_j \rangle$).
The central values and corresponding uncertainties are determined from bootstrap re-sampling over twirls. 
Due to the filtering  procedure, $\langle \hat{Z}_j\rangle$ for each qubit can have a different number of contributing twirls, $N^{(\text{meas})}_j$.
For each sample in the bootstrap ensemble, $N^{(\text{meas})}_j$ random integers with replacement $\{ x \} \in \{1,2,\ldots, N^{(\text{meas})}_j\}$ are generated, with the prediction for the error-free physics expectation value for that sample given by
\begin{equation}
\left.{\overline{\langle \hat{Z}_j \rangle}_{\text{pred}}}\right\rvert_{\text{phys}} \ = \ \left( \sum_{i \in \{ x \}} \left.{\langle \hat{Z}_j \rangle^{(i)}_{\text{meas}}} \right\rvert_{\text{phys}} \right) \times  \left ( \sum_{i \in \{ x \}} \left . \frac{\langle \hat{Z}_j \rangle_{\text{pred}}}{\langle \hat{Z}_j \rangle^{(i)}_{\text{meas}}}\right \rvert_{\text{mit}} \right )\ ,
\label{eq:bootsrap}
\end{equation}
where the superscript $(i)$ labels the twirl.
This is performed for the wavepacket and for the vacuum evolution, with the vacuum subtracted chiral condensate given by
\begin{equation}
\overline{ {\cal X}_j } \ = \ (-1)^j \left ( \left.\overline{\langle \hat{Z}_j \rangle}_{\text{pred}} \right\rvert_{\text{phys}}^{(\text{WP})} \ - \ \left. \overline{\langle \hat{Z}_j \rangle}_{\text{pred}} \right\rvert_{\text{phys}}^{(\text{Vac})}  \right ) \ .
\end{equation}
This process is repeated $N_{\text{Boot}}$ times, with $N_{\text{Boot}}$ large enough for the mean and standard deviation of the bootstrap ensemble $\{ \ \overline{ {\cal X}_j }\ \}$ to have converged.
This mean and standard deviation are used to produce the points with error bars in Figs.~\ref{fig:srcmvac_evol} and~\ref{fig:src_vac_evol}.\footnote{The two sums in Eq.~\eqref{eq:bootsrap} compute the mean of the bootstrap sample. 
If instead the median is used, larger error bars are found.
This is likely due to there being correlations in the tails of both the ensembles of physics and mitigation measurements that are captured by the mean, but suppressed 
by the median.}

We have found that larger angles in the circuits lead to larger systematic errors, independent of circuit depth.
This is likely due to cross-talk errors between gates acting on neighboring qubits when large rotations are applied.
These kinds of errors are not corrected by ODR.
Thus, there is a trade-off between increased number of Trotter steps with smaller angles, and the associated increased circuit depth.
A full determination of this trade off remains to be explored.

\begin{figure}[!htb]
    \centering
    \includegraphics[width=\columnwidth]{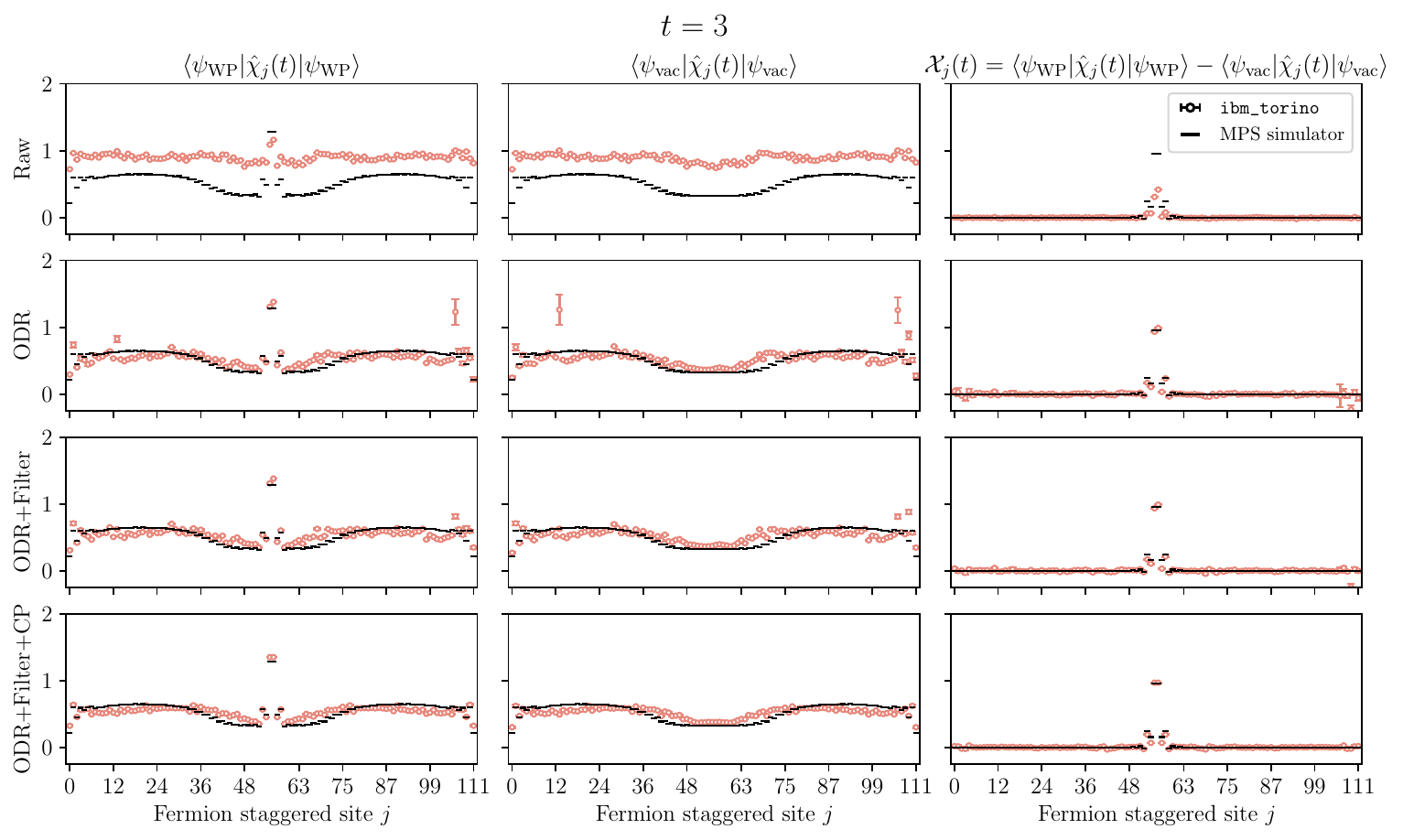}
    \includegraphics[width=\columnwidth]{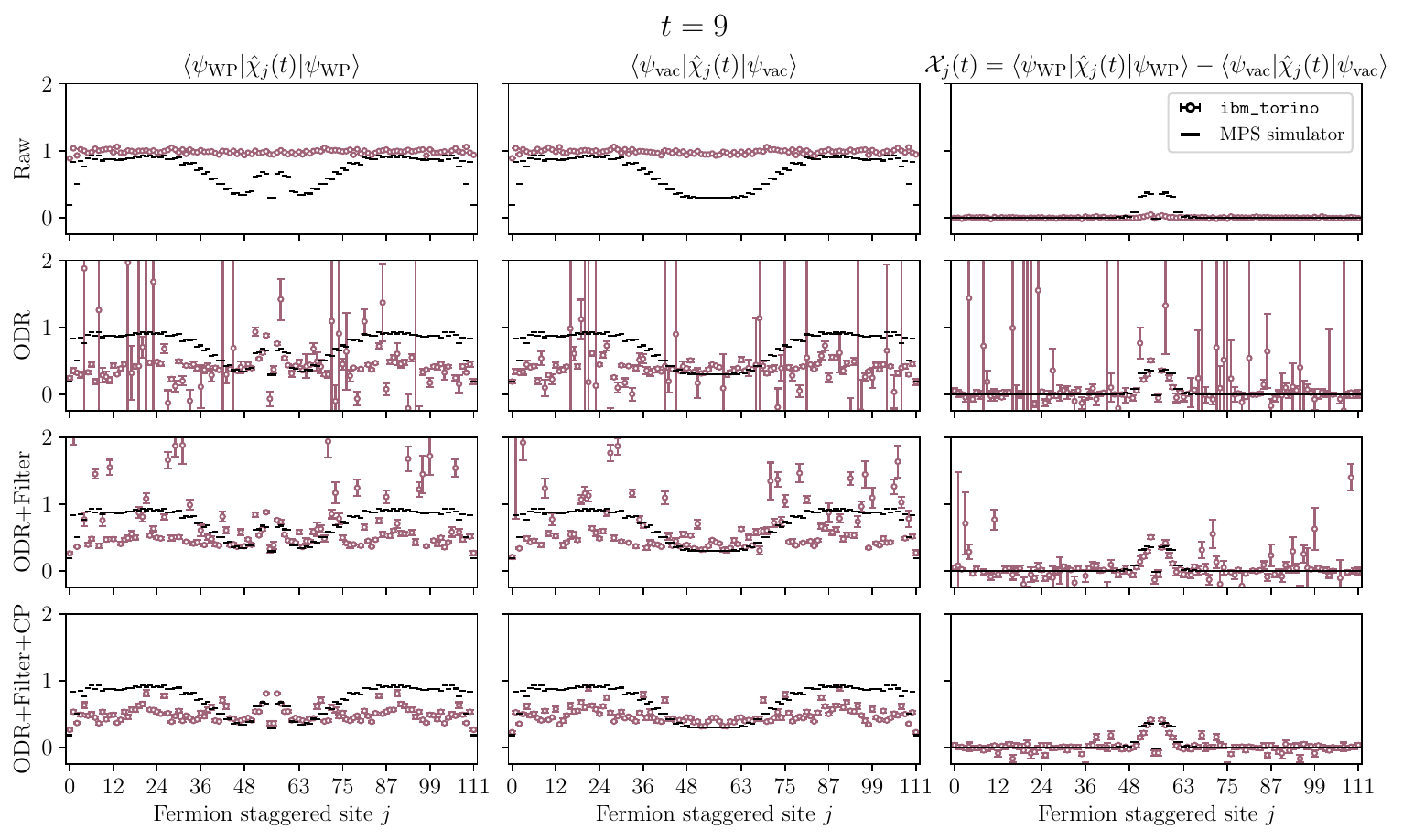}
    \caption{The results obtained from {\tt ibm\_torino} after different stages of error mitigation. Results for $t=3$ and $t=9$ are shown for the chiral condensate of the wavepacket and vacuum evolution, and their subtraction. The four rows give: raw results, after applying ODR, after filtering out the decohered qubits, and after CP averaging. Note that after the filtering procedure, different qubits have different numbers of twirls contributing (and hence different sized errors bars).}
    \label{fig:mitigation_stages_IBM}
\end{figure}

The different stages of error mitigation are displayed in Fig.~\ref{fig:mitigation_stages_IBM}.
Two times, $t=3$ (CNOT depth 120) and $t=9$ (CNOT depth 270), are chosen for the purpose of demonstration. 
Note that in these plots, the decohered value of the chiral condensate is $\langle \hat{\chi}_j\rangle =1$ (with $\mathcal{X}_j =0$).
The first row of Fig.~\ref{fig:mitigation_stages_IBM} shows the ``raw'' results obtained from the device (with dynamical decoupling and readout error mitigation) after averaging over all Pauli twirls.
The device errors for the wavepacket and vacuum evolution outside of the wavepacket region are very similar, and cancel to a large degree in forming the subtraction in ${\cal X}_j(t)$.
It is striking that, for $t=9$, there is no discernible sign of the presence of a wavepacket in the raw results.
The second row of Fig.~\ref{fig:mitigation_stages_IBM} shows the effect of applying ODR.
This helps recover the chiral condensate, being more effective for $t=3$ than $t=9$, but can also lead to large error bars when the qubit is close to being completely decohered ($\langle\hat{Z}_j\rangle_{\text{meas}}\vert_{\text{mit}}$ close to zero).
The third row of Fig.~\ref{fig:mitigation_stages_IBM} shows the effects of filtering out runs where $\langle \hat{Z}_j \rangle_{\text{meas}}/\langle \hat{Z}_j \rangle_{\text{pred}}\vert_{\text{mit}} < 0.01$.
This removes most of the runs contributing to the large error bars, and is more significant for $t=9$ than $t=3$.
It also leads to different numbers of twirls surviving the filtering for different qubits.
Sometimes only a small number survive, compromising the assumption of a depolarizing channel for ODR (and also compromising the error estimates from bootstrap re-sampling).
The fourth row of Fig.~\ref{fig:mitigation_stages_IBM} shows the effects of using the CP symmetry to combine the measurements of $\langle \hat{Z}_j \rangle$ and $-\langle \hat{Z}_{2L-1-j} \rangle$.
This reduces the effects of poorly performing qubits, and gives the final results presented in Figs.~\ref{fig:IBMresultsMPS}, \ref{fig:srcmvac_evol}, and~\ref{fig:src_vac_evol}.

%

\section{Tables}
\label{app:tables}
\noindent
In this appendix, the tabulated numbers plotted in Figs.~\ref{fig:srcmvac_evol} and~\ref{fig:src_vac_evol}, for both MPS simulators and {\tt ibm\_torino}, are displayed. Due to the slow convergence of the MPS results for late times (as discussed in App.~\ref{app:MPSSim}), the precision for the columns showing $\langle \psi_{\rm WP}\vert \ \hat{\chi}_j(t) \  \vert \psi_{\rm WP} \rangle$ and $\langle \psi_{\rm vac}\vert \ \hat{\chi}_j(t) \  \vert \psi_{\rm vac} \rangle$ can be estimated for each time: for $t=1$ (Table~\ref{tab:tablet1}) below $10^{-10}$, $t=2$ (Table~\ref{tab:tablet2}) below $10^{-10}$, $t=3$ (Table~\ref{tab:tablet3}) around $10^{-6}$, $t=4$ (Table~\ref{tab:tablet4}) around $10^{-4}$, $t=5$ (Table~\ref{tab:tablet5}) around $10^{-3}$, $t=6$ (Table~\ref{tab:tablet6}) around $10^{-3}$, $t=7$ (Table~\ref{tab:tablet7}) around $10^{-3}$, $t=8$ (Table~\ref{tab:tablet8}) around $10^{-2}$, $t=9$ (Table~\ref{tab:tablet9}) around $10^{-2}$, $t=10$ (Table~\ref{tab:tablet10}) around $10^{-2}$, $t=11$ (Table~\ref{tab:tablet11}) around $10^{-2}$, $t=12$ (Table~\ref{tab:tablet12}) around $10^{-2}$, $t=13$ (Table~\ref{tab:tablet13}) around $10^{-2}$, and $t=14$ (Table~\ref{tab:tablet14}) around $10^{-2}$.
\begin{table}[!htb]
\renewcommand{\arraystretch}{1}
\scalebox{0.95}{
\begin{tabularx}{\textwidth}{| c || Y | Y | Y | Y | Y | Y |}
 \hline
\multirow{2}{*}{\parbox{2.5cm}{Fermion staggered site $j$}} &	\multicolumn{2}{c|}{\large $\langle \psi_{\rm WP}\vert \ \hat{\chi}_j(t=1) \  \vert \psi_{\rm WP} \rangle$}		&	\multicolumn{2}{c|}{\large $\langle \psi_{\rm vac}\vert \ \hat{\chi}_j(t=1) \  \vert \psi_{\rm vac} \rangle$}	&	\multicolumn{2}{c|}{\large $\mathcal{X}_j(t=1)$}		\\ \cline{2-7}
	&	MPS	&	{\tt ibm\_torino}	&	MPS	&	{\tt ibm\_torino}	&	MPS	&	{\tt ibm\_torino}		\\ \hline \hline
0	&	0.213	&	0.273(10)	&	0.213	&	0.270(11)	&	0.000	&	0.002(14)	\\
1	&	0.386	&	0.522(11)	&	0.386	&	0.510(12)	&	0.000	&	0.013(16)	\\
2	&	0.3360	&	0.4186(58)	&	0.3360	&	0.4110(55)	&	0.0000	&	0.0074(80)	\\
3	&	0.336	&	0.510(14)	&	0.336	&	0.515(13)	&	0.000	&	-0.004(17)	\\
4	&	0.3364	&	0.4674(83)	&	0.3364	&	0.4878(99)	&	0.000	&	-0.021(11)	\\
5	&	0.3716	&	0.4859(76)	&	0.3716	&	0.4925(77)	&	0.0000	&	-0.0068(90)	\\
\hline
\end{tabularx}}
\caption{Numerical values for the $t=1$ chiral condensate of the wavepacket, the vacuum, and their subtraction, from the MPS simulator and {\tt ibm\_torino}, as shown in Figs.~\ref{fig:srcmvac_evol} and~\ref{fig:src_vac_evol} (only the first half of the lattice is shown due to CP symmetry).}
 \label{tab:tablet1}
\end{table}
\begin{table}[!htb]
\renewcommand{\arraystretch}{1}
\scalebox{0.95}{\begin{tabularx}{\textwidth}{| c || Y | Y | Y | Y | Y | Y |}
 \hline
\multirow{2}{*}{\parbox{2.5cm}{Fermion staggered site $j$}} &	\multicolumn{2}{c|}{\large $\langle \psi_{\rm WP}\vert \ \hat{\chi}_j(t=2) \  \vert \psi_{\rm WP} \rangle$}		&	\multicolumn{2}{c|}{\large $\langle \psi_{\rm vac}\vert \ \hat{\chi}_j(t=2) \  \vert \psi_{\rm vac} \rangle$}	&	\multicolumn{2}{c|}{\large $\mathcal{X}_j(t=2)$}		\\ \cline{2-7}
	&	MPS	&	{\tt ibm\_torino}	&	MPS	&	{\tt ibm\_torino}	&	MPS	&	{\tt ibm\_torino}		\\ \hline \hline
0	&	0.193	&	0.287(10)	&	0.193	&	0.278(10)	&	0.000	&	0.009(14)	\\
1	&	0.4420	&	0.6254(95)	&	0.4420	&	0.6106(99)	&	0.000	&	0.015(12)	\\
2	&	0.3580	&	0.4712(59)	&	0.3580	&	0.4743(63)	&	0.0000	&	-0.0028(83)	\\
3	&	0.4992	&	0.6406(97)	&	0.499	&	0.639(10)	&	0.000	&	0.002(11)	\\
4	&	0.4912	&	0.6371(70)	&	0.4912	&	0.6285(63)	&	0.0000	&	0.0084(82)	\\
5	&	0.5138	&	0.6365(79)	&	0.5138	&	0.6444(86)	&	0.000	&	-0.008(10)	\\
\hline
\end{tabularx}}
\caption{Numerical values for the $t=2$ chiral condensate of the wavepacket, the vacuum, and their subtraction, from the MPS simulator and {\tt ibm\_torino}, as shown in Figs.~\ref{fig:srcmvac_evol} and~\ref{fig:src_vac_evol} (only the first half of the lattice is shown due to CP symmetry).}
 \label{tab:tablet2}
\end{table}
\begin{table}[!htb]
\renewcommand{\arraystretch}{1}
\scalebox{0.95}{\begin{tabularx}{\textwidth}{| c || Y | Y | Y | Y | Y | Y |}
 \hline
\multirow{2}{*}{\parbox{2.5cm}{Fermion staggered site $j$}} &	\multicolumn{2}{c|}{\large $\langle \psi_{\rm WP}\vert \ \hat{\chi}_j(t=3) \  \vert \psi_{\rm WP} \rangle$}		&	\multicolumn{2}{c|}{\large $\langle \psi_{\rm vac}\vert \ \hat{\chi}_j(t=3) \  \vert \psi_{\rm vac} \rangle$}	&	\multicolumn{2}{c|}{\large $\mathcal{X}_j(t=3)$}		\\ \cline{2-7}
	&	MPS	&	{\tt ibm\_torino}	&	MPS	&	{\tt ibm\_torino}	&	MPS	&	{\tt ibm\_torino}		\\ \hline \hline
0	&	0.213	&	0.326(12)	&	0.213	&	0.304(13)	&	0.000	&	0.022(17)	\\
1	&	0.604	&	0.640(15)	&	0.604	&	0.623(16)	&	0.000	&	0.017(21)	\\
2	&	0.4454	&	0.4572(90)	&	0.4454	&	0.4729(92)	&	0.000	&	-0.015(12)	\\
3	&	0.595	&	0.564(12)	&	0.595	&	0.581(13)	&	0.000	&	-0.018(17)	\\
4	&	0.565	&	0.571(11)	&	0.565	&	0.540(10)	&	0.000	&	0.032(13)	\\
5	&	0.597	&	0.579(15)	&	0.597	&	0.572(14)	&	0.000	&	0.007(17)	\\
\hline
\end{tabularx}}
\caption{Numerical values for the $t=3$ chiral condensate of the wavepacket, the vacuum, and their subtraction, from the MPS simulator and {\tt ibm\_torino}, as shown in Figs.~\ref{fig:srcmvac_evol} and~\ref{fig:src_vac_evol} (only the first half of the lattice is shown due to CP symmetry).}
 \label{tab:tablet3}
\end{table}
\begin{table}[!htb]
\renewcommand{\arraystretch}{1}
\scalebox{0.95}{\begin{tabularx}{\textwidth}{| c || Y | Y | Y | Y | Y | Y |}
 \hline
\multirow{2}{*}{\parbox{2.5cm}{Fermion staggered site $j$}} &	\multicolumn{2}{c|}{\large $\langle \psi_{\rm WP}\vert \ \hat{\chi}_j(t=4) \  \vert \psi_{\rm WP} \rangle$}		&	\multicolumn{2}{c|}{\large $\langle \psi_{\rm vac}\vert \ \hat{\chi}_j(t=4) \  \vert \psi_{\rm vac} \rangle$}	&	\multicolumn{2}{c|}{\large $\mathcal{X}_j(t=4)$}		\\ \cline{2-7}
	&	MPS	&	{\tt ibm\_torino}	&	MPS	&	{\tt ibm\_torino}	&	MPS	&	{\tt ibm\_torino}		\\ \hline \hline
0	&	0.165	&	0.270(13)	&	0.165	&	0.248(13)	&	0.000	&	0.021(18)	\\
1	&	0.557	&	0.700(13)	&	0.557	&	0.696(12)	&	0.000	&	0.004(18)	\\
2	&	0.3967	&	0.5149(81)	&	0.3967	&	0.5116(88)	&	0.000	&	0.003(12)	\\
3	&	0.604	&	0.633(11)	&	0.604	&	0.601(11)	&	0.000	&	0.032(14)	\\
4	&	0.558	&	0.620(13)	&	0.558	&	0.584(12)	&	0.000	&	0.036(16)	\\
5	&	0.630	&	0.566(15)	&	0.630	&	0.567(15)	&	0.000	&	-0.000(18)	\\
\hline
\end{tabularx}}
\caption{Numerical values for the $t=4$ chiral condensate of the wavepacket, the vacuum, and their subtraction, from the MPS simulator and {\tt ibm\_torino}, as shown in Figs.~\ref{fig:srcmvac_evol} and~\ref{fig:src_vac_evol} (only the first half of the lattice is shown due to CP symmetry).}
 \label{tab:tablet4}
\end{table}
\begin{table}[!htb]
\renewcommand{\arraystretch}{1}
\scalebox{0.95}{\begin{tabularx}{\textwidth}{| c || Y | Y | Y | Y | Y | Y |}
 \hline
\multirow{2}{*}{\parbox{2.5cm}{Fermion staggered site $j$}} &	\multicolumn{2}{c|}{\large $\langle \psi_{\rm WP}\vert \ \hat{\chi}_j(t=5) \  \vert \psi_{\rm WP} \rangle$}		&	\multicolumn{2}{c|}{\large $\langle \psi_{\rm vac}\vert \ \hat{\chi}_j(t=5) \  \vert \psi_{\rm vac} \rangle$}	&	\multicolumn{2}{c|}{\large $\mathcal{X}_j(t=5)$}		\\ \cline{2-7}
	&	MPS	&	{\tt ibm\_torino}	&	MPS	&	{\tt ibm\_torino}	&	MPS	&	{\tt ibm\_torino}		\\ \hline \hline
0	&	0.145	&	0.264(16)	&	0.145	&	0.286(15)	&	0.000	&	-0.023(20)	\\
1	&	0.841	&	0.651(20)	&	0.841	&	0.692(22)	&	0.000	&	-0.041(27)	\\
2	&	0.6258	&	0.4054(70)	&	0.6258	&	0.3913(66)	&	0.0000	&	0.0144(97)	\\
3	&	0.929	&	0.609(13)	&	0.929	&	0.601(13)	&	0.000	&	0.008(16)	\\
4	&	0.940	&	0.489(15)	&	0.940	&	0.510(15)	&	0.000	&	-0.020(18)	\\
5	&	0.772	&	0.448(15)	&	0.772	&	0.445(14)	&	0.000	&	0.002(18)	\\
\hline
\end{tabularx}}
\caption{Numerical values for the $t=5$ chiral condensate of the wavepacket, the vacuum, and their subtraction, from the MPS simulator and {\tt ibm\_torino}, as shown in Figs.~\ref{fig:srcmvac_evol} and~\ref{fig:src_vac_evol} (only the first half of the lattice is shown due to CP symmetry).}
 \label{tab:tablet5}
\end{table}
\begin{table}[!htb]
\renewcommand{\arraystretch}{1}
\scalebox{0.95}{\begin{tabularx}{\textwidth}{| c || Y | Y | Y | Y | Y | Y |}
 \hline
\multirow{2}{*}{\parbox{2.5cm}{Fermion staggered site $j$}} &	\multicolumn{2}{c|}{\large $\langle \psi_{\rm WP}\vert \ \hat{\chi}_j(t=6) \  \vert \psi_{\rm WP} \rangle$}		&	\multicolumn{2}{c|}{\large $\langle \psi_{\rm vac}\vert \ \hat{\chi}_j(t=6) \  \vert \psi_{\rm vac} \rangle$}	&	\multicolumn{2}{c|}{\large $\mathcal{X}_j(t=6)$}		\\ \cline{2-7}
	&	MPS	&	{\tt ibm\_torino}	&	MPS	&	{\tt ibm\_torino}	&	MPS	&	{\tt ibm\_torino}		\\ \hline \hline
0	&	0.194	&	0.318(14)	&	0.194	&	0.310(16)	&	0.000	&	0.008(21)	\\
1	&	0.704	&	0.831(19)	&	0.704	&	0.848(22)	&	0.000	&	-0.018(28)	\\
2	&	0.4283	&	0.4284(68)	&	0.4283	&	0.4412(63)	&	0.0000	&	-0.0134(88)	\\
3	&	0.654	&	0.634(14)	&	0.654	&	0.657(14)	&	0.000	&	-0.022(15)	\\
4	&	0.604	&	0.540(14)	&	0.604	&	0.619(17)	&	0.000	&	-0.078(18)	\\
5	&	0.646	&	0.482(14)	&	0.646	&	0.470(15)	&	0.000	&	0.013(19)	\\
\hline
\end{tabularx}}
\caption{Numerical values for the $t=6$ chiral condensate of the wavepacket, the vacuum, and their subtraction, from the MPS simulator and {\tt ibm\_torino}, as shown in Figs.~\ref{fig:srcmvac_evol} and~\ref{fig:src_vac_evol} (only the first half of the lattice is shown due to CP symmetry).}
 \label{tab:tablet6}
\end{table}
\begin{table}[!htb]
\renewcommand{\arraystretch}{1}
\scalebox{0.95}{\begin{tabularx}{\textwidth}{| c || Y | Y | Y | Y | Y | Y |}
 \hline
\multirow{2}{*}{\parbox{2.5cm}{Fermion staggered site $j$}} &	\multicolumn{2}{c|}{\large $\langle \psi_{\rm WP}\vert \ \hat{\chi}_j(t=7) \  \vert \psi_{\rm WP} \rangle$}		&	\multicolumn{2}{c|}{\large $\langle \psi_{\rm vac}\vert \ \hat{\chi}_j(t=7) \  \vert \psi_{\rm vac} \rangle$}	&	\multicolumn{2}{c|}{\large $\mathcal{X}_j(t=7)$}		\\ \cline{2-7}
	&	MPS	&	{\tt ibm\_torino}	&	MPS	&	{\tt ibm\_torino}	&	MPS	&	{\tt ibm\_torino}		\\ \hline \hline
0	&	0.193	&	0.329(16)	&	0.193	&	0.327(15)	&	0.000	&	0.003(21)	\\
1	&	0.909	&	0.782(29)	&	0.909	&	0.744(26)	&	0.000	&	0.040(33)	\\
2	&	0.6241	&	0.3999(72)	&	0.6241	&	0.3610(48)	&	0.0000	&	0.0393(84)	\\
3	&	1.030	&	0.565(15)	&	1.030	&	0.560(15)	&	0.000	&	0.005(20)	\\
4	&	0.977	&	0.516(17)	&	0.977	&	0.601(19)	&	0.000	&	-0.087(25)	\\
5	&	0.981	&	0.432(12)	&	0.981	&	0.436(13)	&	0.000	&	-0.003(16)	\\
\hline
\end{tabularx}}
\caption{Numerical values for the $t=7$ chiral condensate of the wavepacket, the vacuum, and their subtraction, from the MPS simulator and {\tt ibm\_torino}, as shown in Figs.~\ref{fig:srcmvac_evol} and~\ref{fig:src_vac_evol} (only the first half of the lattice is shown due to CP symmetry).}
 \label{tab:tablet7}
\end{table}
\begin{table}[!htb]
\renewcommand{\arraystretch}{1}
\scalebox{0.95}{\begin{tabularx}{\textwidth}{| c || Y | Y | Y | Y | Y | Y |}
 \hline
\multirow{2}{*}{\parbox{2.5cm}{Fermion staggered site $j$}} &	\multicolumn{2}{c|}{\large $\langle \psi_{\rm WP}\vert \ \hat{\chi}_j(t=8) \  \vert \psi_{\rm WP} \rangle$}		&	\multicolumn{2}{c|}{\large $\langle \psi_{\rm vac}\vert \ \hat{\chi}_j(t=8) \  \vert \psi_{\rm vac} \rangle$}	&	\multicolumn{2}{c|}{\large $\mathcal{X}_j(t=8)$}		\\ \cline{2-7}
	&	MPS	&	{\tt ibm\_torino}	&	MPS	&	{\tt ibm\_torino}	&	MPS	&	{\tt ibm\_torino}		\\ \hline \hline
0	&	0.184	&	0.344(20)	&	0.184	&	0.241(13)	&	0.000	&	0.104(23)	\\
1	&	0.791	&	0.825(34)	&	0.791	&	0.692(19)	&	0.000	&	0.133(38)	\\
2	&	0.456	&	0.488(22)	&	0.4564	&	0.3586(74)	&	0.000	&	0.129(24)	\\
3	&	0.774	&	0.585(18)	&	0.774	&	0.543(12)	&	0.000	&	0.041(20)	\\
4	&	0.689	&	0.607(25)	&	0.689	&	0.597(16)	&	0.000	&	0.009(29)	\\
5	&	0.697	&	0.454(14)	&	0.6969	&	0.4378(97)	&	0.000	&	0.017(16)	\\
\hline
\end{tabularx}}
\caption{Numerical values for the $t=8$ chiral condensate of the wavepacket, the vacuum, and their subtraction, from the MPS simulator and {\tt ibm\_torino}, as shown in Figs.~\ref{fig:srcmvac_evol} and~\ref{fig:src_vac_evol} (only the first half of the lattice is shown due to CP symmetry).}
 \label{tab:tablet8}
\end{table}
\begin{table}[!htb]
\renewcommand{\arraystretch}{1}
\scalebox{0.95}{\begin{tabularx}{\textwidth}{| c || Y | Y | Y | Y | Y | Y |}
 \hline
\multirow{2}{*}{\parbox{2.5cm}{Fermion staggered site $j$}} &	\multicolumn{2}{c|}{\large $\langle \psi_{\rm WP}\vert \ \hat{\chi}_j(t=9) \  \vert \psi_{\rm WP} \rangle$}		&	\multicolumn{2}{c|}{\large $\langle \psi_{\rm vac}\vert \ \hat{\chi}_j(t=9) \  \vert \psi_{\rm vac} \rangle$}	&	\multicolumn{2}{c|}{\large $\mathcal{X}_j(t=9)$}		\\ \cline{2-7}
	&	MPS	&	{\tt ibm\_torino}	&	MPS	&	{\tt ibm\_torino}	&	MPS	&	{\tt ibm\_torino}		\\ \hline \hline
0	&	0.185	&	0.266(19)	&	0.185	&	0.230(20)	&	0.000	&	0.037(28)	\\
1	&	0.831	&	0.535(28)	&	0.831	&	0.526(25)	&	0.000	&	0.007(37)	\\
2	&	0.508	&	0.367(12)	&	0.508	&	0.356(12)	&	0.000	&	0.012(19)	\\
3	&	0.842	&	0.491(29)	&	0.842	&	0.499(24)	&	0.000	&	-0.009(33)	\\
4	&	0.763	&	0.633(44)	&	0.763	&	0.619(40)	&	0.000	&	0.014(62)	\\
5	&	0.883	&	0.474(29)	&	0.883	&	0.480(26)	&	0.000	&	-0.006(33)	\\
\hline
\end{tabularx}}
\caption{Numerical values for the $t=9$ chiral condensate of the wavepacket, the vacuum, and their subtraction, from the MPS simulator and {\tt ibm\_torino}, as shown in Figs.~\ref{fig:srcmvac_evol} and~\ref{fig:src_vac_evol} (only the first half of the lattice is shown due to CP symmetry).}
 \label{tab:tablet9}
\end{table}
\begin{table}[!htb]
\renewcommand{\arraystretch}{1}
\scalebox{0.95}{\begin{tabularx}{\textwidth}{| c || Y | Y | Y | Y | Y | Y |}
 \hline
\multirow{2}{*}{\parbox{2.5cm}{Fermion staggered site $j$}} &	\multicolumn{2}{c|}{\large $\langle \psi_{\rm WP}\vert \ \hat{\chi}_j(t=10) \  \vert \psi_{\rm WP} \rangle$}		&	\multicolumn{2}{c|}{\large $\langle \psi_{\rm vac}\vert \ \hat{\chi}_j(t=10) \  \vert \psi_{\rm vac} \rangle$}	&	\multicolumn{2}{c|}{\large $\mathcal{X}_j(t=10)$}		\\ \cline{2-7}
	&	MPS	&	{\tt ibm\_torino}	&	MPS	&	{\tt ibm\_torino}	&	MPS	&	{\tt ibm\_torino}		\\ \hline \hline
0	&	0.199	&	0.328(21)	&	0.199	&	0.259(22)	&	0.000	&	0.068(30)	\\
1	&	0.850	&	0.521(26)	&	0.850	&	0.467(32)	&	0.000	&	0.055(42)	\\
2	&	0.436	&	0.377(14)	&	0.436	&	0.403(14)	&	0.000	&	-0.026(20)	\\
3	&	0.837	&	0.402(20)	&	0.837	&	0.428(23)	&	0.000	&	-0.027(32)	\\
4	&	0.677	&	0.419(21)	&	0.677	&	0.377(17)	&	0.000	&	0.042(25)	\\
5	&	0.696	&	0.420(18)	&	0.696	&	0.417(17)	&	0.000	&	0.003(21)	\\
\hline
\end{tabularx}}
\caption{Numerical values for the $t=10$ chiral condensate of the wavepacket, the vacuum, and their subtraction, from the MPS simulator and {\tt ibm\_torino}, as shown in Figs.~\ref{fig:srcmvac_evol} and~\ref{fig:src_vac_evol} (only the first half of the lattice is shown due to CP symmetry).}
 \label{tab:tablet10}
\end{table}
\begin{table}[!htb]
\renewcommand{\arraystretch}{1}
\scalebox{0.95}{\begin{tabularx}{\textwidth}{| c || Y | Y | Y | Y | Y | Y |}
 \hline
\multirow{2}{*}{\parbox{2.5cm}{Fermion staggered site $j$}} &	\multicolumn{2}{c|}{\large $\langle \psi_{\rm WP}\vert \ \hat{\chi}_j(t=11) \  \vert \psi_{\rm WP} \rangle$}		&	\multicolumn{2}{c|}{\large $\langle \psi_{\rm vac}\vert \ \hat{\chi}_j(t=11) \  \vert \psi_{\rm vac} \rangle$}	&	\multicolumn{2}{c|}{\large $\mathcal{X}_j(t=11)$}		\\ \cline{2-7}
	&	MPS	&	{\tt ibm\_torino}	&	MPS	&	{\tt ibm\_torino}	&	MPS	&	{\tt ibm\_torino}		\\ \hline \hline
0	&	0.238	&	0.294(16)	&	0.238	&	0.275(17)	&	0.000	&	0.021(23)	\\
1	&	0.912	&	0.473(23)	&	0.912	&	0.494(30)	&	0.000	&	-0.020(37)	\\
2	&	0.479	&	0.372(12)	&	0.479	&	0.365(15)	&	0.000	&	0.008(19)	\\
3	&	0.765	&	0.364(20)	&	0.765	&	0.364(22)	&	0.000	&	-0.000(30)	\\
4	&	0.768	&	0.487(28)	&	0.768	&	0.411(24)	&	0.000	&	0.077(28)	\\
5	&	0.793	&	0.391(18)	&	0.793	&	0.423(15)	&	0.000	&	-0.033(19)	\\
\hline
\end{tabularx}}
\caption{Numerical values for the $t=11$ chiral condensate of the wavepacket, the vacuum, and their subtraction, from the MPS simulator and {\tt ibm\_torino}, as shown in Figs.~\ref{fig:srcmvac_evol} and~\ref{fig:src_vac_evol} (only the first half of the lattice is shown due to CP symmetry).}
 \label{tab:tablet11}
\end{table}
\begin{table}[!htb]
\renewcommand{\arraystretch}{1}
\scalebox{0.95}{\begin{tabularx}{\textwidth}{| c || Y | Y | Y | Y | Y | Y |}
 \hline
\multirow{2}{*}{\parbox{2.5cm}{Fermion staggered site $j$}} &	\multicolumn{2}{c|}{\large $\langle \psi_{\rm WP}\vert \ \hat{\chi}_j(t=12) \  \vert \psi_{\rm WP} \rangle$}		&	\multicolumn{2}{c|}{\large $\langle \psi_{\rm vac}\vert \ \hat{\chi}_j(t=12) \  \vert \psi_{\rm vac} \rangle$}	&	\multicolumn{2}{c|}{\large $\mathcal{X}_j(t=12)$}		\\ \cline{2-7}
	&	MPS	&	{\tt ibm\_torino}	&	MPS	&	{\tt ibm\_torino}	&	MPS	&	{\tt ibm\_torino}		\\ \hline \hline
0	&	0.214	&	0.396(41)	&	0.214	&	0.313(36)	&	0.000	&	0.082(54)	\\
1	&	0.838	&	0.505(36)	&	0.838	&	0.440(29)	&	0.000	&	0.066(44)	\\
2	&	0.419	&	0.369(11)	&	0.419	&	0.377(13)	&	0.000	&	-0.008(16)	\\
3	&	0.879	&	0.360(25)	&	0.879	&	0.406(29)	&	0.000	&	-0.049(39)	\\
4	&	0.675	&	0.401(22)	&	0.675	&	0.457(25)	&	0.000	&	-0.056(34)	\\
5	&	0.730	&	0.361(12)	&	0.730	&	0.412(17)	&	0.000	&	-0.051(19)	\\
\hline
\end{tabularx}}
\caption{Numerical values for the $t=12$ chiral condensate of the wavepacket, the vacuum, and their subtraction, from the MPS simulator and {\tt ibm\_torino}, as shown in Figs.~\ref{fig:srcmvac_evol} and~\ref{fig:src_vac_evol} (only the first half of the lattice is shown due to CP symmetry).}
 \label{tab:tablet12}
\end{table}
\begin{table}[!htb]
\renewcommand{\arraystretch}{1}
\scalebox{0.95}{\begin{tabularx}{\textwidth}{| c || Y | Y | Y | Y | Y | Y |}
 \hline
\multirow{2}{*}{\parbox{2.5cm}{Fermion staggered site $j$}} &	\multicolumn{2}{c|}{\large $\langle \psi_{\rm WP}\vert \ \hat{\chi}_j(t=13) \  \vert \psi_{\rm WP} \rangle$}		&	\multicolumn{2}{c|}{\large $\langle \psi_{\rm vac}\vert \ \hat{\chi}_j(t=13) \  \vert \psi_{\rm vac} \rangle$}	&	\multicolumn{2}{c|}{\large $\mathcal{X}_j(t=13)$}		\\ \cline{2-7}
	&	MPS	&	{\tt ibm\_torino}	&	MPS	&	{\tt ibm\_torino}	&	MPS	&	{\tt ibm\_torino}		\\ \hline \hline
0	&	0.217	&	0.250(19)	&	0.217	&	0.290(26)	&	0.000	&	-0.040(32)	\\
1	&	0.929	&	0.407(20)	&	0.929	&	0.432(26)	&	0.000	&	-0.024(35)	\\
2	&	0.496	&	0.365(19)	&	0.4956	&	0.3520(99)	&	0.000	&	0.012(23)	\\
3	&	0.817	&	0.355(18)	&	0.817	&	0.338(19)	&	0.000	&	0.018(25)	\\
4	&	0.818	&	0.441(30)	&	0.818	&	0.542(53)	&	0.000	&	-0.100(60)	\\
5	&	0.7731	&	0.3620(95)	&	0.773	&	0.382(14)	&	0.000	&	-0.021(17)	\\
\hline
\end{tabularx}}
\caption{Numerical values for the $t=13$ chiral condensate of the wavepacket, the vacuum, and their subtraction, from the MPS simulator and {\tt ibm\_torino}, as shown in Figs.~\ref{fig:srcmvac_evol} and~\ref{fig:src_vac_evol} (only the first half of the lattice is shown due to CP symmetry).}
 \label{tab:tablet13}
\end{table}
\begin{table}[!htb]
\renewcommand{\arraystretch}{1}
\scalebox{0.95}{\begin{tabularx}{\textwidth}{| c || Y | Y | Y | Y | Y | Y |}
 \hline
\multirow{2}{*}{\parbox{2.5cm}{Fermion staggered site $j$}} &	\multicolumn{2}{c|}{\large $\langle \psi_{\rm WP}\vert \ \hat{\chi}_j(t=14) \  \vert \psi_{\rm WP} \rangle$}		&	\multicolumn{2}{c|}{\large $\langle \psi_{\rm vac}\vert \ \hat{\chi}_j(t=14) \  \vert \psi_{\rm vac} \rangle$}	&	\multicolumn{2}{c|}{\large $\mathcal{X}_j(t=14)$}		\\ \cline{2-7}
	&	MPS	&	{\tt ibm\_torino}	&	MPS	&	{\tt ibm\_torino}	&	MPS	&	{\tt ibm\_torino}		\\ \hline \hline
0	&	0.216	&	0.250(25)	&	0.216	&	0.253(25)	&	0.000	&	-0.003(36)	\\
1	&	0.822	&	0.458(25)	&	0.822	&	0.439(23)	&	0.000	&	0.019(32)	\\
2	&	0.435	&	0.333(12)	&	0.435	&	0.375(13)	&	0.000	&	-0.042(18)	\\
3	&	0.928	&	0.313(24)	&	0.928	&	0.352(21)	&	0.000	&	-0.038(32)	\\
4	&	0.669	&	0.392(18)	&	0.669	&	0.441(28)	&	0.000	&	-0.048(34)	\\
5	&	0.749	&	0.387(13)	&	0.749	&	0.392(14)	&	0.000	&	-0.006(17)	\\
\hline
\end{tabularx}}
\caption{Numerical values for the $t=14$ chiral condensate of the wavepacket, the vacuum, and their subtraction, from the MPS simulator and {\tt ibm\_torino}, as shown in Figs.~\ref{fig:srcmvac_evol} and~\ref{fig:src_vac_evol} (only the first half of the lattice is shown due to CP symmetry).}
 \label{tab:tablet14}
\end{table}

\FloatBarrier
\bibliography{bibi}

\end{document}